%% file: ms.tex
\documentclass[preprint,12pt]{aastex}

\shorttitle{PBRS Outflows}
\shortauthors{Tobin et al.}

\newcommand{\kms}{\mbox{km s$^{-1}$}}

\begin{document}

%\title{Characterizing the Youngest Herschel-detected Protostars II. CO ($J=1\rightarrow0$) Outflows and Far-infrared Spectroscopy}
\title{Characterizing the Youngest Herschel-detected Protostars II. Molecular Outflows from the Millimeter and the Far-infrared\footnotemark}
\author{John J. Tobin\altaffilmark{2,3}, Amelia M. Stutz\altaffilmark{4}, P. Manoj\altaffilmark{5},
S. Thomas Megeath\altaffilmark{6}, Agata Karska\altaffilmark{7}, Zsofia Nagy\altaffilmark{6}, Friedrich Wyrowski\altaffilmark{8}, 
William Fischer\altaffilmark{6,9}, Dan M. Watson\altaffilmark{10},
Thomas Stanke\altaffilmark{11}}

\begin{abstract}

We present CARMA CO ($J=1\rightarrow0$) observations and \textit{Herschel} PACS spectroscopy, characterizing
the outflow properties toward extremely young and deeply embedded protostars in the Orion
molecular clouds. The sample comprises a subset of the Orion protostars known as the 
PACS Bright Red Sources (PBRS) (Stutz et al.). We observed 14 PBRS with CARMA and 8 of 
these 14 with \textit{Herschel}, acquiring full spectral scans 
from 55~\micron\ to 200~\micron. Outflows are detected 
in CO ($J=1\rightarrow0$) from 8 of 14 PBRS, with two additional 
tentative detections; outflows are also detected from the
outbursting protostar HOPS 223 (V2775 Ori) and the Class I protostar HOPS 68. 
The outflows have a range of morphologies, some are
spatially compact, $<$10000 AU in extent, while others extend beyond the primary beam.
The outflow velocities 
and morphologies are consistent with being dominated
by intermediate inclination angles (80\degr~$\ge$~$i$~$\ge$20\degr). This 
confirms the interpretation of the very red 24~\micron\ to 70~\micron\ colors of the PBRS 
as a signpost of high envelope densities, with only one (possibly two)
cases of the red colors resulting from edge-on inclinations.
We detect high-J (J$_{up}$~$>$~13) CO lines and/or H$_2$O lines from 5 of 8 PBRS and only for
those with detected CO outflows.
The far-infrared CO rotation temperatures of the detected PBRS are marginally
colder ($\sim$230~K) than those observed for most protostars ($\sim$300~K), and only one
of these 5 PBRS has detected [OI] 63~\micron\ emission. The high envelope densities could
be obscuring some [OI] emission and cause a $\sim$20~K reduction to the CO
rotation temperatures.

\end{abstract}

%The upper limits for [OI] luminosity are
%not significantly lower than other protostars with comparable bolometric luminosities and
%the high-J CO luminosities are consistent with other observed samples.

\footnotetext[1]{{\it Herschel} is an ESA space observatory with science instruments provided by European-led Principal Investigator consortia and with important participation from NASA.}
\altaffiltext{2}{Veni Fellow, Leiden Observatory, Leiden University, P.O. Box 9513, 2300-RA Leiden, The Netherlands; tobin@strw.leidenuniv.nl}
\altaffiltext{3}{National Radio Astronomy Observatory, Charlottesville, VA 22903, USA}
\altaffiltext{4}{Max-Planck-Institut f\"ur Astronomie, D-69117 Heidelberg, Germany}
\altaffiltext{5}{Department of Astronomy and Astrophysics, Tata Institute of Fundamental Research, Colaba, Mumbai 400005, India}
\altaffiltext{6}{Ritter Astrophysical Research Center, Department of Physics and Astronomy, University of Toledo, Toledo, OH 43560}
\altaffiltext{7}{Centre for Astronomy, Nicolaus Copernicus University, Faculty of Physics, Astronomy and Informatics, Grudziadzka 5, PL-87100 Torun, Poland}
\altaffiltext{8}{Max-Planck-Institut f\"ur Radioastronomie, Auf dem H\"ugel 69, 53121, Bonn, Germany}
\altaffiltext{9}{NASA Goddard Space Flight Center, 8800 Greenbelt Road, Greenbelt, MD 20771}
\altaffiltext{10}{Department of Physics and Astronomy, University of Rochester, Rochester, NY 14627}
\altaffiltext{11}{European Southern Observatory, 85748 Garching bei M\"unchen, Germany}

\section{Introduction}

The earliest stage of the star formation process is characterized by a
dense, infalling envelope of gas and dust surrounding a nascent protostar. 
This early phase, in particular, is known to be associated with powerful outflows
 \citep{arce2007,frank2014}. These outflows may ultimately play a role in 
halting the mass infall process and dispersing the envelope \citep{arce2006}, thereby contributing
to the overall low efficiency of the star formation process \citep{offner2014}. These outflows
develop rapidly and with velocities of $\sim$10 - 100 \kms\ the outflows may propagate by 0.1 pc in 10,000 yr - 1,000 yr
timescales. Therefore, outflows are important to characterize at the youngest possible ages
in order to understand their early evolution.

The youngest identified protostars are known as Class 0 sources \citep{andre1993}; they 
are distinguished from more-evolved Class I sources by their cold bolometric temperatures
(T$_{bol}$ $<$ 70 K; \citep{myers1993}) and/or ratio of submillimeter luminosity (L$_{submm}$) 
to bolometric luminosity 
($L_{bol}$) being $>$ 0.5\%. These diagnostics indicate that Class 0 sources typically have 
denser and more massive infalling envelopes than Class I sources. In addition to the Class 0
sources, an earlier phase of the star formation process has been postulated, the 
first hydrostatic cores \citep[FHSC; e.g.,][]{larson1969}. A number of candidate FHSCs
have been identified \citep{enoch2010,chen2010, pineda2011, schnee2012}; moreover, candidate
FHSCs have quite low luminosities and bear some similarity to the \textit{Spitzer}-identified very 
low-luminosity sources \citep[VeLLOs][]{young2004,dunham2006}. The exact nature of the 
VeLLOs and candidate FHSCs remains unclear as it is difficult to distinguish
bonafide FHSCs from sources that will go on to form very low mass stars \citep{dunham2014}.

As part of the \textit{Herschel} Orion Protostar Survey (HOPS) 
\citep[e.g.,][]{fischer2010,stanke2010,ali2010,manoj2013,furlan2016}, a sample of
19 protostars with bright 70 \micron\ and 160 \micron\ emission and correspondingly faint or undetected
(8 sources) 24 \micron\ emission were detected in the Orion star forming region \citep[][ hereafter ST13]{stutz2013}. 
We refer to these 
protostars
as the PACS Bright Red Sources (PBRS);
of the 19 PBRS, 12 were first identified as protostars 
by \textit{Herschel} and 7 \textit{Spitzer}-identified protostars 
also fulfilled the 24~\micron\ to 70~\micron\ color criteria (ST13).
The PBRS are \textit{not} low-luminosity like the VeLLOs and candidate FHSCs;
they have bolometric luminosities (L$_{bol}$) ranging between 0.65 L$_{\sun}$ 
and 30.6 L$_{\sun}$, with a median L$_{bol}$ of $\sim$3 L$_{\sun}$. Thus, the PBRS are the largest sample
of extremely young protostars with typical luminosities;
the median luminosity of Class 0 protostars is
 3.5~L$_{\sun}$ in Orion and 1.4 L$_{\sun}$ in the nearby clouds \citep{dunham2014}.
While the PBRS have only been 
well-characterized in Orion, similar examples are present in more nearby clouds (e.g., VLA 1623, IRAS 16293-2422),
and \citet{sadavoy2014} identified several protostars in Perseus that were not classified as protostars in 
\textit{Spitzer} or undetected at 24~\micron\ \citep[i.e., HH211-mms][]{rebull2007}.

%$[24\,\mu{\rm m}]-[70\,\mu{\rm m}]$ colors  (in log ($\lambda$F$_{\lambda}$ space) redder than 1.65.

We further characterized the envelopes of 14 PBRS using observations of the 2.9 mm dust continuum
\citep{tobin2015}; that study, hereafter Paper I, focused specifically
on the most deeply embedded and \textit{Herschel}-identified sources. 
The observed PBRS were all detected and found to have among the largest 2.9 mm luminosities of known Class 0
protostars. We  also found that 6 out of 14 have visibility amplitudes that are flat within increasing
uv-distance. The flat visibility amplitudes indicate that the 2.9 mm emission is very concentrated, and this finding,
together with the high 2.9 mm luminosities, confirms that most PBRS have dense envelopes. 
This corroborates the interpretation of the
spectral energy 
distribution (SED) model comparisons in ST13. The characterization of the PBRS from both the SEDs and
millimeter continuum have led us to conclude that the PBRS may be among the youngest Class 0 objects.
If the PBRS represent a distinct portion of early
Class 0 evolution, as suggested by ST13, then the relative numbers of PBRS to Class 0 sources in Orion indicates
that a `PBRS phase' could last $\sim$25,000 yr. This estimate assumes that the Class 0 phase 
lasts $\sim$150,000 yr \citep{dunham2014}.

A remaining source of uncertainty in the interpretation of the PBRS as the youngest Class 0 
protostars is their unknown disk/envelope inclination angles with respect to the plane of 
the sky. There is a degeneracy 
between high envelope densities versus high (nearly edge-on) inclinations that could not be mitigated 
due to the lack of emission shortward of 10 \micron\ toward most PBRS \citep[e.g.,][]{whitney2003,furlan2016}.
Assuming that outflows are perpendicular to the disk or envelope midplanes,
observations of outflows to constrain their orientations (e.g., in molecular
lines) are an excellent way to estimate disk/envelope inclinations and further
constrain the envelope properties.
Furthermore, if the PBRS are among the youngest Class 0 protostars, then the sample as a whole
represents an opportunity to examine the outflow properties of the youngest protostars.

The jets and outflows from protostars are detected with a variety of complementary methods
and the types of outflows and the ways to detect them also vary with evolution.
Collimated jets detected in optical or near-infrared line emission are typically associated
with more evolved Class I or Class II sources \citep[e.g., HH111]{reipurth1997,reipurth2010}, while Class 0 protostars
typically have a molecular outflow observable in only millimeter lines of CO and other molecules 
\citep{arce2007, frank2014}. However, this does not 
mean there is no collimated jet emission, just that it may be undetectable due to high levels
of obscuration. The molecular outflow emission toward some low-mass protostars
 has an angular dependence of velocity, with low-velocity material 
at the edges of the outflow cavity and velocities as high as $\sim$ 100 \kms\ along the main axis of the
outflow \citep[e.g.,][]{santiago2009,hirano2011}. Jet-like features can also be
seen in shock-tracing molecules such as SiO and SO \citep[e.g.,][]{lee2008,lee2009}. The velocity
gradients along the outflow axis also offer crucial information 
of disk-protostar orientation \citep[e.g.,][]{cabrit1986, lee2000}.

Far-infrared spectroscopy with the \textit{Infrared Space Observatory} and
the \textit{Herschel Space Observatory} has also been found to be an 
excellent probe of the physical conditions of outflows from young stars. 
The high-J CO (J$_u$ $>$ 13) and H$_2$O transitions, in addition
to OH and [OI] transitions, probe the warm and hot outflow conditions on scales 
very near the protostar and the jet driving source \citep[e.g., ][]{vankempen2010,karska2013,
green2013,manoj2013}. The lines are thought to be excited primarily by shocks \citep{manoj2013}, with
UV radiation photo-dissociating H$_2$O, causing lower abundances relative to non-irradiated shock models
\citep{karska2014}. 

The initial development of the outflows and their subsequent breakout from their
surrounding envelopes are still quite uncertain. Outflows have also been detected
from VeLLOs and candidate FHSCs \citep{dunham2011, pineda2011, 
schnee2012, tobin2015}. Theory has predicted that such young objects can indeed produce 
the slow outflows ($\sim$2 - 7 \kms) that have been observed \citep{price2012}, and the outflows may 
develop prior to the formation of a rotationally-supported accretion disk 
\citep[e.g.,][]{li2013,li2014}. However, it is still uncertain how quickly
more powerful outflows emerge in protostars; do the outflows have a steady growth in power
as the source luminosity (from accretion) increases or do they only become powerful
once a certain threshold in luminosity is reached?

In order to examine the outflow conditions from the youngest known Class 0 protostars,
we have obtained interferometric observations of the CO ($J=1\rightarrow0$) molecular line and
far-infrared spectroscopy with the \textit{Herschel Space Observatory} toward the
PBRS in the Orion A and B molecular clouds. The youth and number of PBRS sources 
in Orion offers an unique opportunity to examine the
properties of outflows toward objects that are consistent with being among the
youngest protostars. Furthermore,
spectrally and spatially resolved observations of the molecular outflows toward these protostars
will enable us to constrain the range of possible inclination angles of the protostellar sources, ensuring that
their characterization as the youngest protostars is not strongly influenced by orientation.

We have observed 14 PBRS (from the full sample of 19 cataloged by \citet{stutz2013} and Paper I)
with the Combined Array for Research in Millimeter-wave Astronomy (CARMA), 
focusing on the \textit{Herschel}-detected PBRS sample. We observed the protostars in both the dust
continuum and spectral line emission to examine the envelope and outflow properties of these sources. We discuss
the observations in Section 2, our outflow results from CO ($J=1\rightarrow0$) and \textit{Herschel} spectroscopy
are presented in Section 3, we discuss the results in Section 4, and summarize our main conclusions in Section 5.

\section{Observations and Data Reduction}

\subsection{CARMA Observations}

We conducted observations toward 14 out of 19 of the PBRS identified in ST13 
with CARMA in the D-configuration  ($\sim$5\arcsec\ resolution) during late 2012 and early 2014 and follow-up
observations in C-configuration ($\sim$2 \arcsec\ resolution) for some in early 2014.
The observations were conducted with the main CARMA array comprised of 6 - 10.4 m and
9 - 6.1 m antennas. We observed two or three sources per track and
configured the correlator with four 500 MHz continuum windows, two 8 MHz windows to observe
para-NH$_2$D ($J=1_{11}\rightarrow1_{01}$) and C$^{18}$O ($J=1\rightarrow~0$), and 
the two 31 MHz windows for observation of $^{13}$CO ($J=1\rightarrow0$) and $^{12}$CO ($J=1\rightarrow0$).
The C-configuration observations had five 500MHz continuum windows because we did not observe
para-NH$_2$D in that configuration. The continuum observations were presented in \citet{tobin2015}
and here we will present only the $^{12}$CO ($J=1\rightarrow0$) results because
other lines did not yield strong detections.
Our sensitivity is typically 0.15 Jy beam$^{-1}$ channel $^{-1}$ 
for the CO ($J=1\rightarrow0$) in 0.5 \kms\ channels. We used standard procedures within the MIRIAD software package \citep{sault1995}
to edit, reduce, and image the data; all maps were reconstructed with natural weighting.
The CARMA observation log is given in Table 1. The absolute flux
calibration uncertainty is $\sim$10-20\%. The largest angular scale that can be
recovered from observations is $\sim$20\arcsec; we estimate this number to be twice the
minimum baseline length.

\subsection{\textit{Herschel} PACS Spectroscopy Observations}

We also observed 8 PBRS sources with the Photodetector Array Camera and Spectrometer \citep[PACS;][]{poglitsch2010}
on the \textit{Herschel Space Observatory} \citep{pilbratt2010} as part of program OT2\_jtobin\_2; we also observed 
the Class I protostar HOPS 347. The PACS spectrometer is a far-infrared integral field spectrograph
with a 5$\times$5 spaxel (spatial pixel) foot print and spaxel sizes of 9\arcsec, for more information see \citet{poglitsch2010}.

We conducted full range scans of the entire spectral range
from $\sim$55~\micron\ to $\sim$ 200~\micron\ in standard chop-nod mode. Table 2 lists
the observations dates and observations ids for the observed sources.
The PACS range scan spectra were reduced using HIPE 13.0 SPG v11.1.0, 
calibration tree version 56. The root-mean-squared absolute flux calibration uncertainty of
the PACS spectra is $\sim$12\%.

The line spectroscopy observations of the [OI] 63.18~\micron\ transition were conducted
in unchopped mode. The unchopped mode uses separately defined off positions away from
the cloud to prevent corrupting the [OI] line with a contaminated off position in chop-nod mode.
This mode was necessary because
extended [OI] emission is very prevalent in the Orion molecular cloud. The use of unchopped mode
will, however, result in foreground/background [OI] emission on the surrounding molecular cloud
being preserved, in addition to that of the protostar itself.
These observations
were taken in bright line mode, which has less redundancy at each wavelength than faint line mode.
The data used in this paper are the from the default archive reduction
from science product generation version 12.1.0 and utilizing PACS calibration tree version 65.

In this paper, we are making use of the flux densities derived from the central 
spaxel, corrected for the point spread function losses. For flat-fielding, we use
the observed relative spectral response function (RSRF) rather that the telescope
background method.

\subsection{Magellan Near-infrared Observations}

We observed the source HOPS 68 with the Magellan Baade telescope, located at Las Campanas in Chile on
2009 January 17. The observations were conducted with the Persson Auxiliary Nasmyth Infrared Camera
\citep[PANIC, ][]{martini2004}, which has 2\arcmin\ $\times$ 2\arcmin\ field of view on a 
1024 $\times$ 1024 pixel detector. HOPS 68 was observed in Ks-band using a 3$\times$3 dither pattern with
20 second integrations at each dither position and 15\arcsec\ steps between dither positions. The sky image
was constructed from a median combination of the on-source frames thereby losing some large-scale emission.
The data were reduced using the Image Reduction and Analysis Facility (IRAF) using standard methods
for near-infrared imaging observations; see \citet{tobin2010} for a description of the methods used.

\subsection{Sample and Sub-samples}

The observations and results presented in this paper are based on sub-samples of the PBRS sample
presented in ST13. ST13 identified 18 sources with $[24\,\mu{\rm m}]-[70\,\mu{\rm m}]$ colors 
(in log ($\lambda$F$_{\lambda}$) space) redder than 1.65. Of this sample, 11 were first discovered with
\textit{Herschel} observations and 7 were previously known HOPS protostars from the \textit{Spitzer} surveys
of the region that met the redness criteria. Furthermore, an additional PBRS (135003) was not included in ST13, but was first presented in Paper I,
bringing the total number of PBRS to 19. We list the full sample of PBRS in Table 3 and identify those that have been followed-up with
CARMA and \textit{Herschel} PACS Spectroscopy. The CARMA follow-up concentrated primarily on sources that
had not been previously identified by \textit{Spitzer} as protostars due to their deeply embedded nature, rendering them
faint or undetected at 24~\micron. The \textit{Herschel} PACS spectroscopy then concentrated on the \textit{Herschel}-identified 
PBRS that had been found in the HOPS data that had been analyzed prior to
the \textit{Herschel} Open Time 2 proposal deadline. Thus, our source follow-up
is not homogeneous, but there is enough overlap in order to identify characteristic
trends within the sample and sub-samples which we will detail in the following sections.

\section{Results}

We have compiled a significant amount of data to further characterize
the PBRS and their outflow properties. We will first discuss the cold 
molecular outflows probed by CARMA CO ($J=1\rightarrow0$) and probe scales beyond 
those examined by CARMA using \textit{Spitzer}
4.5~\micron\ emission. Lastly, we will discuss the results for the 
warm and hot components of the molecular outflows
with \textit{Herschel} PACS spectroscopy and place the properties 
of the PBRS outflows in the context of larger protostar samples observed with far-infrared 
spectroscopy. While the three datasets do not cover the same samples (see Table 3) and the
spatial scales examined are different, they all contribute to a deeper 
understanding of the PBRS than when considered on their own. We will attempt to concentrate
on overarching trends in the following discussion of results and the discussion of individual sources can be
found in the Appendix.

\subsection{Molecular Outflows}

The $^{12}$CO ($J=1\rightarrow0$) molecular line was observed to examine the outflow activity toward
each source; this is the canonical tracer of outflowing gas toward protostellar objects \citep{snell1980}.
Outflows are generally characterized by distinct red and blue-shifted emission located on either
side of the protostellar source, modulo inclination effects. The pervasiveness of CO in the Orion molecular cloud
complicates analysis of outflows. Emission at $\pm$2 \kms\ around the systemic velocity cannot be analyzed due to 
the $^{12}$CO ($J=1\rightarrow0$) emission being resolved-out due to confusion with the extended molecular 
cloud. Therefore, we are generally only able to detect outflow features that have velocities high enough to emit
outside the $\pm$2 \kms\ velocity range.

\subsubsection{Detections and Morphologies}
\begin{figure}

\begin{center}
\includegraphics[scale=0.4,trim=3cm 1.25cm 5cm 1cm, clip=true]{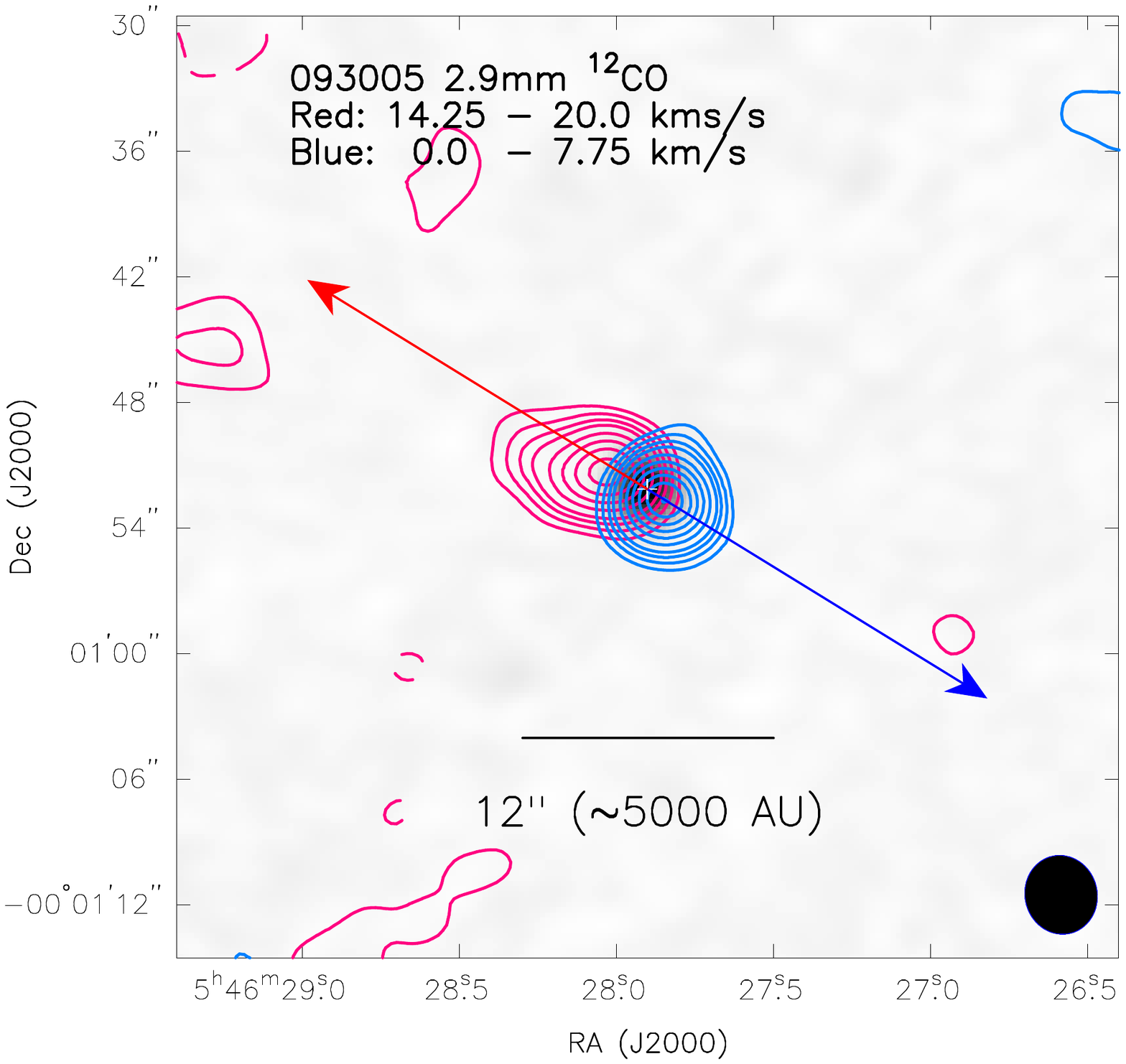}
\includegraphics[scale=0.4,trim=3cm 1.25cm 5cm 1cm, clip=true]{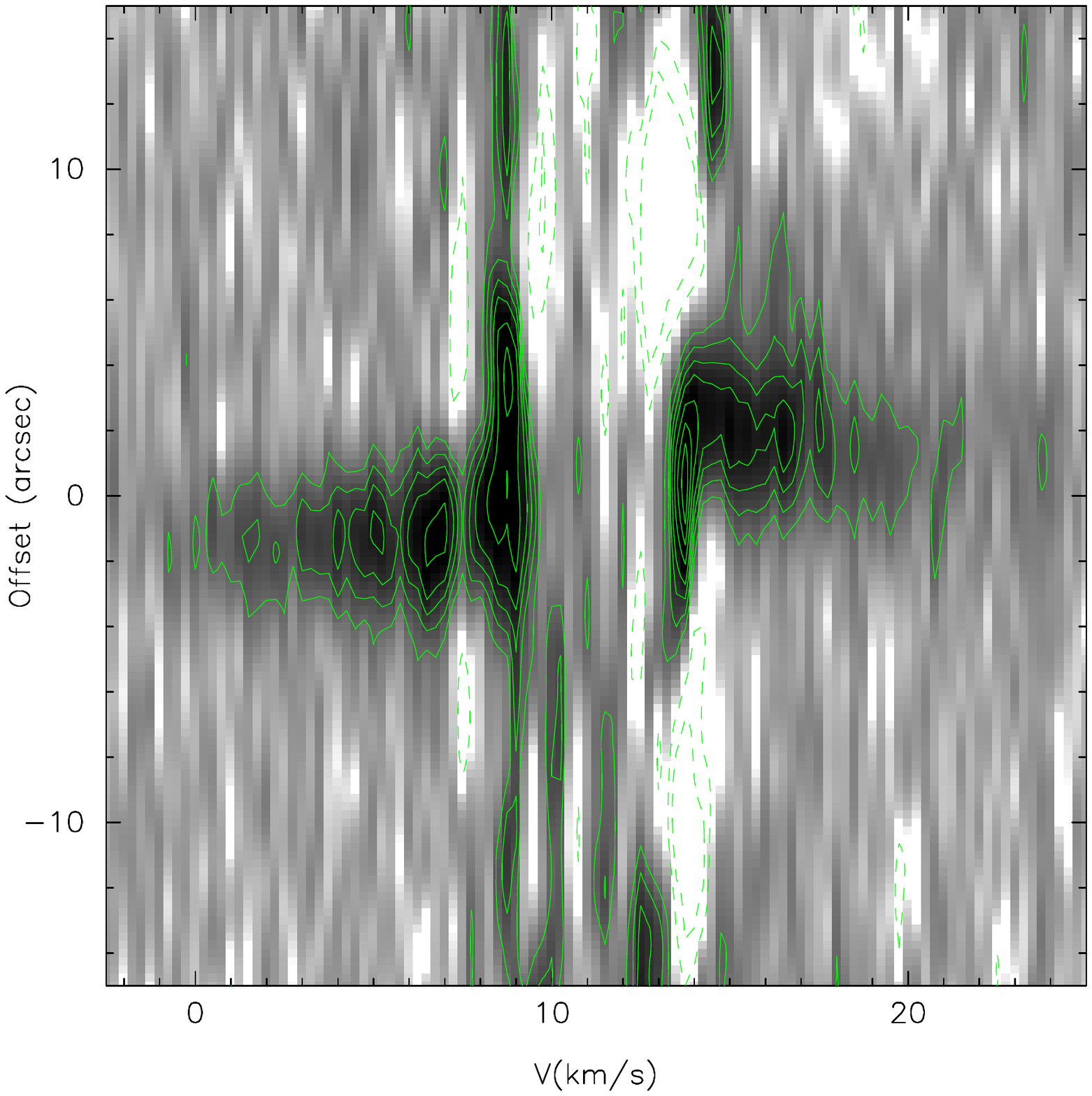}

\end{center}
\caption{PBRS 093005 -- Red and blue-shifted $^{12}$CO ($J=1\rightarrow0$) integrated 
intensity contours are overlaid on the 2.9 mm continuum image from combined D and C configuration
data. 
The compact outflow is centered on the continuum source; 
the position velocity plot on the right shows the
velocity distribution of the $^{12}$CO emission along the red 
and blue vectors marking the outflow axis. The cross marks the position of the
protostar inferred from the 2.9~mm continuum. The PV
plot shows that the emission is compact and confined to regions near the source.
 The contours in the integrated intensity plot
are [$\pm$6, 9, 12, 15, 20, 30, 40, 50, 60] $\times$ $\sigma$ for the blue and the red; 
$\sigma_{red}$ = 0.84 K and $\sigma_{blue}$ = 0.72 K.
The PV plot contours are [-6, -3, 3, 5, 7, 9, 12, 15, 18, 21, 24, 27, 30] 
$\times$ $\sigma$ and $\sigma$ = 0.65 K.
 }
\label{093005}
\end{figure}

\begin{figure}

\begin{center}
\includegraphics[scale=0.4,trim=3cm 1.25cm 5cm 1cm, clip=true]{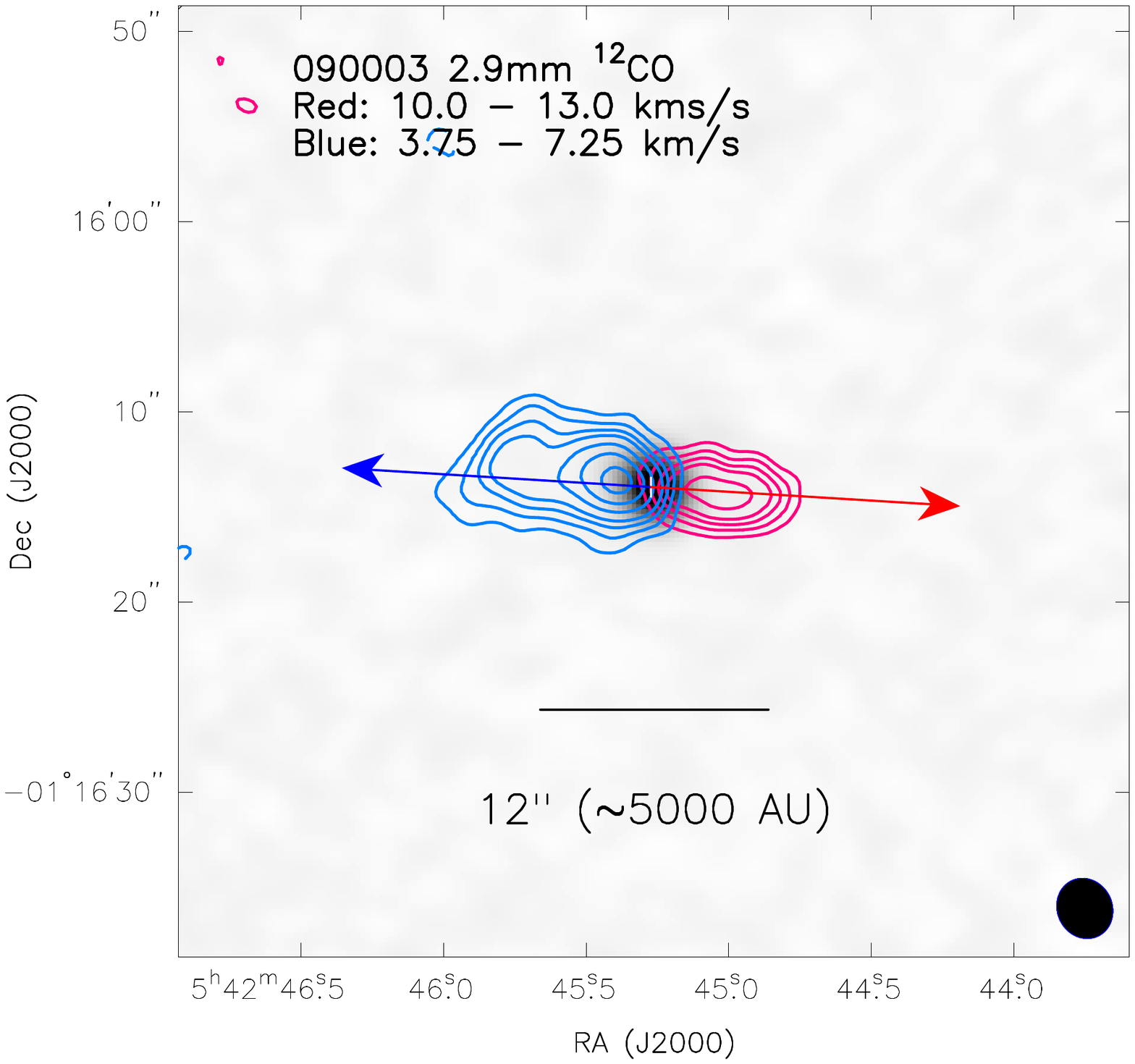}
\includegraphics[scale=0.4,trim=3cm 1.25cm 5cm 1cm, clip=true]{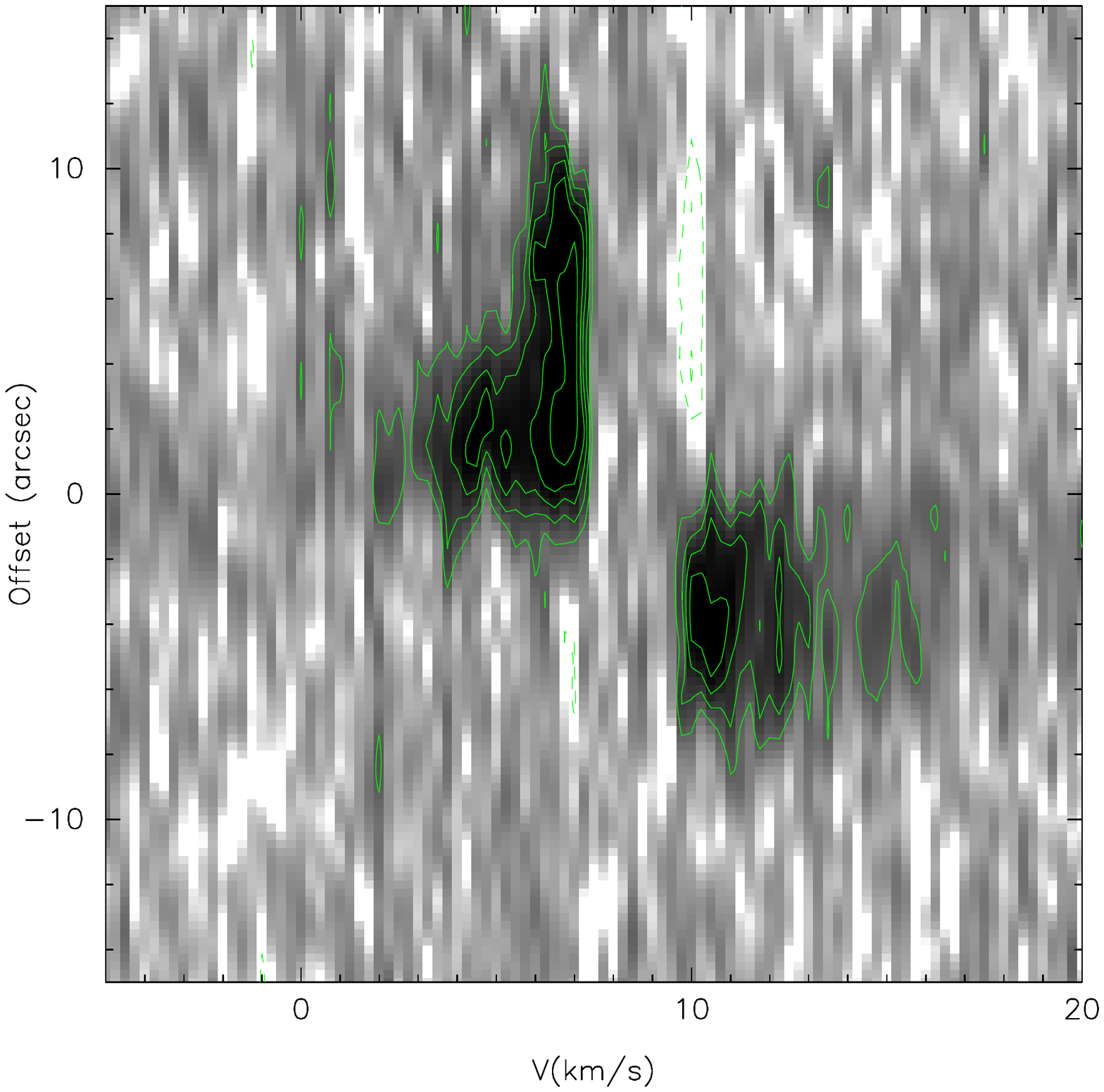}
\end{center}
\caption{PBRS 090003 -- Same as Figure \ref{093005}; the outflow toward this source is similar 
to 093005 in that it is compact and collimated. But,
there is a bit more spatial separation between the high-velocity components.
The contours in the integrated intensity plot
are [$\pm$6, 9, 12, 15 ,20, 25, 30, 35, 40, 45, 50] $\times$ $\sigma$ for the blue and the red; 
$\sigma_{red}$ = 0.58 K and $\sigma_{blue}$ = 0.66 K.
The PV plot contours are [-6, -3, 3, 5, 7, 9, 12, 15, 18, 21, 24, 27, 30] $\times$ $\sigma$ and $\sigma$ = 0.4 K.
}
\label{090003}
\end{figure}

\begin{figure}

\begin{center}
\includegraphics[scale=0.4,trim=3cm 1.25cm 5cm 1cm, clip=true]{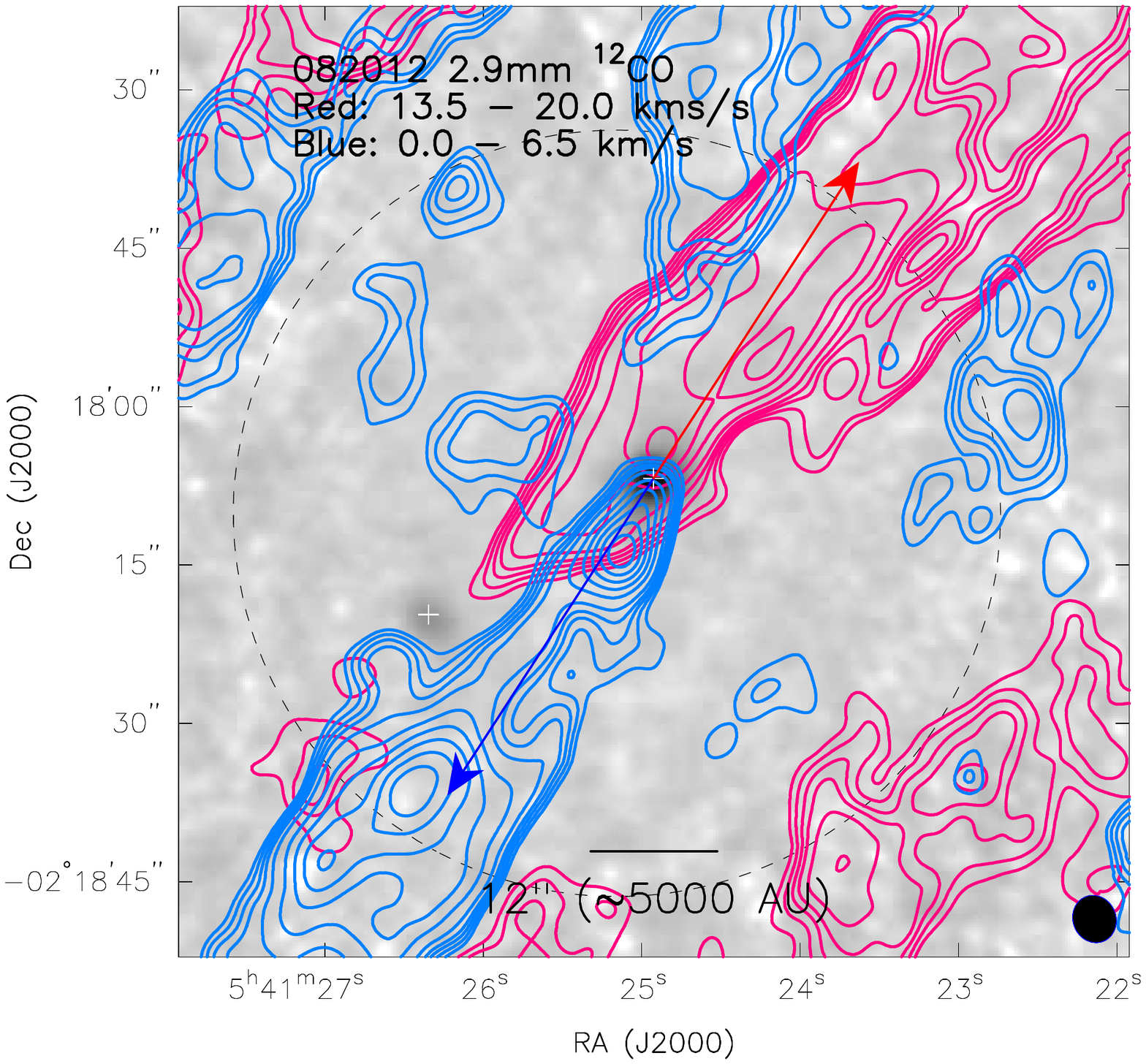}
\includegraphics[scale=0.4,trim=3cm 1.25cm 5cm 1cm, clip=true]{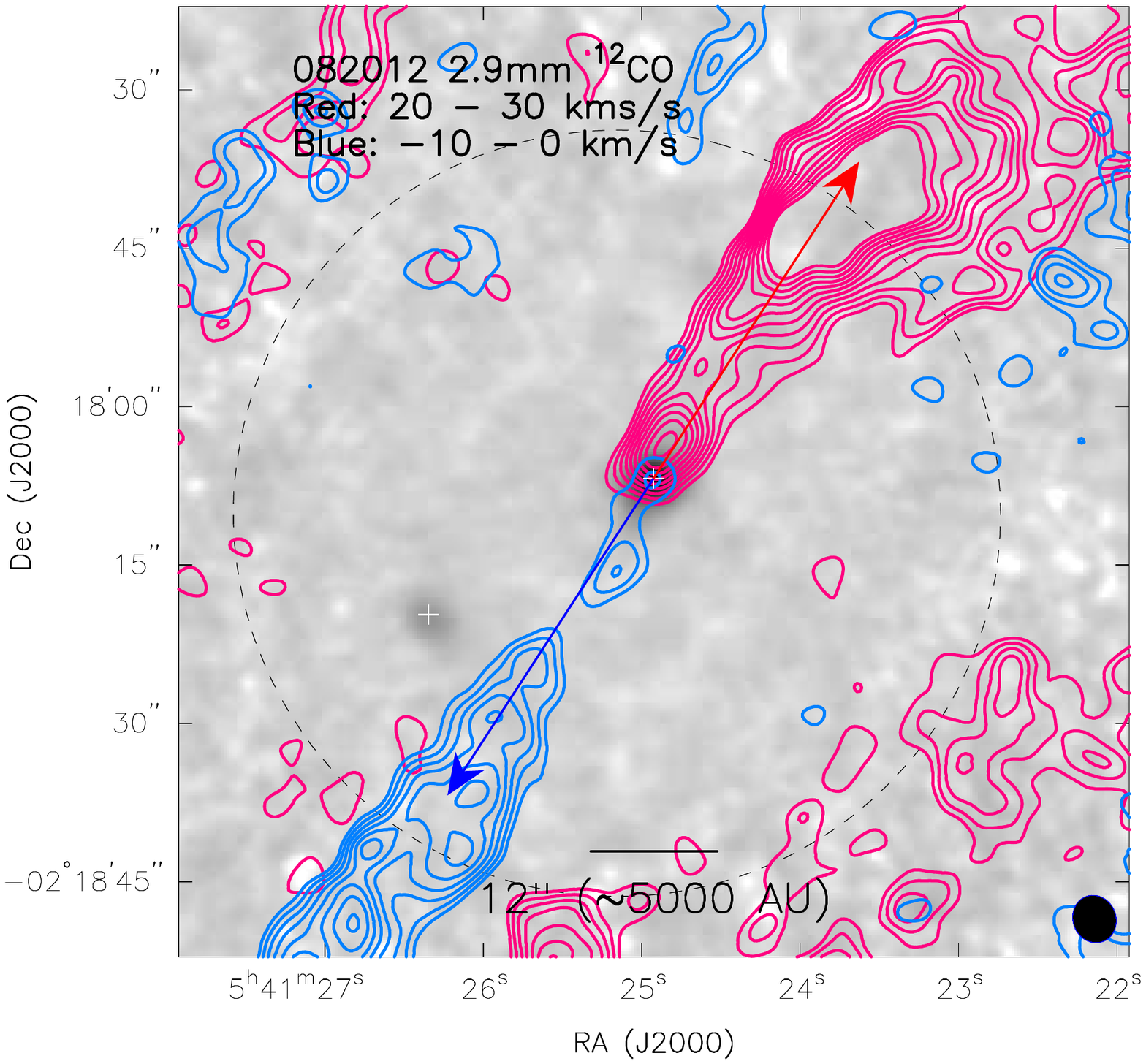}
\includegraphics[scale=0.4,trim=3cm 1.25cm 5cm 1cm, clip=true]{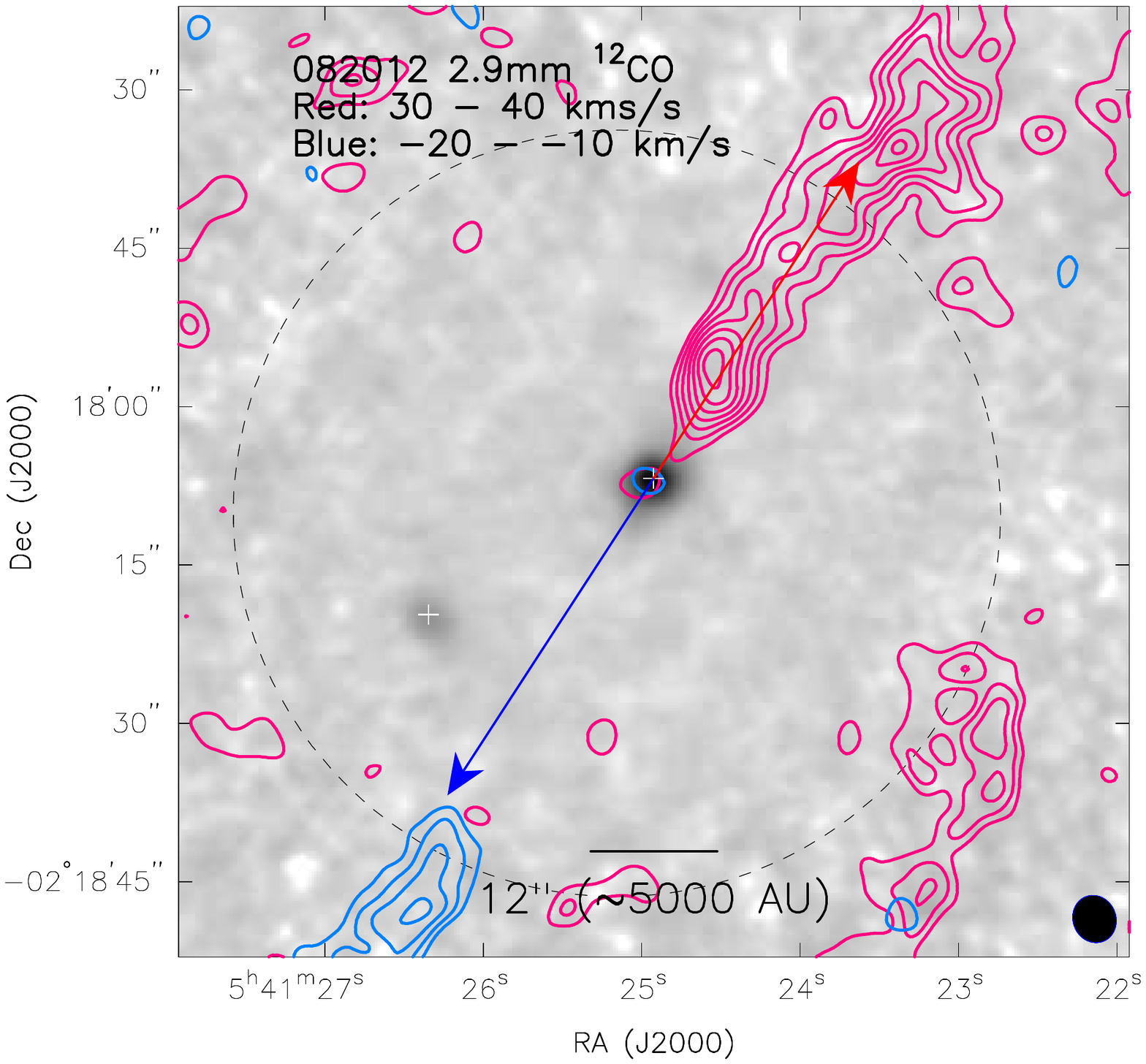}
\includegraphics[scale=0.4,trim=3cm 1.25cm 5cm 1cm, clip=true]{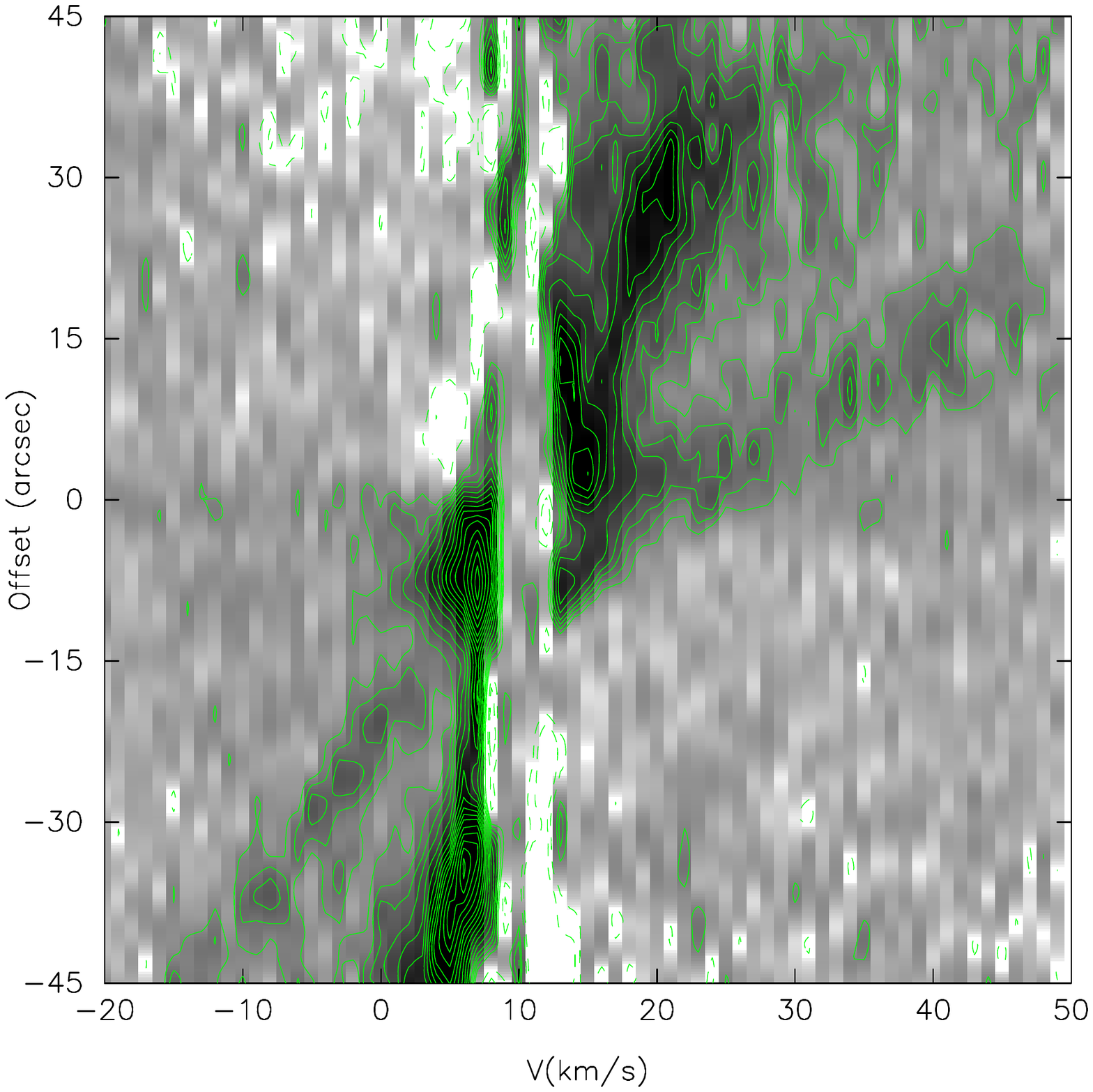}

\end{center}
\caption{PBRS 082012 -- Same as Figure \ref{093005}, showing both 082012 
(center cross) and HOPS 372 (southeast cross). The outflow from 082012 is a very strong, 
collimated outflow that is extended to a large distance away from the protostar.
We show three plots at low (0 - 20 \kms), moderate (20 - 30 \kms; -10 - 0 \kms), and high (30 - 40 \kms; -20 - -10 \kms). 
The blue-shifted side disappears at velocities higher than
-10 \kms and the outflow becomes more narrow and jet like at the higher velocities. 
The low velocity emission appears to trace the combination of an outflow from HOPS 372 and 082012, with 082012 being dominant.
Then at higher velocities, the 082012 outflow is most apparent and HOPS 372 does not seem to contribute.
The PV plot clearly shows the high and low velocity components of the outflow.
The contours in the low velocity plot are [$\pm$6, 9, 12, 15 ,20, 30, 40, 50, 60, 70] $\times$ 
$\sigma$ for the blue and the red; $\sigma_{red}$ = 1.27 K and $\sigma_{blue}$ = 1.27 K. 
For the remaining plots the contours are [-6, -4, 4, 6, 8, ... ,24]$\sigma$; 
$\sigma_{red}$ and $\sigma_{blue}$ = 1.56 K.
The PV plot contours are [-6, -3, 3, 5, 7, 9, 12, 15, 18, 21, 24, 27, 30, 35] 
$\times$ $\sigma$ and $\sigma$ = 0.35 K. The half-power point of
the primary beam is plotted as the dashed circle.
}
\label{082012}
\end{figure}

\begin{figure}

\begin{center}
\includegraphics[scale=0.4,trim=3cm 1.25cm 5cm 1cm, clip=true]{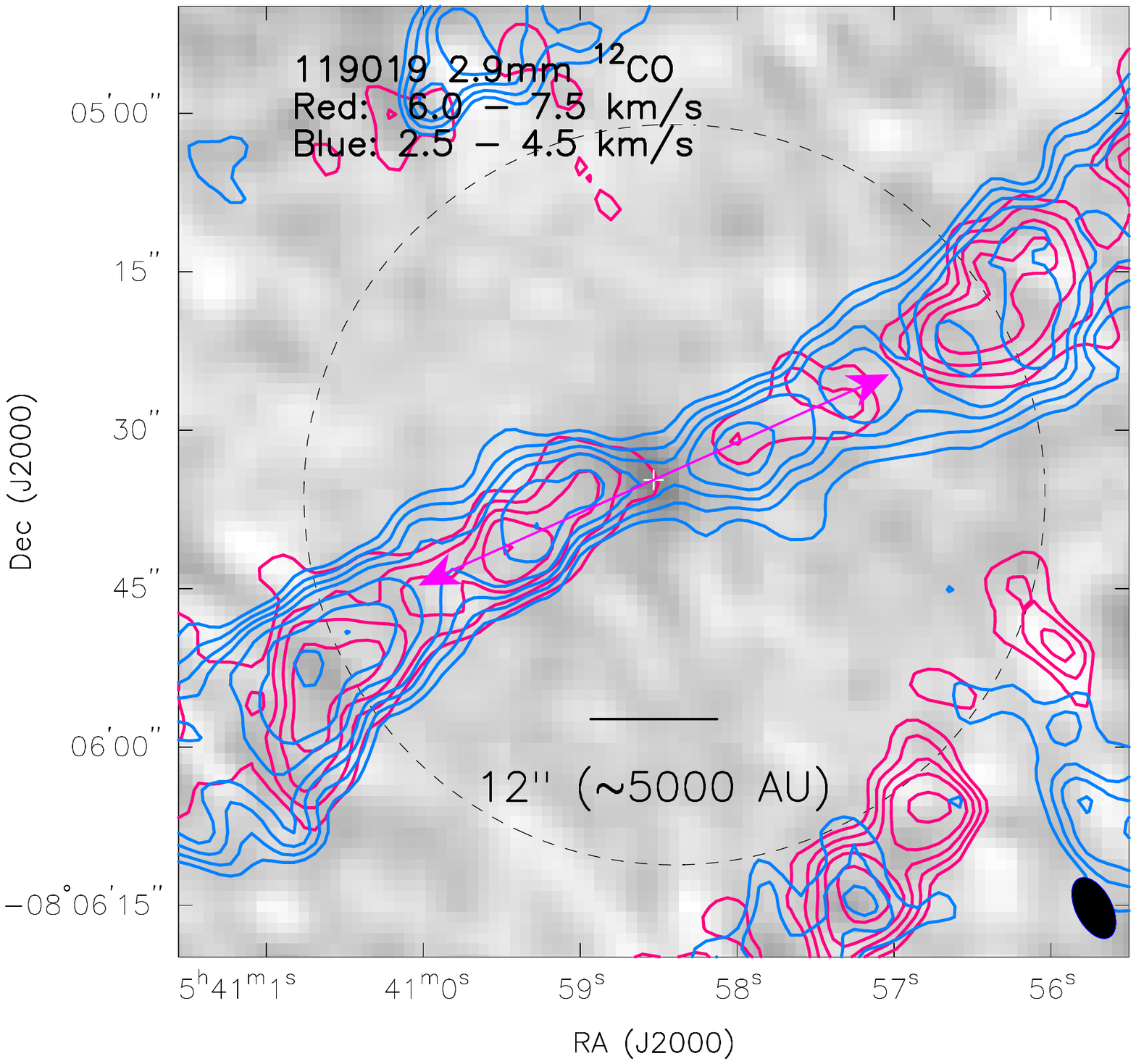}
\includegraphics[scale=0.4,trim=3cm 1.25cm 5cm 1cm, clip=true]{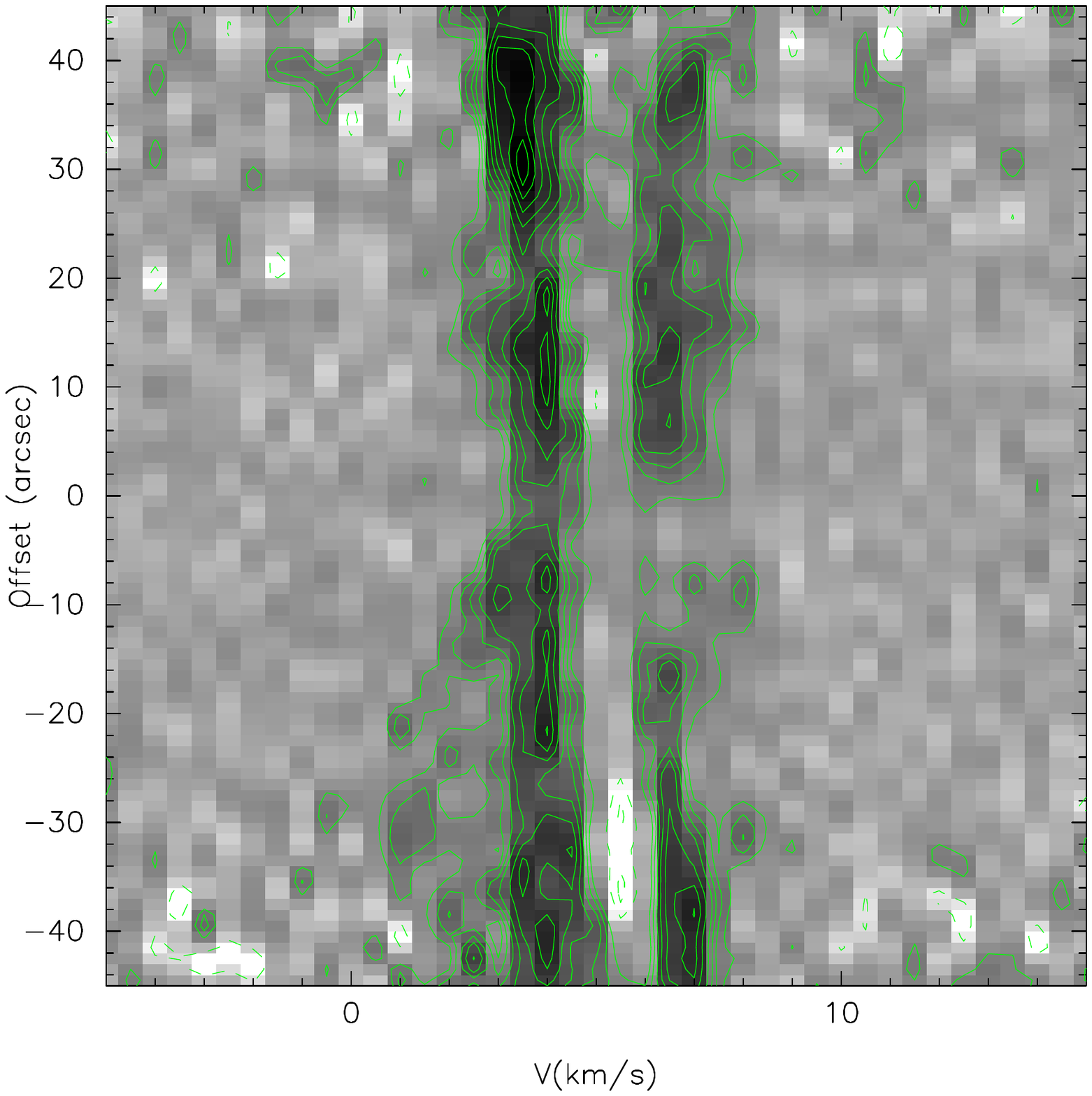}
\end{center}
\caption{PBRS 119019 -- Same as Figure \ref{093005}, but for D-configuration only. 
The outflow toward 119019 appears nearly in the plane of the sky, this is because the outflow
is very extended and the red and blue-shifted emission have spatial overlap in all locations
with detected emission. The full extent 
goes beyond the CARMA primary beam (dashed circle). The PV diagram further shows the low velocities and overlap of
red and blue-shifted emission along the outflow.
The contours in the integrated intensity plot are 
[$\pm$6, 9, 12, 15, 20, 25, 30, 35, 40, 45, 50, 60, 70] $\times$ $\sigma$ for the blue and the red; 
$\sigma_{red}$ = 0.48 K and $\sigma_{blue}$ = 0.54 K.
The PV plot contours are [-6, -3, 3, 5, 7, 9, 12, 15, 18, 21, 24, 27, 30] $\times$ $\sigma$ and $\sigma$ = 0.4 K.
}
\label{119019}
\end{figure}

\begin{figure}

\begin{center}
\includegraphics[scale=0.4,trim=3cm 1.25cm 5cm 1cm, clip=true]{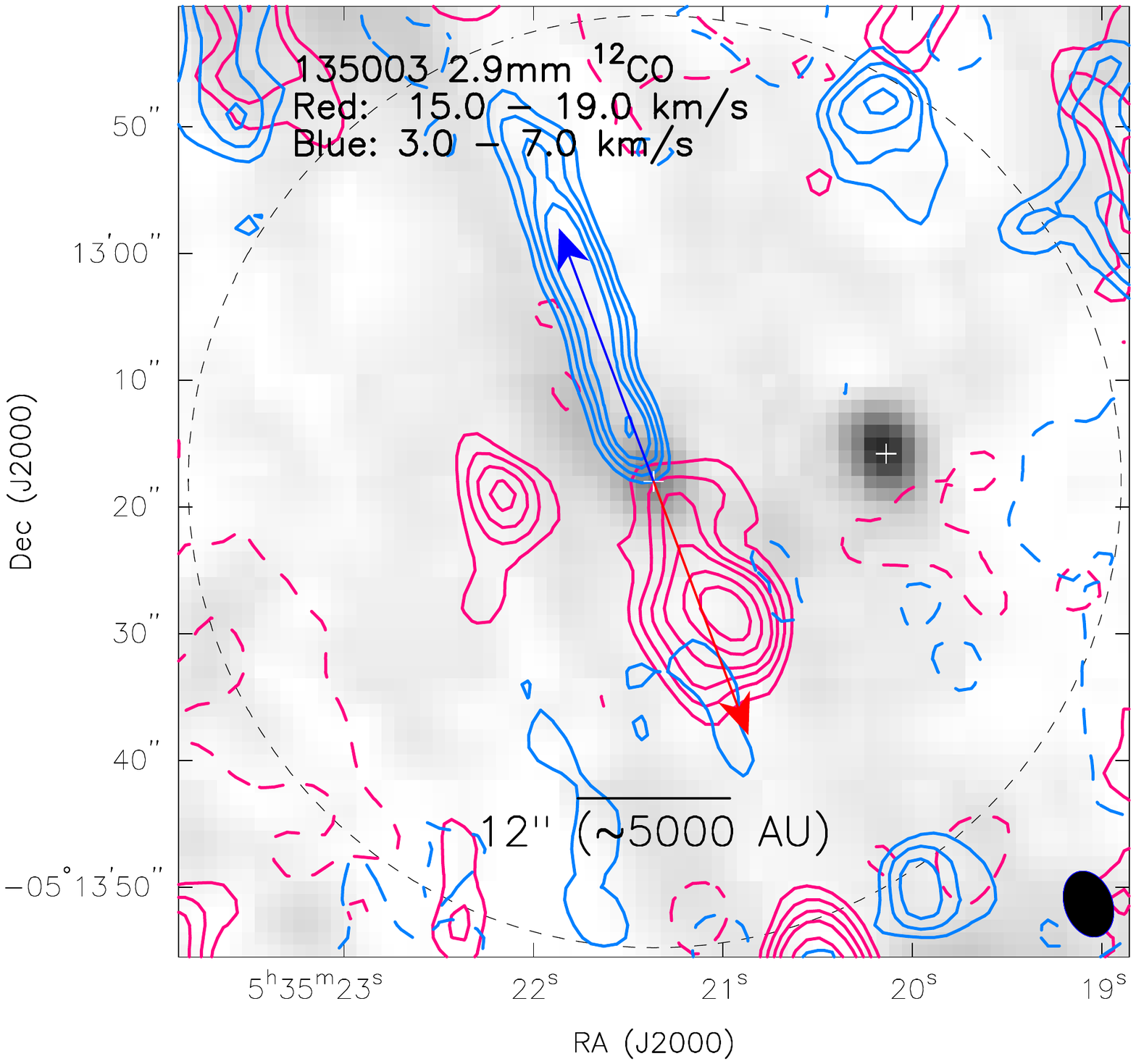}
\includegraphics[scale=0.4,trim=3cm 1.25cm 5cm 1cm, clip=true]{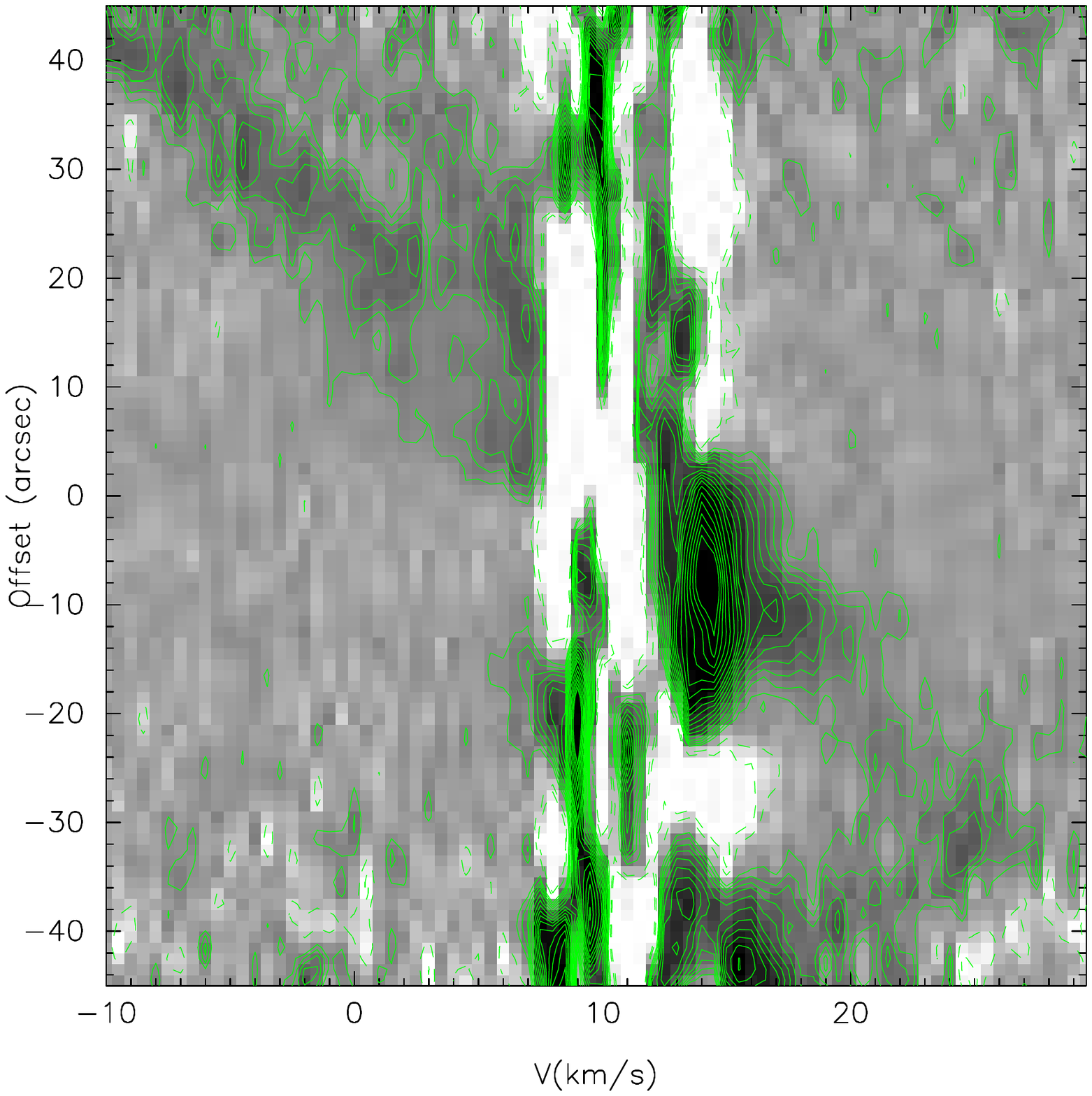}

\end{center}
\caption{PBRS 135003-- Same as Figure \ref{093005}, but for D-configuration only. 
The continuum peak of PBRS 135003 is in the center of the image and marked with 
a cross. HOPS 59 is also present in this field and marked with the cross in the 
western part of the image. the contours 
in the integrated intensity map start at $\pm$10$\sigma$ and increase in 5$\sigma$ 
intervals for the blue and the red-shifted 
contours start are [-6, 6, 9, 12, 15, 20,..., 60] $\times$ $\sigma$; $\sigma_{red}$, $\sigma_{blue}$  = 0.86 K.
The PV plot contours are [-6, -3, 3, 5, 7, 9, 12, 15, 18, 21, 24, 27, 30, 35,..., 60] $\times$ $\sigma$ and $\sigma$ = 0.75 K. The half-power point of
the primary beam is plotted as the dashed circle.
}
\label{135003}
\end{figure}

\begin{figure}

\begin{center}
\includegraphics[scale=0.4,trim=3cm 1.25cm 5cm 1cm, clip=true]{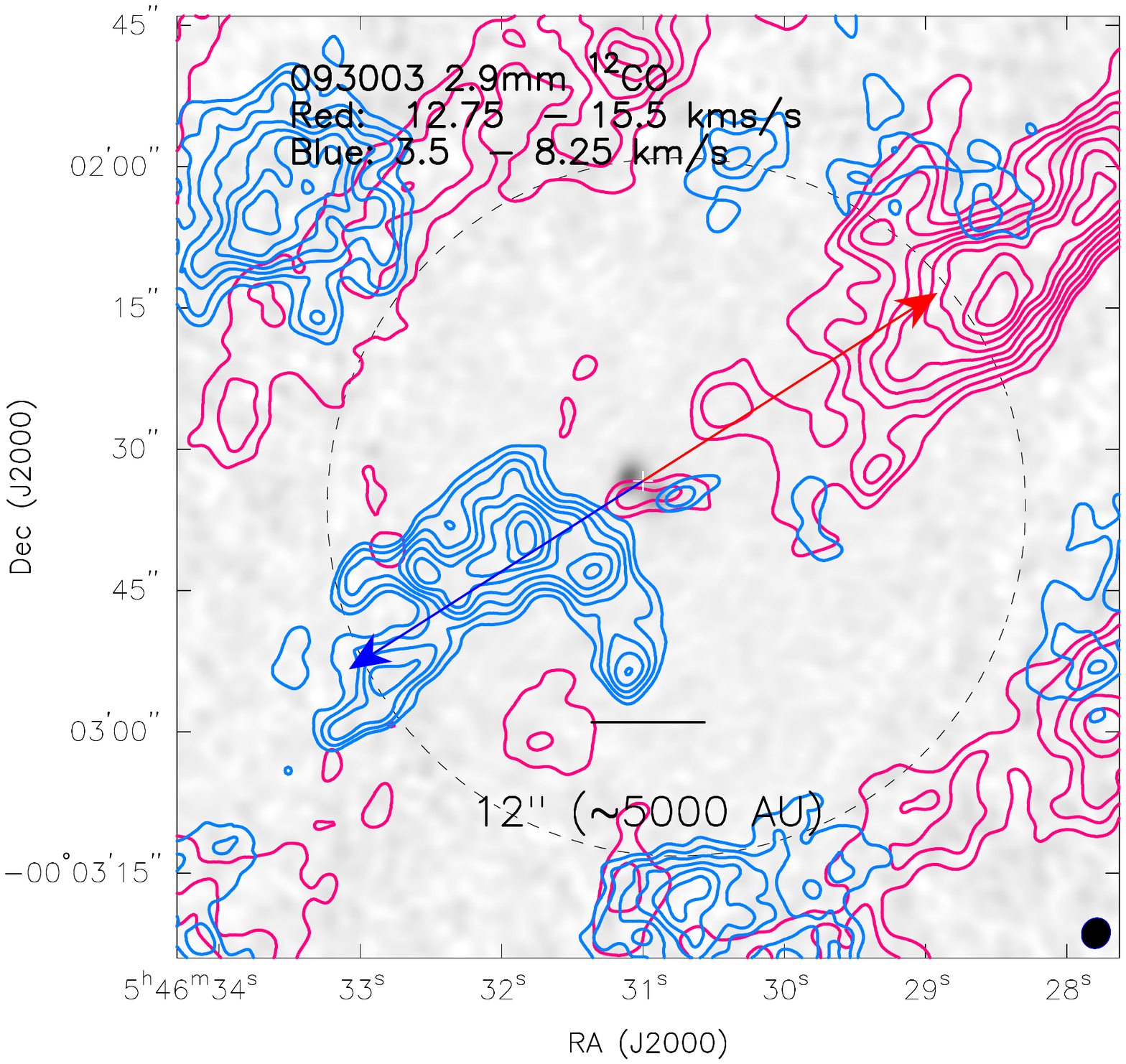}
\includegraphics[scale=0.4,trim=3cm 1.25cm 5cm 1cm, clip=true]{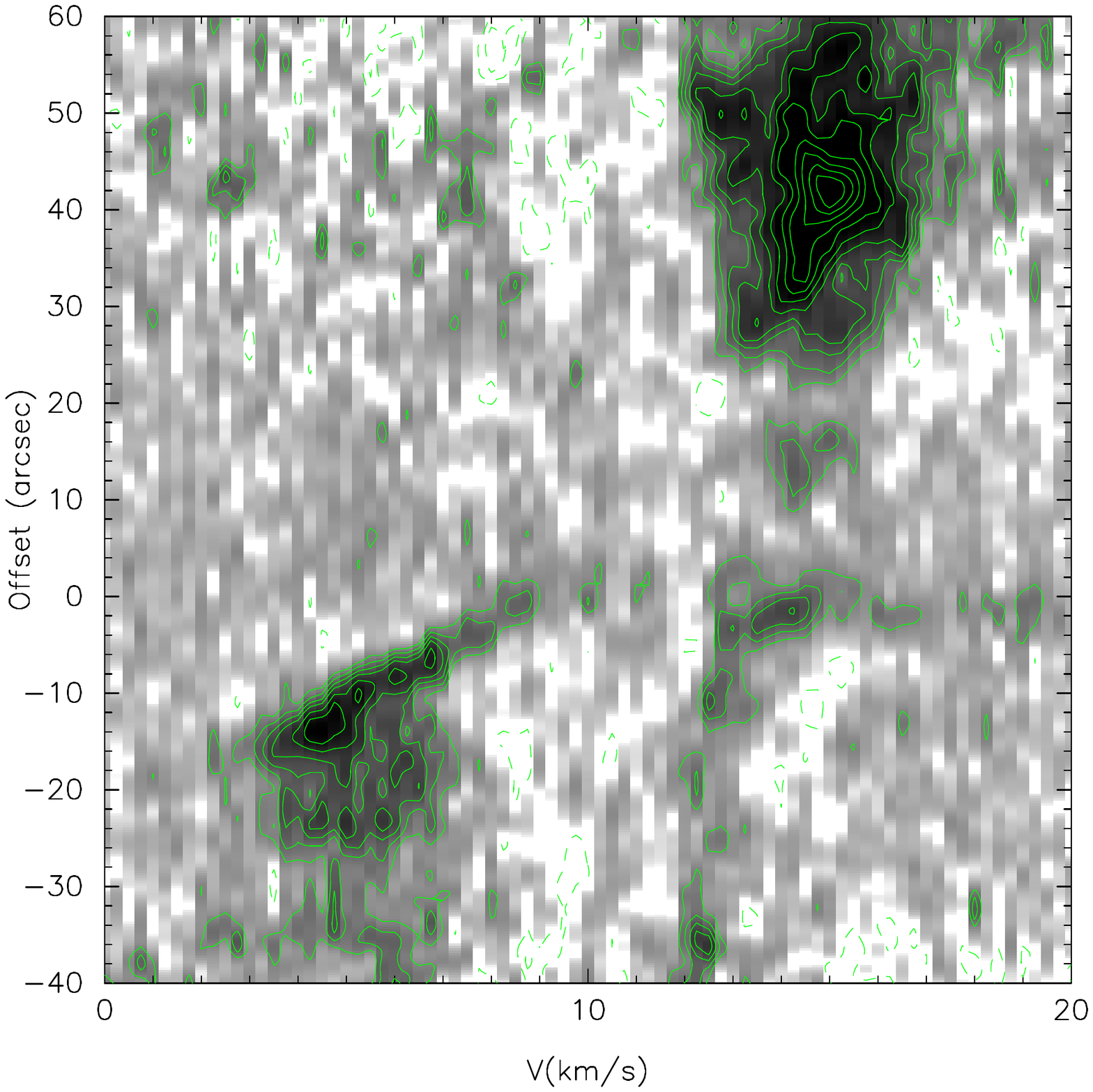}

\end{center}
\caption{PBRS HOPS 373 -- Same as Figure \ref{093005}, the outflow toward HOPS 373 is wide 
and was known to have a large outflow from single-dish studies \citep{gibb2000}. Very near
the protostar there is compact blue-shifted and red-shifted emission in the opposite directions
as compared to the larger outflow; this may be a second outflow from the binary source.
The PV plot shows the blue-shifted component on either side of the protostar and the red-shifted
component is evident at a large distance from the protostar; note that most of the red-shifted 
lobe is outside the primary beam of CARMA. The contours in the line map start 
at $\pm$10$\sigma$ and increase in 5$\sigma$ intervals for the blue and the red-shifted 
contours start at $\pm$20$\sigma$ and increase in 10$\sigma$; $\sigma_{red}$ = 0.93 K and 
$\sigma_{blue}$ = 1.21 K. The PV plot contours are [-6, -3, 3, 5, 7, 9, 12, 15, 18, 21, 
24, 27, 30, 35,..., 60] $\times$ $\sigma$ and $\sigma$ = 0.75 K. The half-power point of
the primary beam is plotted as the dashed circle.
}
\label{HOPS373}
\end{figure}

\begin{figure}

\begin{center}
\includegraphics[scale=0.4,trim=3cm 1.25cm 5cm 1cm, clip=true]{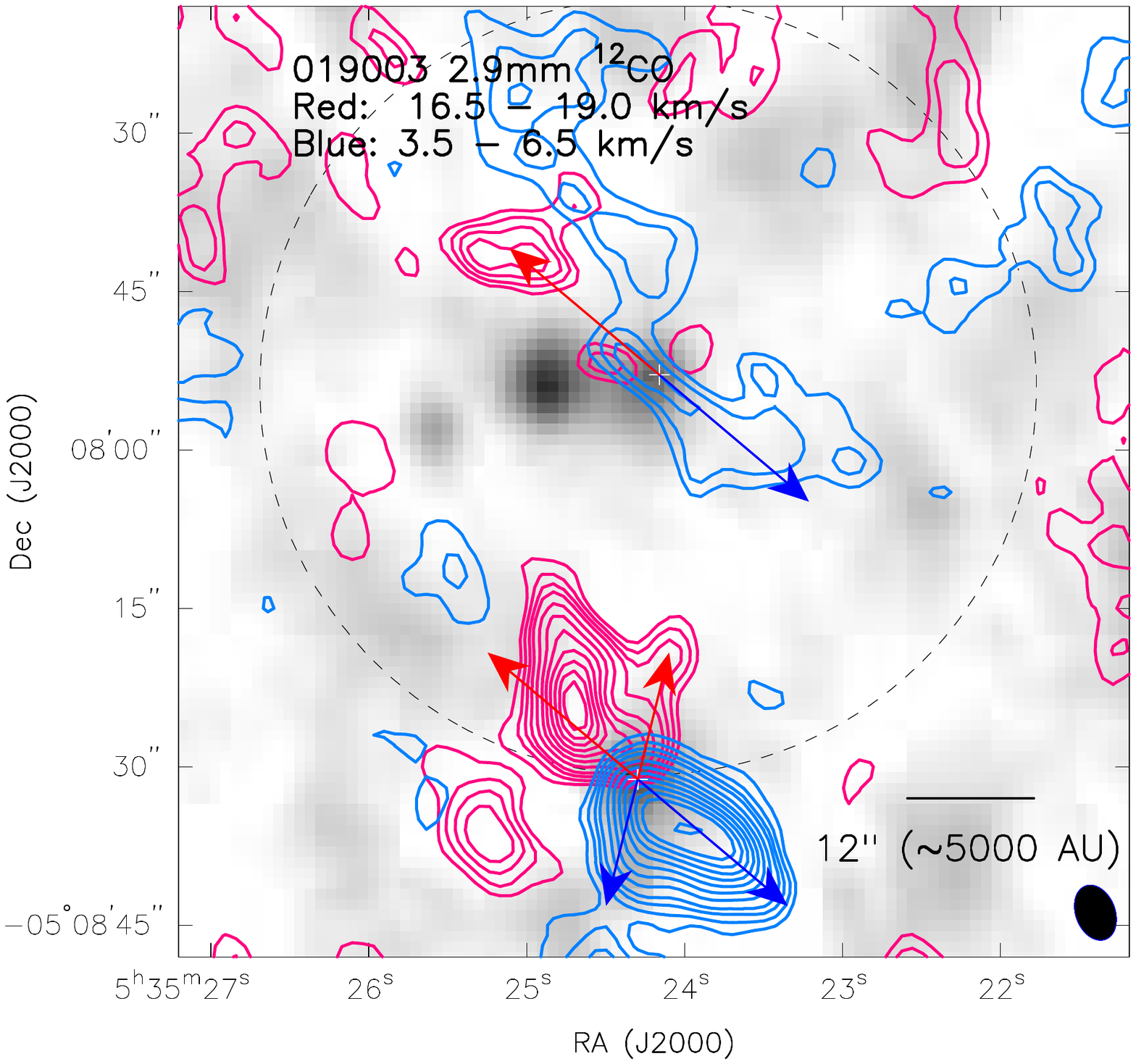}
\includegraphics[scale=0.4,trim=3cm 1.25cm 5cm 1cm, clip=true]{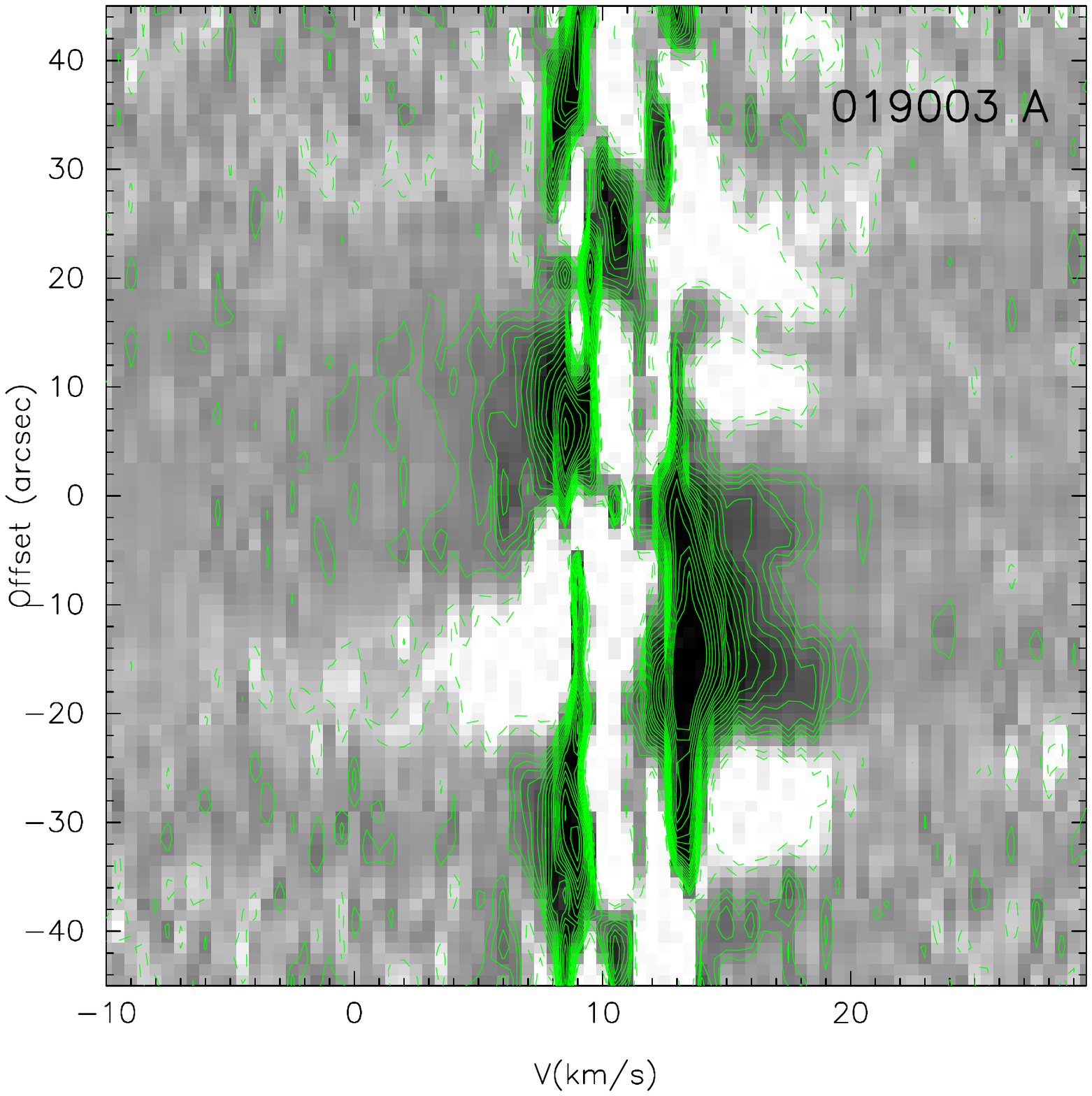}

\end{center}
\caption{PBRS 019003-- Same as Figure \ref{093005}, with data from D-configuration only. HOPS 68 is located in the southern part of the map and is shown in 
more detail in Figure \ref{HOPS68}. The contours in the line map start at $\pm$10$\sigma$ and increase in 5$\sigma$ intervals for the blue and the red-shifted 
contours start at $\pm$20$\sigma$ and increase in 10$\sigma$; $\sigma_{red}$ = 1.6 K and $\sigma_{blue}$ = 1.56 K.
The PV plot contours are [-6, -3, 3, 5, 7, 9, 12, 15, 18, 21, 24, 27, 30, 35,..., 60] $\times$ $\sigma$ and $\sigma$ = 0.75 K. The half-power point of
the primary beam is plotted as the dashed circle.
}
\label{019003}
\end{figure}

\begin{figure}

\begin{center}

\includegraphics[scale=0.35,trim=1cm 4cm 1cm 5cm, clip=true]{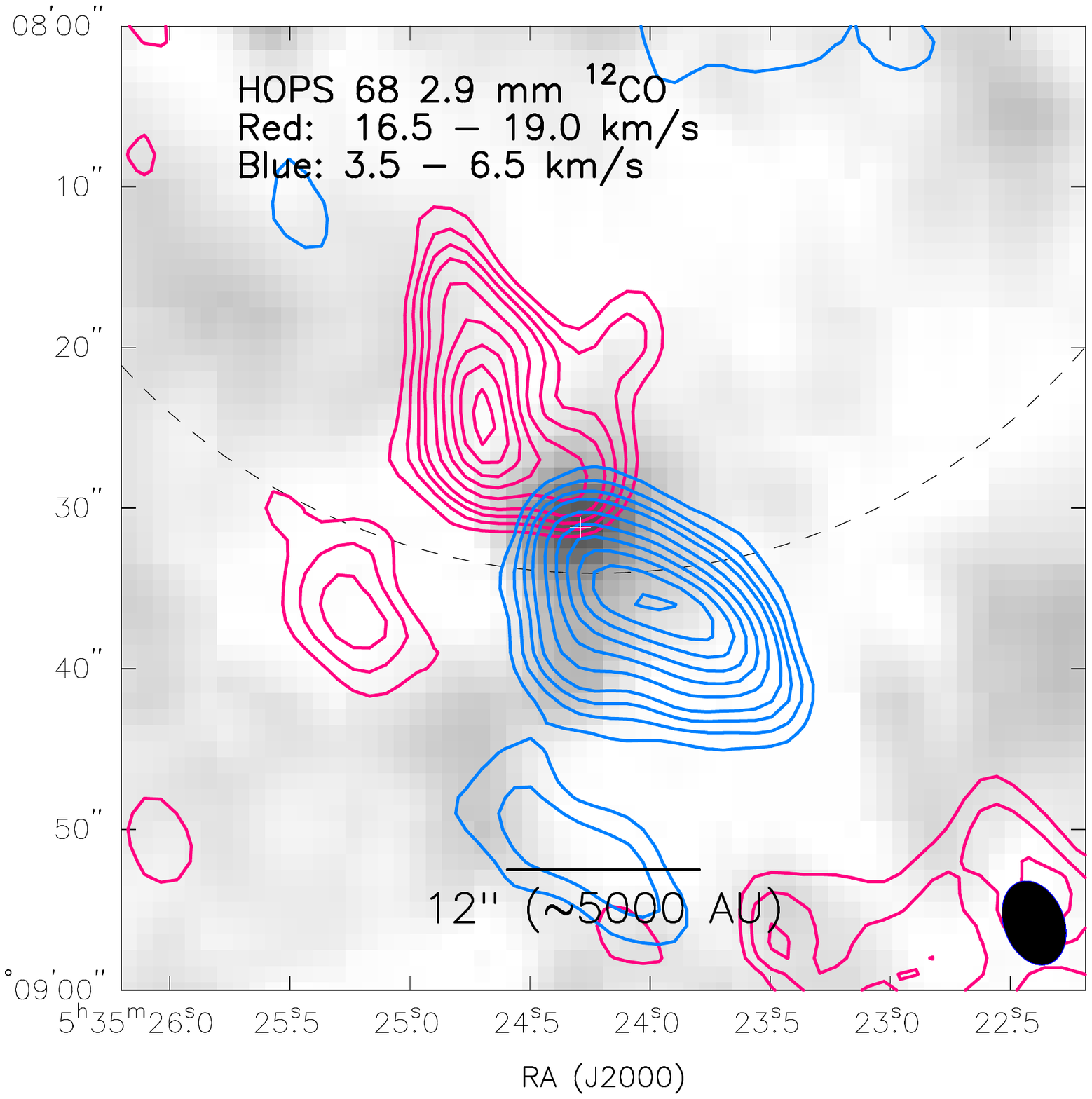}
\includegraphics[scale=0.35,trim=1cm 4cm 1cm 5cm, clip=true]{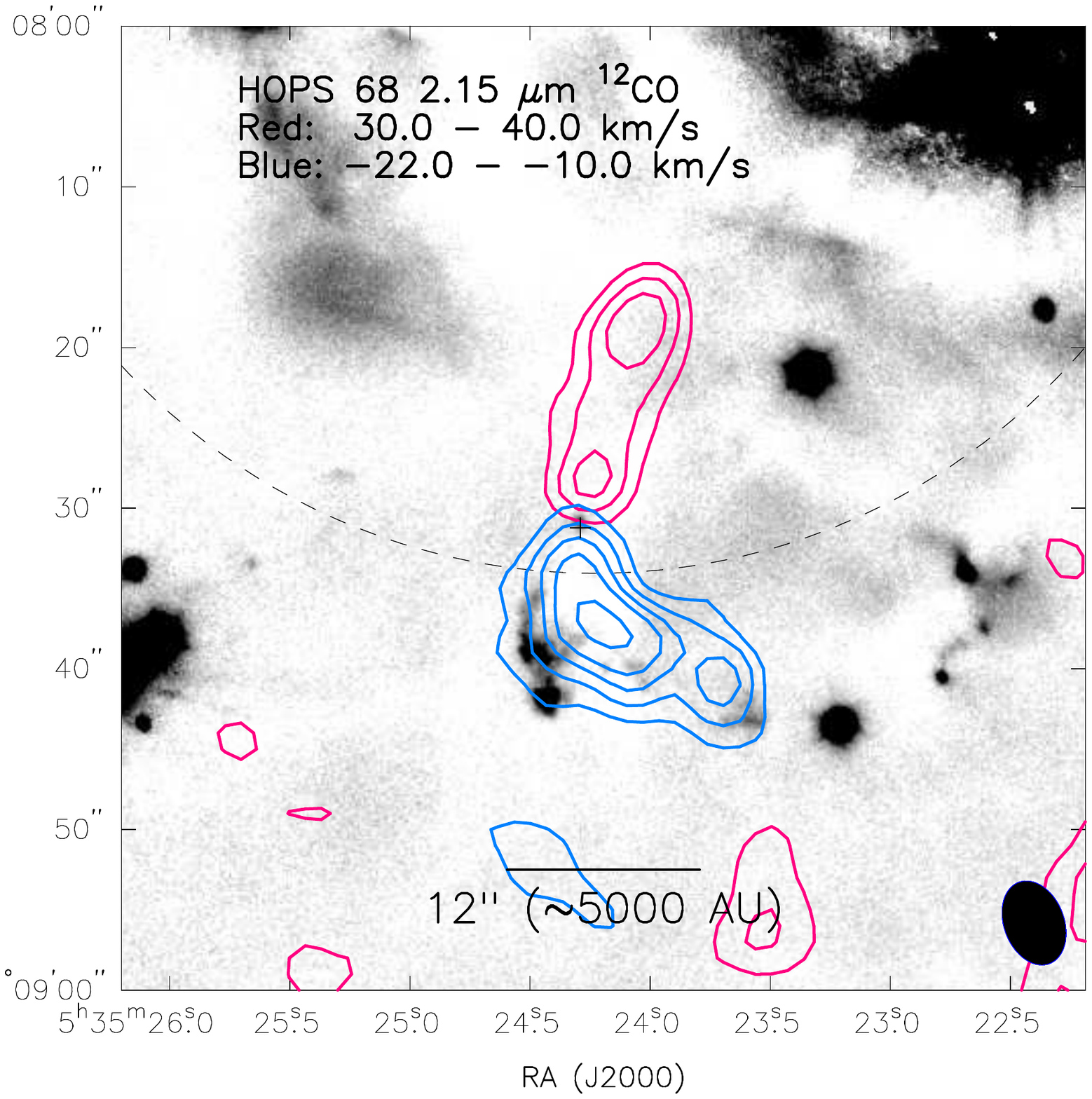}
\includegraphics[scale=0.35,trim=1cm 4cm 1cm 5cm, clip=true]{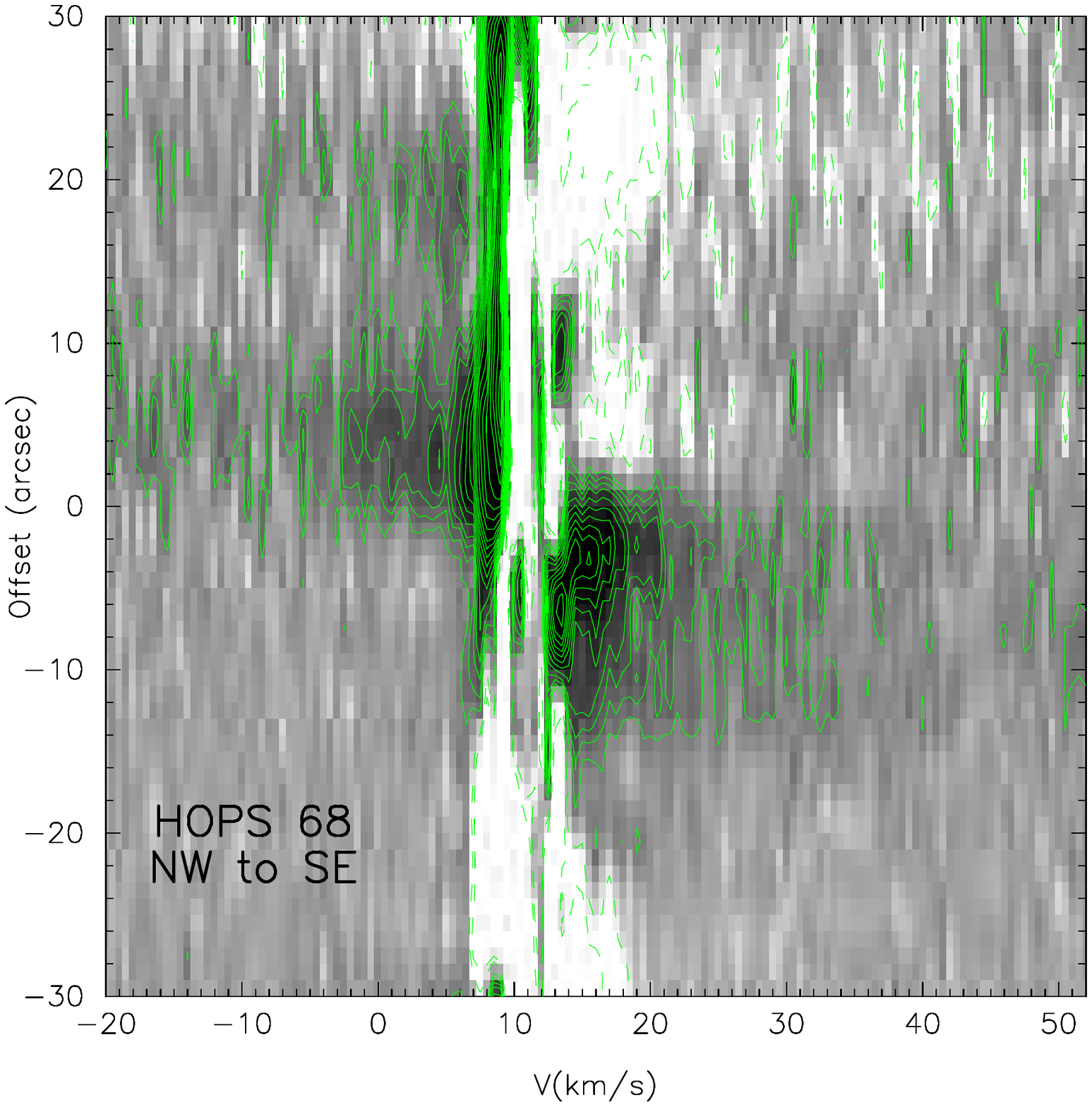}
\includegraphics[scale=0.35,trim=1cm 4cm 1cm 5cm, clip=true]{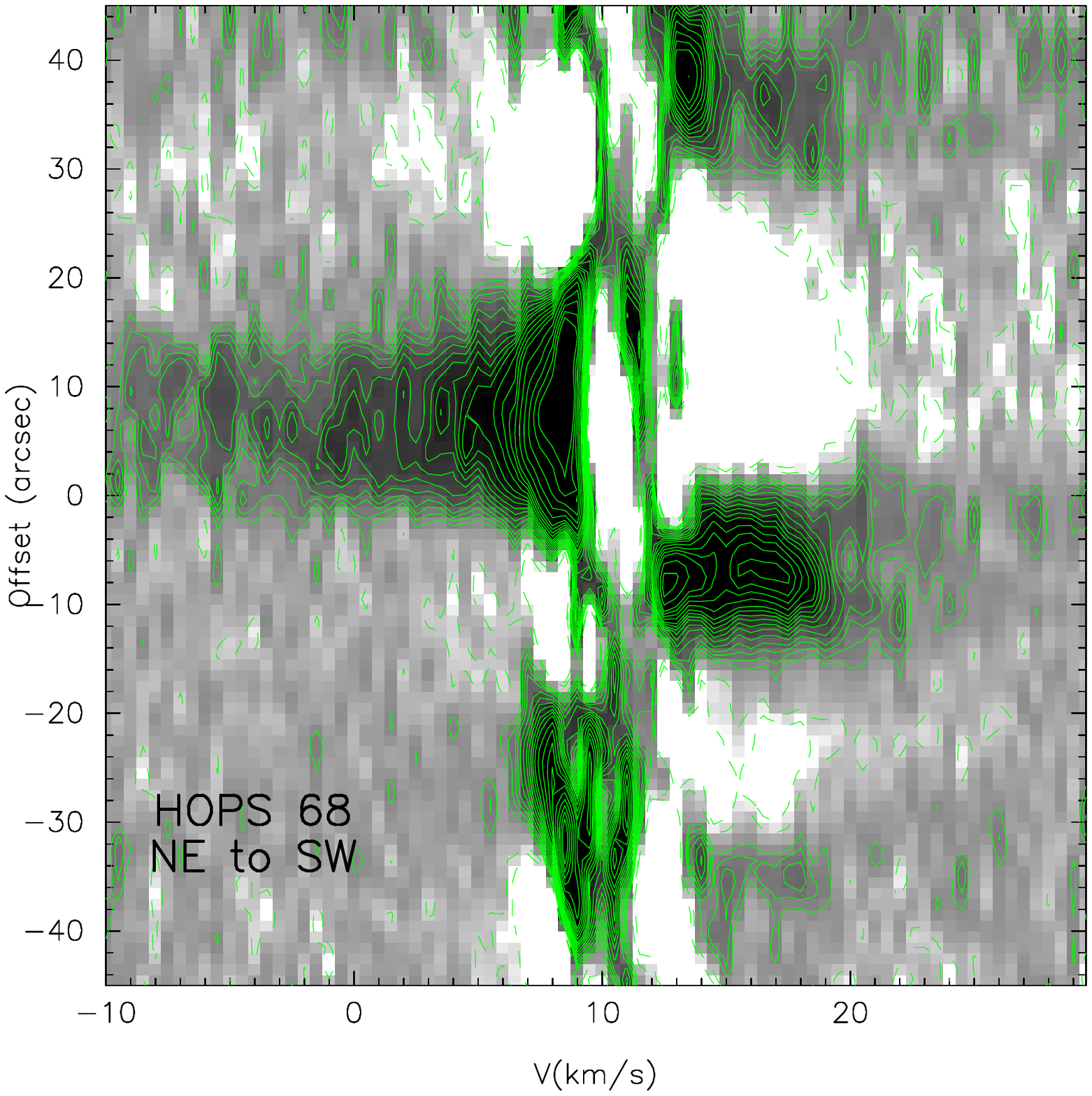}

\end{center}
\caption{HOPS 68-- Zoom in on the HOPS 68 outflow(s). We show plots of the low velocity CO contours (3.5 - 6.5 \kms; 16.5 - 19.5 \kms) overlaid
on the 2.9 mm continuum image and we overlaid the high velocity CO (30 - 40 \kms; -22 - -10 \kms) contours on the Ks-band (2.15 \micron)
image from Magellan PANIC. Between the high and low velocity ranges, the apparent position angle of the outflow changes from about 225\degr\ to 170\degr.
The contours in all plots are plot are [$\pm$6, 9, 12, 15 ,20, 30, 40, 50, 60, 70] $\times$ 
$\sigma$; $\sigma_{red}$ = 1.1 K and 1.39 K for the low and high velocities. For the blue contours,
  $\sigma_{blue}$ = 1.0 K and 1.55 K for the low and high velocities.
The PV plot contours are [-6, -3, 3, 5, 7, 9, 12, 15, 18, 21, 24, 27, 30, 35,..., 60] $\times$ $\sigma$ and $\sigma$ = 0.45 K. The half-power point of
the primary beam is plotted as the dashed arc.
}
\label{HOPS68}
\end{figure}

\begin{figure}

\begin{center}
\includegraphics[scale=0.4,trim=3cm 1.25cm 5cm 1cm, clip=true]{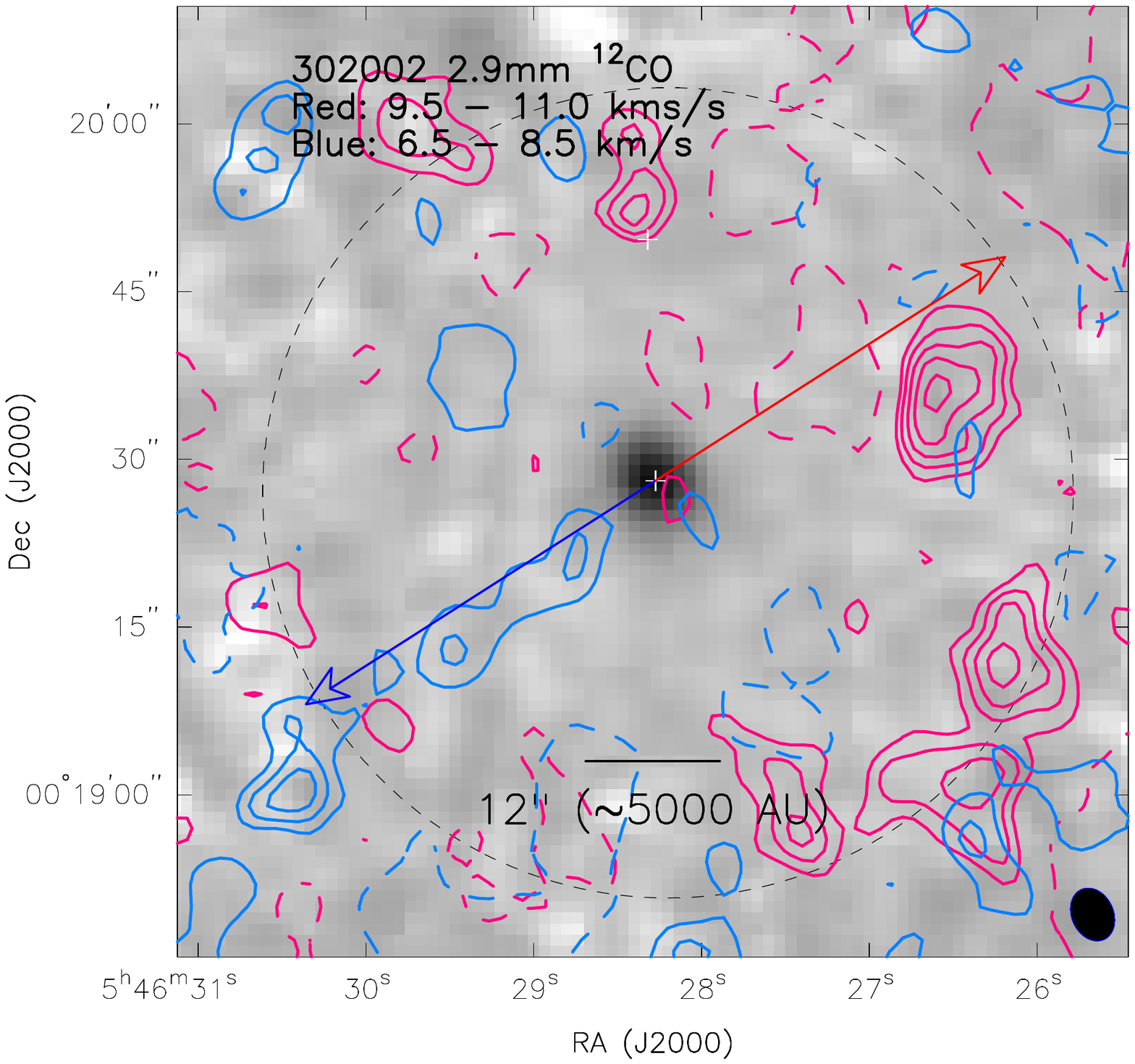}
\includegraphics[scale=0.4,trim=3cm 1.25cm 5cm 1cm, clip=true]{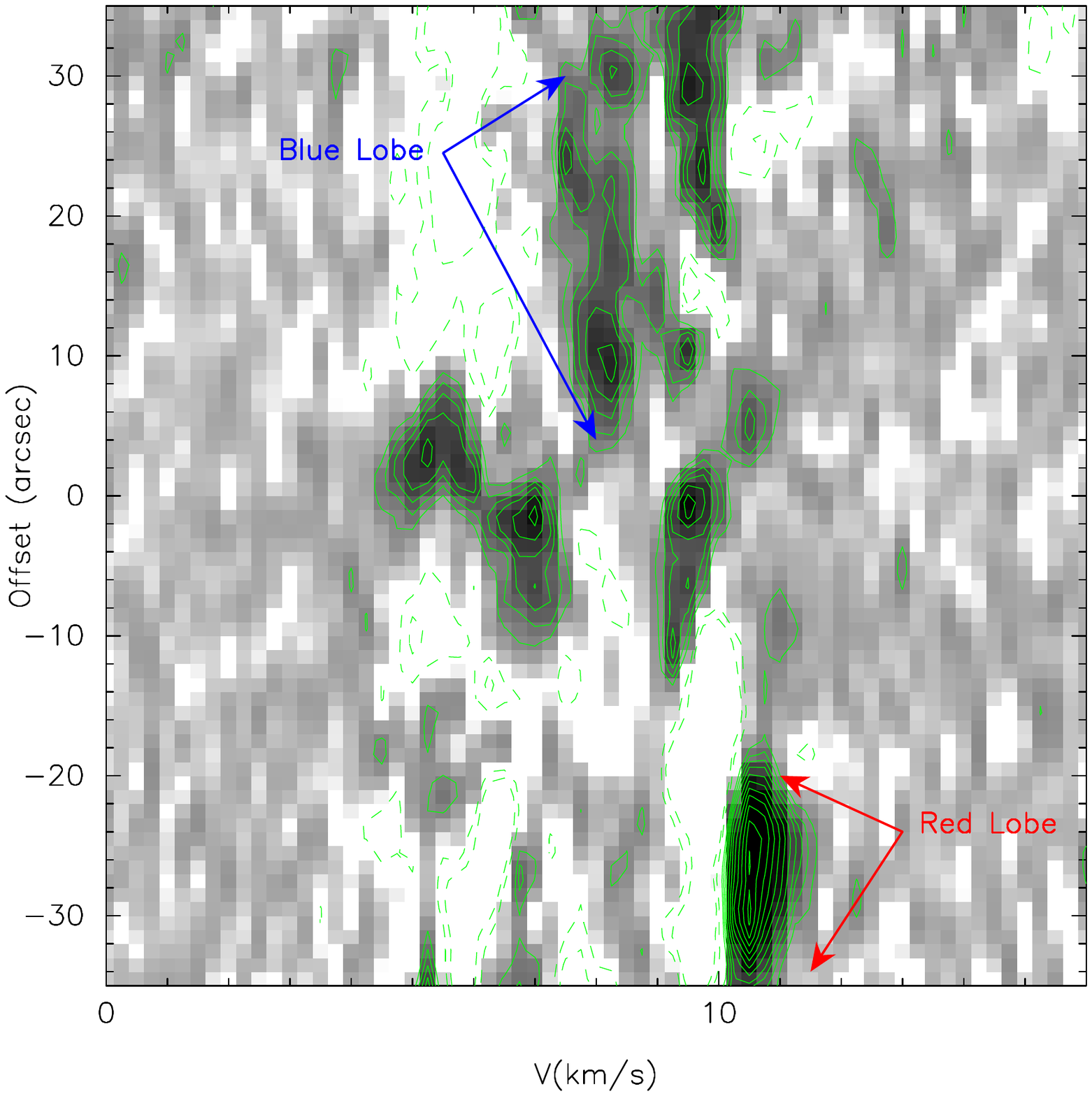}

\end{center}
\caption{PBRS 302002 -- Same as Figure \ref{093005}, but with D-configuration only. This source has some evidence of an outflow, 
however it is at low signal to noise and the blue and red-shifted lobes are not
exactly in a linear configuration. The PV plot shows that the outflow emission has quite low intensity and the plot
is also more complex than others because there appears to be a foreground cloud at $\sim$6 \kms\ which results in more resolved-out
flux. We have labeled the blue and red-shifted lobes that are shown in the integrated intensity plot.
North of 302002, HOPS 331 is not detected in the 2.9 mm continuum, but is marked with a cross.
There is red-shifted CO emission associated with HOPS 331, appearing to be an outflow from this source. The contours 
start at $\pm$10$\sigma$ and increase in 5$\sigma$ intervals; 
$\sigma_{red}$ = 0.41 K and $\sigma_{blue}$ = 0.46 K. The half-power point of
the primary beam is plotted as the dashed circle.  }
\label{302002}
\end{figure}

\begin{figure}

\begin{center}
\includegraphics[scale=0.3,trim=4cm 1.25cm 6cm 1cm, clip=true]{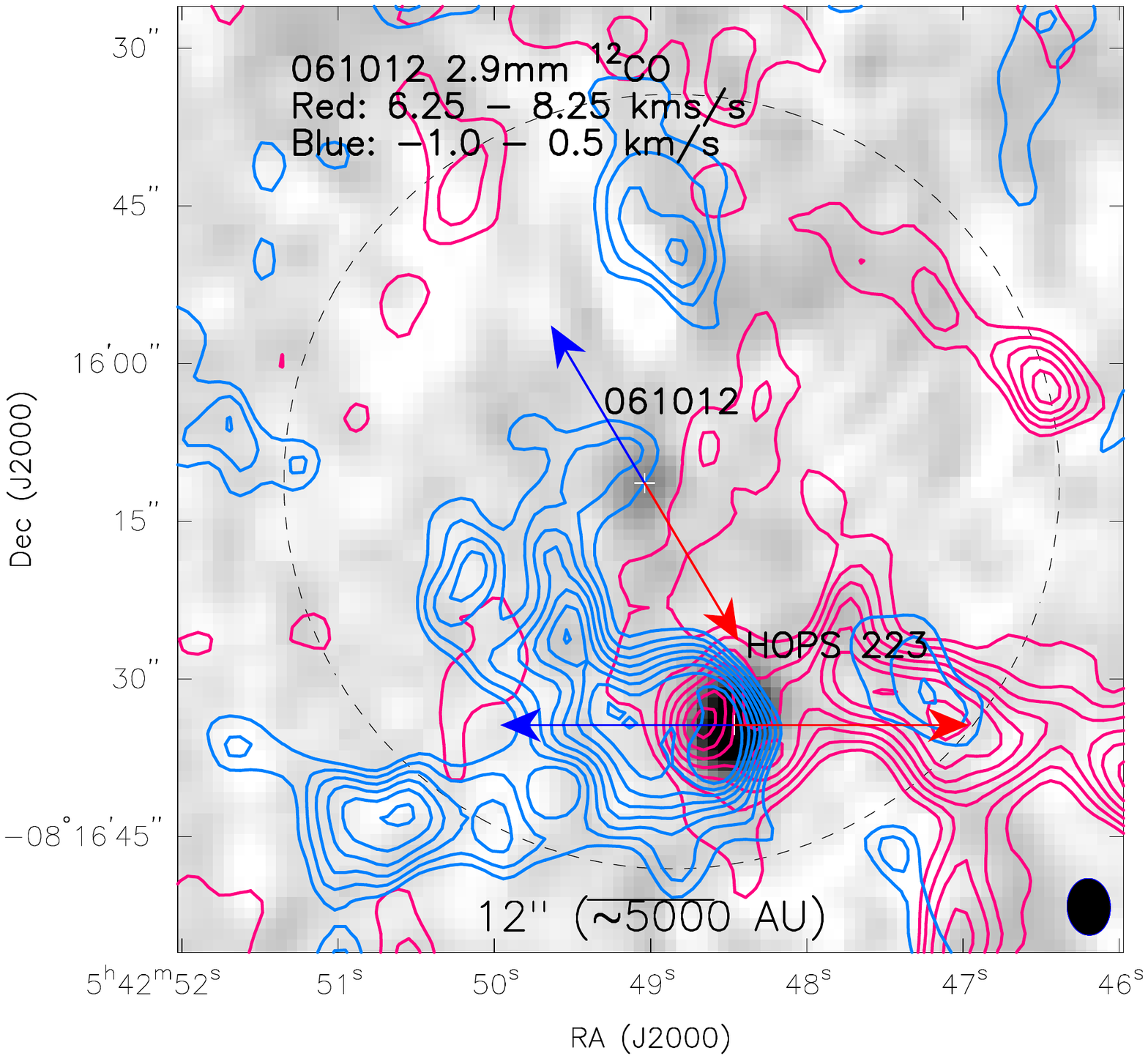}
\includegraphics[scale=0.3,trim=4cm 1.25cm 6cm 1cm, clip=true]{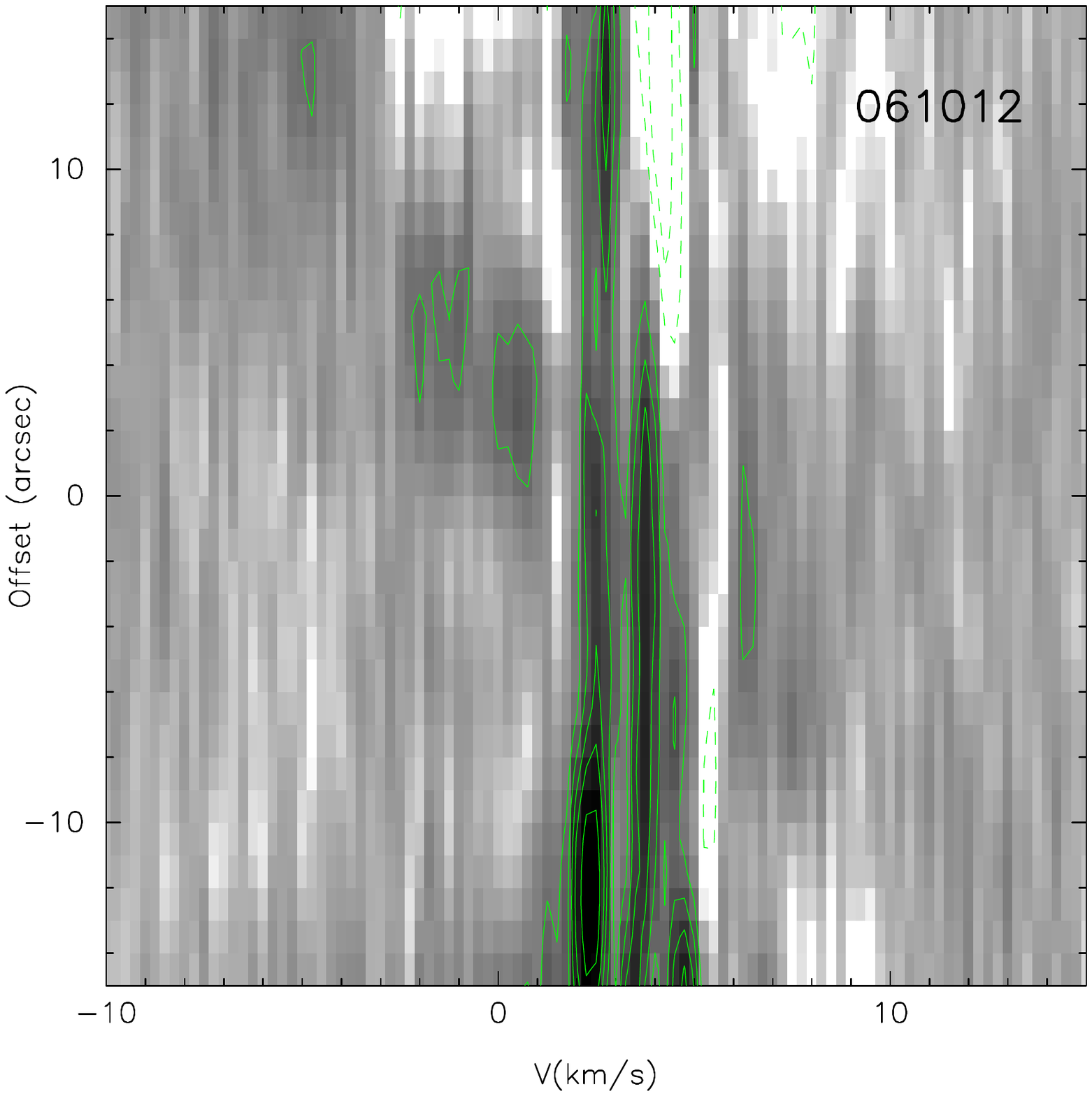}
\includegraphics[scale=0.3,trim=4cm 1.25cm 6cm 1cm, clip=true]{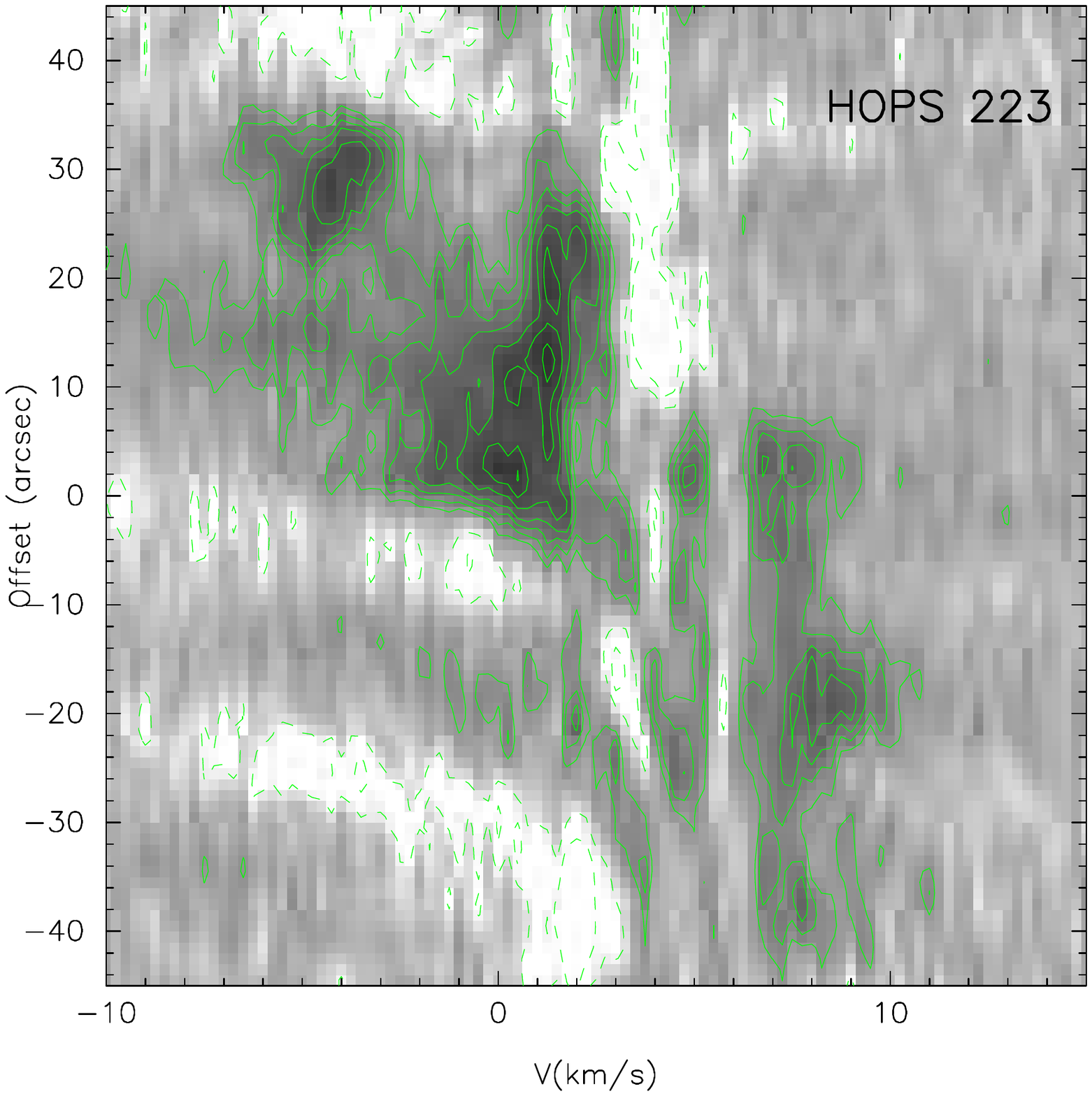}

\end{center}
\caption{PBRS 061012 and HOPS 223 -- Same as Figure 1, but for D-configuration data only.
 There is not a clear outflow detection from 061012; however, 
it is strongly confused with the outflow being driven by HOPS 223 and may need higher resolution to disentangle.
The PV plot shows that there is a blue-shifted feature that is along a possible outflow axis for 061012. The outflow
of HOPS 223 looks quite complex in the PV diagram with multiple velocity components; 
note that spatial filtering may be causing this outflow to appear more complex that it truly is.
The contours in the line map start at 10$\sigma$ and increase in 5$\sigma$ intervals for the blue and the red-shifted 
contours start at 10$\sigma$ and increase in 5$\sigma$; $\sigma_{red}$ = 0.44 K and $\sigma_{blue}$ = 0.46 K. Negative contours
 are not drawn for clarity. The PV plot contours are [-6, -3, 3, 5, 7, 9, 12, 15, 18, 21, 24, 27, 30] $\times$ $\sigma$ and $\sigma$ = 0.65 K. The half-power point of
the primary beam is plotted as the dashed circle.
}
\label{061012}
\end{figure}

We detect clear CO outflows toward 7 PBRS sources 093005, 090003, 
082012, 119019, 135003, HOPS 373 and 019004
(Figures \ref{093005}, \ref{090003}, \ref{082012}, \ref{119019}, \ref{135003},
\ref{HOPS373}, \& \ref{019003}), as well as for the Class I source HOPS 68 in the field
of 019003 shown in Figure \ref{HOPS68}. Tentative
detections are found toward 3 additional PBRS 302002, 061012, and HOPS 372 
(Figures \ref{302002}, \ref{061012}, \& \ref{082012}). The HOPS 372 outflow is apparent
in the low-velocity panel of Figure \ref{082012}, but at higher velocities 
the outflow emission is dominated by 082012. We did not detect outflow emission 
toward four PBRS 091015, 091016, 097002, and 082005; however, 
this does not mean that these sources do not have outflows, but that they were not detectable
with our resolution and sensitivity.

The outflows have a variety of morphologies, there is not a typical CO outflow morphology 
toward the PBRS sources. The PBRS 093005
and 090003 have spatially compact outflows, with total lengths of the red and blue-shifted 
lobes being less than 0.05 pc (Figures \ref{093005} \& \ref {090003}).
 The outflows toward 119019, 082012, 135003, and HOPS 373 all extend outside 
the CARMA primary beam, with total lengths greater 
than 0.1 pc (Figures \ref{082012}, \ref{119019}, \ref{135003}, and \ref{HOPS373}).
The outflows toward 082012 and 135003 also have emission extending to velocities
$>$ $\pm$ 10 \kms\ from the systemic velocity with jet-like morphologies.

Toward 061012, there is evidence for an outflow, but this is unclear due to confusion with the wide-angle
outflow of its neighbor HOPS 223 (Figure \ref{061012}). HOPS 223 (also known as V2775 Ori)
is an outbursting Class I source \citep{fischer2012} and this is the first clear detection
of a CO outflow toward this source. However, the \textit{Spitzer} imaging already showed strong
evidence for outflow-associated features.
Toward 302002 (Figure \ref{302002}) there appears to be low-velocity 
$^{12}$CO emission in its vicinity that appears outflow-like, but its detection
is not definitive.

The outflow
toward 119019 is distinct from the other PBRS in that it has a large spatial extent, but
at low-velocities; the full velocity width is only 6 \kms\ for both the red and blue
sides of the outflow. Moreover, the spatial overlap between the redshifted and
blueshifted emission is strong evidence that this source is viewed close to edge-on. 

The non-detections of outflows toward 091015, 091016, 097002, and 082005
could result from the outflows having low-velocities and being confused
with the emission from the molecular cloud. Also, there is a tentative trend
between detectable outflows and L$_{bol}$. The PBRS 119019 was 
the lowest luminosity source (L$_{bol}$ = 1.56 L$_{\sun}$) with a 
clear outflow detection; the tentative outflow detections and 
non-detections have luminosities between 0.65 L$_{\sun}$ and 1.56 L$_{\sun}$.
The outflow properties of individual sources are described in 
more detail in the Appendix.

The outflow from HOPS 68 (Figure \ref{HOPS68}) is worth mentioning because it was 
also found to have quite high velocities, and the relative position angle of 
the red and blue-shifted lobes changes from high to low-velocity.
At low velocities the outflow is oriented northeast to southwest, but at high velocities
the red-shifted side is oriented northwest to southeast while the blue lobe still appears extended
in the same direction as at low velocities. We overlaid the high-velocity CO contours on a Ks-band (2.15 \micron)
image from Magellan PANIC (Figure \ref{HOPS68}) and we see that there are two sets of bow-shock features that overlap with the blue-shifted
CO emission. One set of features is in the southeast direction and the other set is in the south west direction. Thus,
the change in position angle of the CO emission from low to high velocities is likely indicative of two outflows from HOPS 68.

\subsubsection{Outflow Parameters}
We calculate the outflow mass, momentum, and energy following the procedure used 
by \citet{plunkett2013} \citep[based on ][]{bally1999}, and give these values
in Table 4. The analysis by \citet{plunkett2013} uses $^{13}$CO optical depths and 
excitation temperature derived from $^{12}$CO (assuming optically thick
emission) in order to calculate column densities, from which the 
mass, momentum, and energy can be calculated. However, our observations 
did not have enough sensitivity to detect the $^{13}$CO ($J=1\rightarrow0$) outflow emission, we therefore adopted
a $^{12}$CO/$^{13}$CO ratio of 62 \citep{langer1993} and 
divided 
the $^{12}$CO ($J=1\rightarrow0$) intensities by this ratio, under the assumption that the $^{12}$CO emission is 
optically thin at all velocities. This assumption is not valid at all velocities, but probably most reasonable for
the higher velocity ($>\pm$10~\kms)emission. The principal effect will be an underestimate of the 
CO column densities and cause all the outflow parameters to be lower limits. \citet{dunham2014} showed that opacity corrections to the outflow parameters can be up to an order of magnitude; missing flux will also affect the parameters but this is more difficult to quantify since the
low-velocity emission with the highest opacity will be the most severely affected by
spatial filtering.
The $^{13}$CO abundance is taken to be 
N($^{13}$CO) = N(H$_2$)/7$\times$10$^{5}$ \citet{frerking1982} and the excitation temperature is calculated using
the $^{12}$CO brightness temperature and was between 15 K and 40 K in our observations \citep[see Equation 3 in ][]{plunkett2013}. We do not attempt to correct the outflow properties for the effects on inclination.

\begin{figure}

\begin{center}
\includegraphics[scale=0.65]{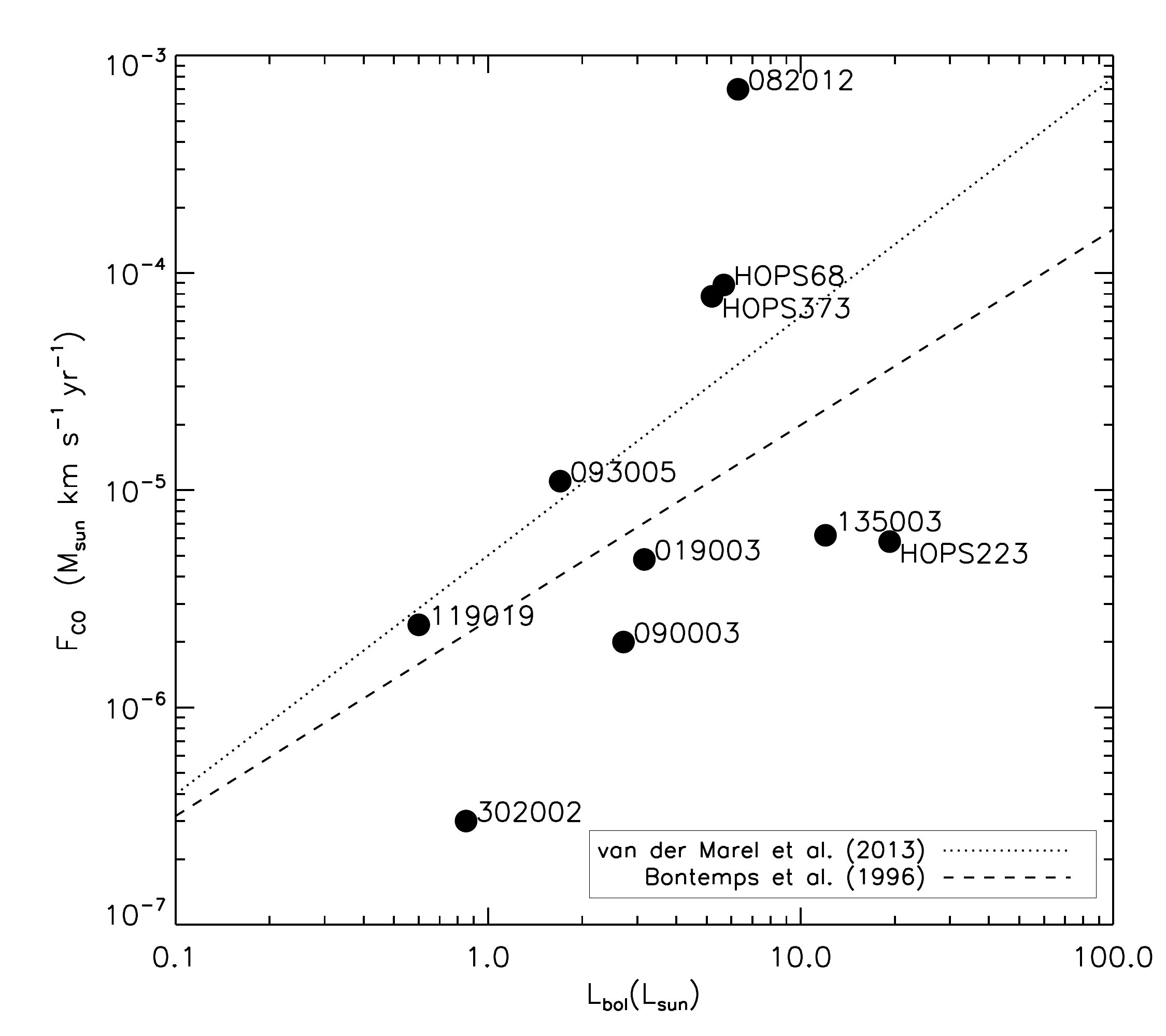}

\end{center}
\caption{Measured outflow forces (F$_{CO}$) versus L$_{bol}$ for all sources with detected outflows. There is no
clear relationship between these two source properties in our data; however, there is a general tendency of high luminosity
sources having greater outflow forces. We plot the relationships that have been found in the literature for larger samples
of objects from \citet{vdmarel2013} (dotted line) and \citet{bontemps1996} (dashed line). The literature relationships utilized single-dish data,
while our data are interferometric; thus, missing flux could cause F$_{CO}$ to be systematically underestimated. 
Furthermore, the Class 0 sources
in the literature have F$_{CO}$ and $L_{bol}$ values that are above the \citet{bontemps1996} relation and the plotted relationships
are principally fit to the Class I protostars.
}
\label{outflow-lbol}
\end{figure}

The observed outflow properties (mass, momenta, energy, and force; see Table 4) 
of the PBRS are generally consistent with results 
from \citet{plunkett2013}; however, 
there is a general tendency for
lower values of mass, momentum, and energy for the PBRS, which could
result from the lack of $^{13}$CO. We also
computed the outflow force (F$_{CO}$) and dynamical time based on the apparent
outflow size and the maximum velocity of observed CO.
We examined the relationship between L$_{bol}$ and F$_{CO}$ in Figure \ref{outflow-lbol}. For the PBRS
with a detected outflow, there is no clear correlation between F$_{CO}$ and L$_{bol}$, but more luminous sources tend to have greater
values of F$_{CO}$. We have also plotted the relationships derived by \citet{bontemps1996} and \citet{vdmarel2013}
for comparison. The relationships were derived from
samples primarily comprised of Class I protostars and Class 0
protostars lie above the relationship and not below; the CO ($J=6\rightarrow5$) measurements from \citet{yildiz2015}
for Class 0 sources are also above the \citet{bontemps1996} relationship.
The relations do go through  our observed the points, but
four PBRS are found below the \citet{bontemps1996} relationship. On the other hand
we use interferometer data without zero-spacings, while the other studies used single-dish maps.
We also did not have $^{13}$ CO detections, making our values lower limits. 
We do not
calculate upper limits for the sources with non-detections because the large 
amount of resolved emission near the source velocities
results in these values having little physical meaning. 
However, their outflow parameters will (at a minimum) be lower
that those measured for 090003/093005.

\subsubsection{Outflow Inclinations}
The outflow inclinations are difficult to precisely measure; 
however, we qualitatively compared our data with the simulations
of \citet{cabrit1986}, which show model PV plots and integrated 
intensity plots for accelerating outflows. 
Model outflows
are shown for a fixed opening angle and outflow length at
inclinations of 5\degr, 30\degr, 50\degr, and 80\degr. As such, 
the uncertainty in our estimates of the outflow inclination is likely $\pm\sim$20\degr.
The outflows of 090003, 093005, HOPS~223, and 019003
are consistent
with an outflow inclinations near 30\degr,
given their compact extent and distribution 
of highest velocities near the source.
The well-collimated
outflows of 082012, 135003, and HOPS~68
appear most
consistent with an 
inclination near 50\degr.
Both the wide-angle outflow toward HOPS 373 and the tenuous outflow toward 
302002 are consistent with 
having inclinations between 50\degr\ and 80\degr; based on their velocity
distributions, HOPS 373 is likely closer to
50\degr, while 302002 is likely closer to 80\degr. The PBRS 061012 
appears to have an outflow, but the data do not lend themselves to
a reasonable estimate of the inclination.
Finally, 119019 is
the only PBRS that is consistent with 
having a near edge-on inclination, as indicated by the CO emission only being detected at low velocities
and the extended spatial overlap of the red and blue-shifted emission
toward 119019. 

We can broadly conclude
that for the PBRS with detected outflows, extreme 
edge-on orientations cannot be the cause of their extremely red
24 \micron\ to 70 \micron\ colors, except for 119019. 
The estimated inclinations for the PBRS are also given in Table 5.
Furthermore, while there is a large degree in uncertainty in the outflow inclinations, it is most likely
that the distribution of inclination angles appears dominated by
intermediate inclinations (80\degr~$\ge$~$i$~$\ge$20\degr). While our numbers are
small, the distribution is likely consistent with a random distribution
of inclinations (the average inclination for a random distribution is 60\degr), which is expected
for a collection of sources whose selection criteria is not particularly biased toward a particular geometric
orientation; as had been a previous concern with respect to the PBRS was they could have simply been edge-on sources
and the outflow data show that this is clearly not the case. Given the uncertainty in the inclination
angles, we have not corrected the derived outflow parameters in Table 4 for this effect.

\subsection{Evidence for Extended Outflows}
\begin{figure}
\figurenum{12a}
\begin{center}

\includegraphics[scale=0.35]{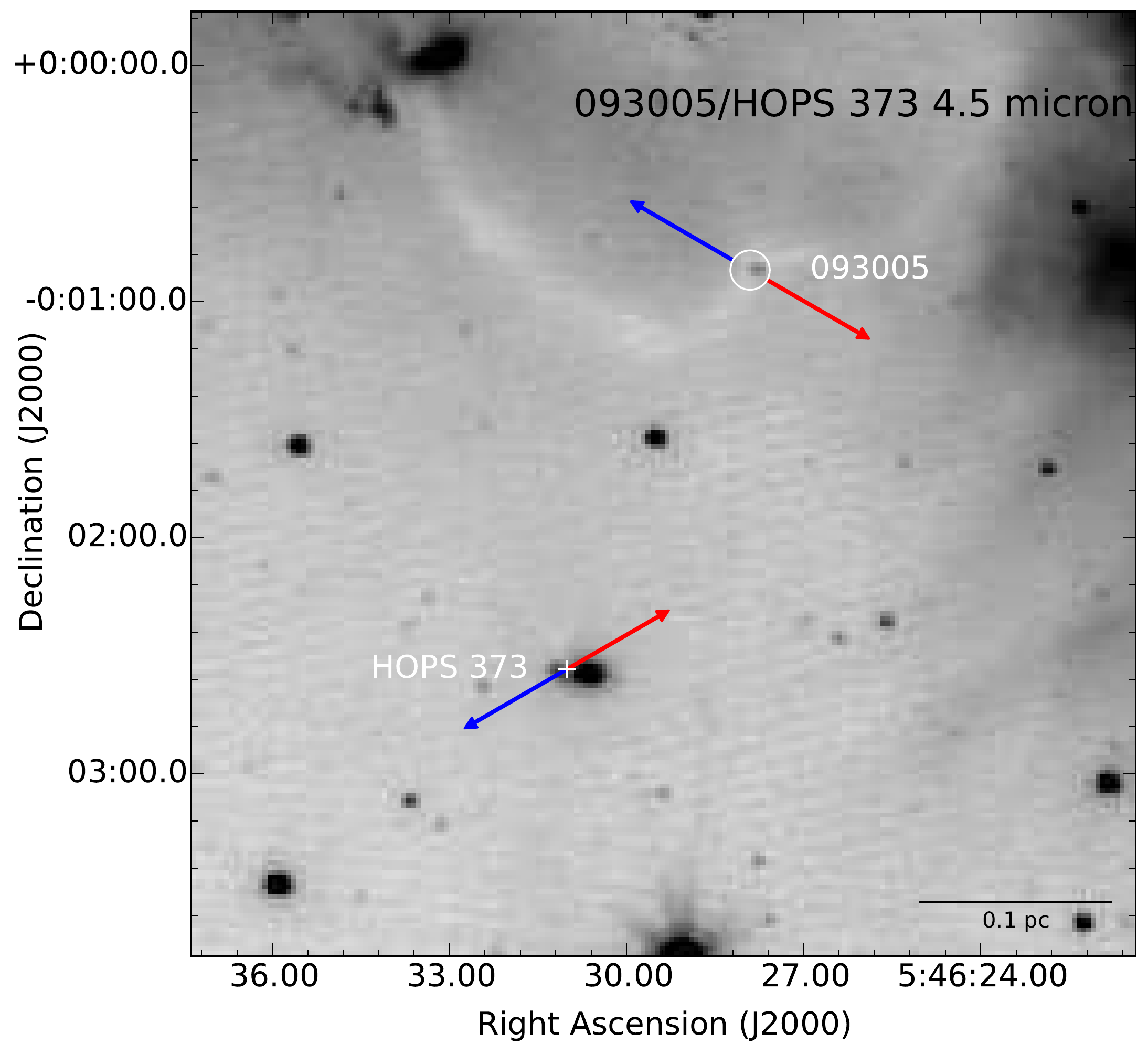}
\includegraphics[scale=0.35]{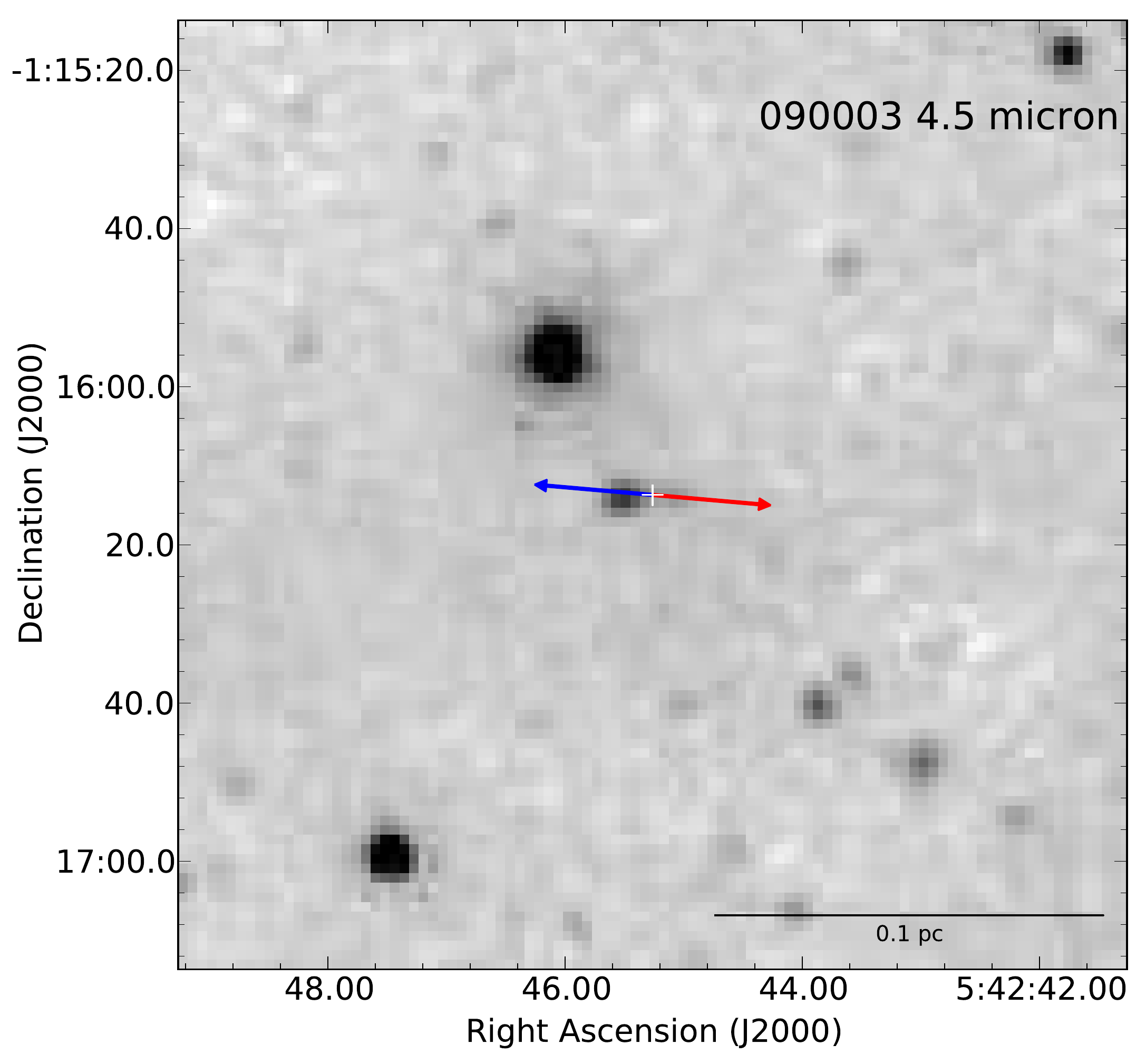}
\includegraphics[scale=0.35]{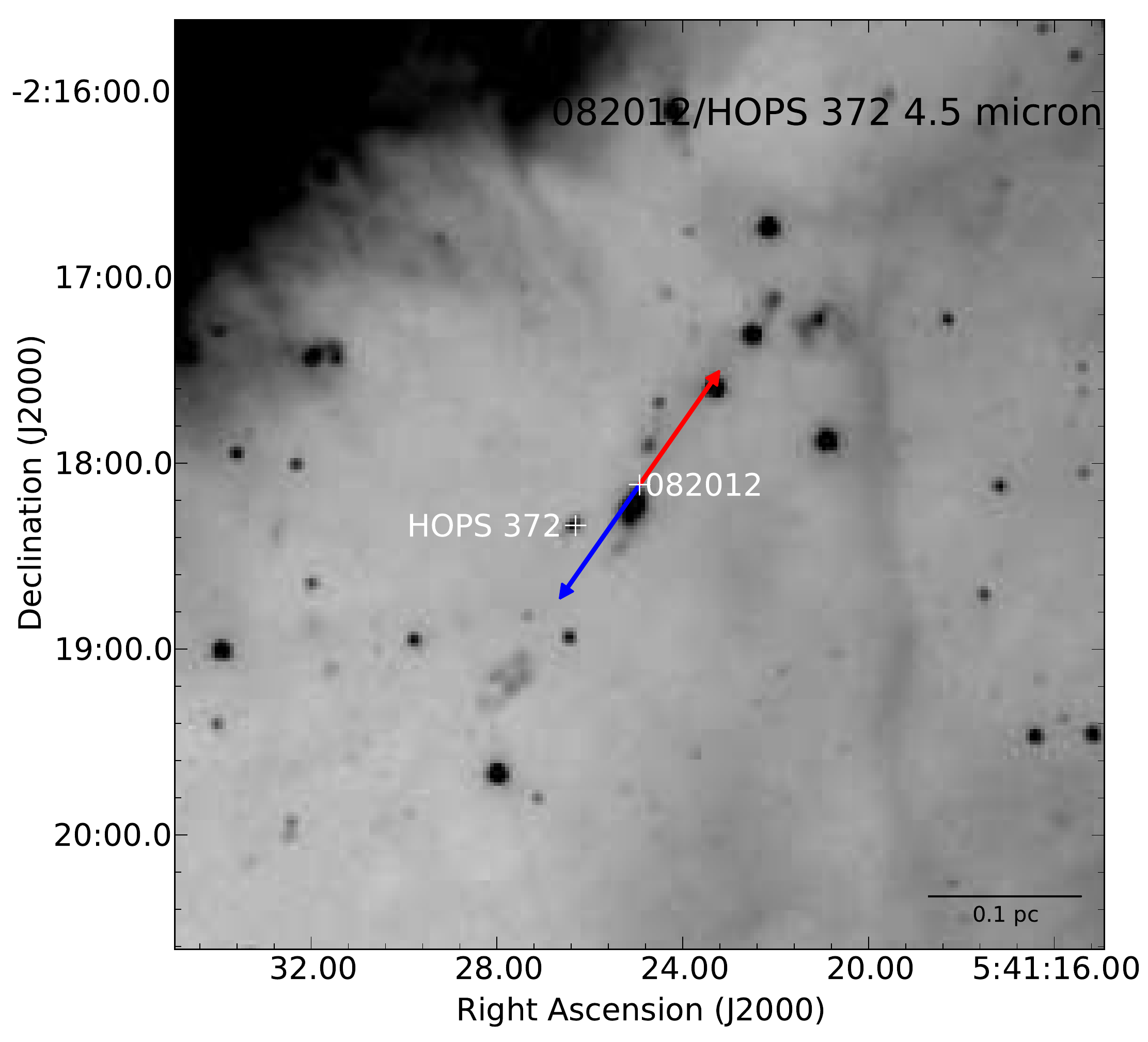}
\includegraphics[scale=0.35]{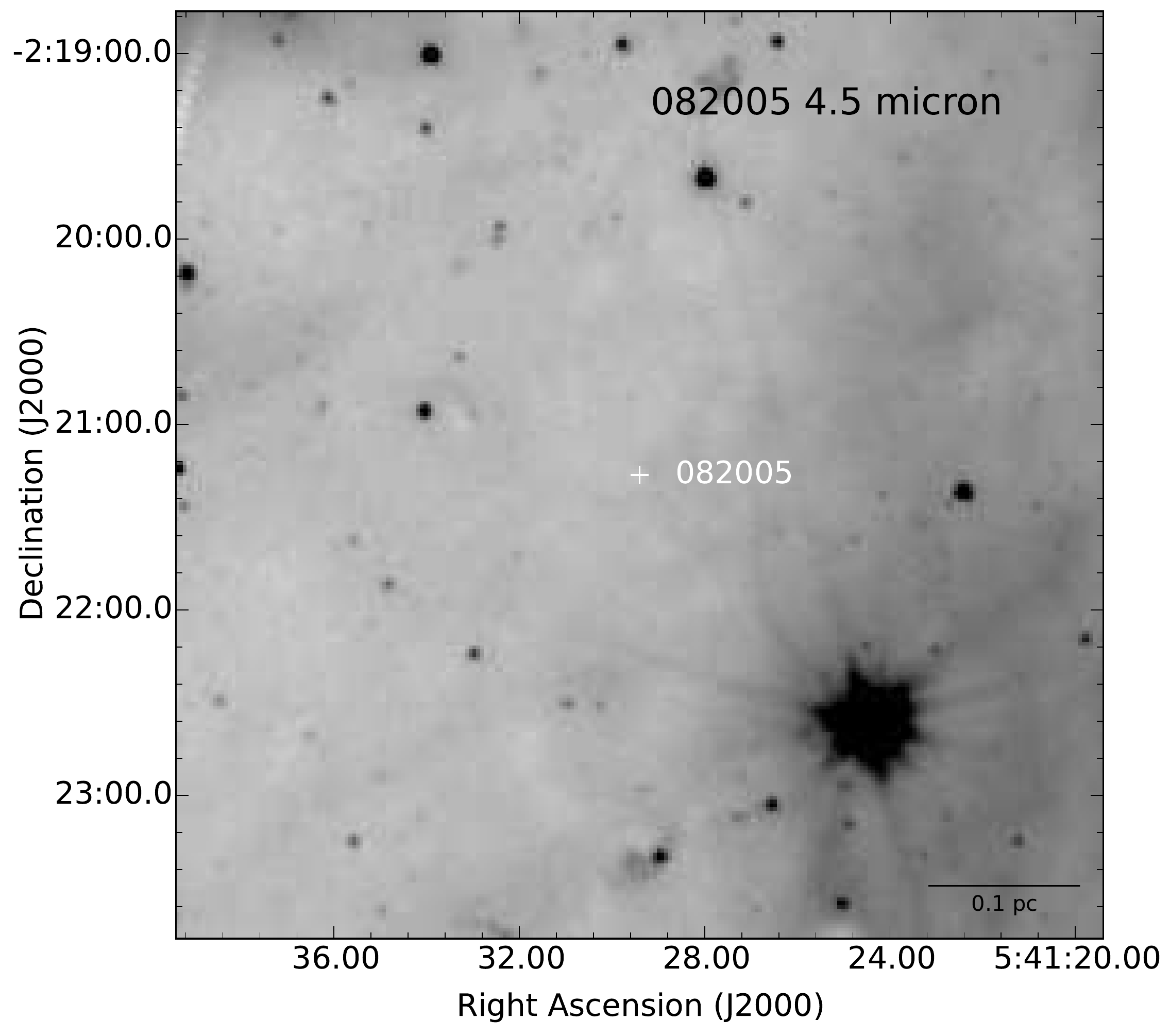}
\end{center}
\caption{\textit{Spitzer} 4.5 \micron\ images of the PBRS sample also having
CO ($J=1\rightarrow0$) observations.
This particular \textit{Spitzer} band has a bright H$_2$ feature commonly associated
with shock-excited outflow emission but is also sensitive to scattered light in the
outflow cavities.
 The protostars positions are marked with either white crosses or small circles and
 the outflow position angles are denoted by the red
and blueshifted arrows. The PBRS 093005, 090003, HOPS 373, and 302002 have compact 4.5 \micron\ emission
near the location of the protostars and no extended H$_2$ knots. The PBRS 061012, HOPS~223, 082012, and 119019
have indications of H$_2$ emission extended $\ga$0.1 pc from the protostars. Moreover, in the
061012 field the protostars HOPS 221 shows another apparent east-west outflow. The protostars
091015, 091016, 097002, and 082005 do not show evidence of any emission shortward of 70 \micron.
The source near the location of 097002 is another young star and not the PBRS (ST13).}
\label{spitzer-a}

\end{figure}

The CARMA $^{12}$CO observations are only sensitive to emission within the 30\arcsec\ (12600 AU) radius primary beam, hence
other observations are needed to determine if the outflows extend to larger scales.
We examined the \textit{Spitzer} 4.5~\micron\ images of all the sources from \citet{megeath2012}.
The emission at 4.5~\micron\ can trace both scattered light in the outflow cavities near the protostars
and shock-excited H$_2$ emission along the outflows. Smooth 4.5~\micron\ emission near
the source is likely indicative of scattered light and knotty or bow shock-like
features along the outflow are likely H$_2$ emission \citep[e.g.,][]{tobin2007}.
Images of the 4.5~\micron\ emission are shown for all the sources in Figure \ref{spitzer-a} and \ref{spitzer-b}.

Toward the sources HOPS 373, 093005, 302002, and 090003 there is 4.5 \micron\ emission
within 0.05 pc of the sources and no apparent evidence for emission out on larger-scales that 
is likely to have originated from these systems. Thus, for 093005, 302002, and 090003 we 
are likely covering the full extent of the outflow with our CO observations; for HOPS 373, however, the
outflow extends out of the primary beam, but perhaps not much further \citep{gibb2000}. 

There are a few cases where the association of the 4.5 \micron\ emission with the outflow is ambiguous. For 135003, there are some knotty
features along the direction of the known outflow, extending $\sim$0.15 pc and was identified as an outflow
candidate (SMZ 1-38) by \citet{stanke2002}. Then in the case of 019003,
we see a feature adjacent to the position of the protostellar source from the 2.9 mm continuum, and
possibly an extended feature in the direction of the blue-shifted outflow lobe. The crowding and number
of imaging artifacts from bright sources make this field difficult to interpret.

We only find clear evidence for 4.5 \micron\ emission extended $>$ 0.1 pc for three sources 
082012, 061012, and 119019. The bow-shock directions or trail of H$_2$ knots
indicate a likely origin from the PBRS source. The emission from 061012 and 119019 appears to extend $\sim$0.3 pc
and the emission from 082012 extends $\sim$0.2 pc. If we assume an outflow propagation speed of 10 - 100 \kms,
then the dynamical time is between 3000 - 30000 yr for 0.3 pc and 2000 - 20000 yr for 0.2 pc. Thus,
even though there is evidence for outflows toward these sources extending
relatively large distances, extreme youth is still likely.

Toward the sources without detected CO ($J=1\rightarrow0$) outflows, 091015, 091016, 082005, and
097002, there is also no evidence for 4.5 \micron\ emission (or emission shortward of 70 \micron) associated with the sources, as shown in
(see Figure \ref{spitzer-a} and \ref{spitzer-b}). Whereas, the sources with compact emission at 4.5~\micron\ also had detections
of CO outflows.

\subsection{Warm/Hot Outflow Gas}

We obtained \textit{Herschel} PACS spectroscopy toward a subset of the PBRS (eight observed with PACS).
This subset samples luminosities between 0.65 L$_{\sun}$ and 12 L$_{\sun}$ and
a variety of $^{12}$CO molecular outflow emission properties; thus, this subsample should
be reasonably representative of the PBRS as a whole. PACS spectroscopy offers a
complementary view of the outflow emission from protostars; rather than the cold, entrained
gas traced in the CO ($J=1\rightarrow$0) line, the PACS lines trace the warm/hot shock-heated 
portion of the outflow concentrated on scales $<$2000~AU.

The continuum-subtracted PACS spectra for all observed sources, extracted
from the central spaxel, are shown in Figure \ref{pacsspectra}. The spectra have
a wide variety of emission line strengths; detections in high-J CO and water are found
toward 5 out of the 8 PBRS. The spectrum toward HOPS 373 is particularly strong and
rich in line emission, detecting CO transitions with J$_{u}$ $>$ 30. Also, lines in the PACS spectrometer
range are detected toward all sources that exhibit a clear outflow in the CO ($J=1\rightarrow0$) transition.
We calculate the total high-J CO luminosities and give their values in Table 6.

Figure \ref{co14-13spectra} shows the non-continuum subtracted CO ($J=14\rightarrow13$) spectra
for the all observed sources. The PBRS 061012 has a tentative detection (2.5$\sigma$) in the CO ($J=14\rightarrow13$)
line, while its detection was not immediately apparent in the full spectrum shown in Figure \ref{pacsspectra}.
However, 061012 does not have detected emission in the 179.5 \micron\ H$_2$O 2$_{12}$-1$_{01}$ line
which typically has a line flux greater than or equal to 
the CO ($J=14\rightarrow13$) line. Thus, the detection toward 061012 is considered tentative.

Observations were also obtained toward all the PBRS in unchopped line spectroscopy observations
of the [OI] 63.18 \micron\ transition. This emission line is thought to be an
tracer of the protostellar jet, perhaps even before the molecular outflow is well-established \citep{hollenbach1989}.
Since these observations were conducted in the unchopped mode, 
extended [OI] emission from the cloud is present in the spectral cubes. This extended [OI] 
emission from the cloud must be subtracted from the data in order to isolate [OI] emission
from the protostar itself. To remove the extended [OI] (and continuum emission), we have
calculated the median intensity at each wavelength in the spectral cube using the 18 edge spaxels. We
also compute the standard deviation of the edge spaxel intensities at each wavelength, this is 
representative of the uncertainty in the background emission subtracted at each wavelength. We use the 
median intensity of the edge spaxels rather than the mean because some spaxels
have very high intensities and the mean would be skewed toward a value larger than most of the
edge spaxel intensities. The background subtracted [OI] spectra are shown toward 
each source in Figure \ref{OI-spectra} as the thick solid line and the standard deviation of the
background at each wavelength is shown as the thin dashed line in Figure \ref{OI-spectra}. 

The only PBRS with a clear detection of the [OI] line is HOPS 373; 019003
at first glance appears to have a detection, but it is $\sim$2$\sigma$ above the uncertainty
of the subtracted background [OI] emission, so this detection is tentative. 
Furthermore, 119019 and 061012 have apparent
peaks at location of the [OI] line; however, both of these are only 2$\sigma$ 
detections above the noise and other features are found in those spectral with the same significance, but
do not correspond to an expected spectral feature. Therefore, 
neither of these sources are regarded as detections.

\citet{nisini2015} showed a sample of protostellar sources with extended [OI] emission
in their jets and outflows. This highlights the possibility that some of the PBRS may have extended
emission along their outflows, and that our subtraction of background [OI] emission from the edge
spaxels may remove [OI] emission from the source. However, we inspected the spectral cubes before and after subtraction
of extended [OI] emission, and we do not detect any enhancement of [OI] emission along the 
outflow directions (for the PBRS with detected outflows), nor do the 8 pixels adjacent to the 
central spaxel show emission after background subtraction toward HOPS 373. Therefore, we conclude
that the well-detected emission toward HOPS 373 is only detected in the central spaxel and
we are not missing extended flux at our sensitivity ($\sigma_{[OI]}$~$\sim$~1~Jy~channel$^{-1}$), and
we are not subtracting off extended emission associated with the PBRS outflows.

We have examined the [OI] line luminosities with respect to larger protostar samples 
from \citep{green2013} and Mottram et al. (2016, submitted). The
$\sim$2$\sigma$ [OI] detections for 119019 and 062012 and 3$\sigma$ upper limits for 093005, 091015, and 091016 
have [OI] luminosities upper limits consistent with the detected range of [OI] luminosities 
for a given L$_{bol}$ \citep[L$_{[OI]}$~=~10$^{-5}$~-~10$^{-2}$~L$_{\sun}$ ; ][]{green2013}. Thus, the [OI] line is not found to be particularly strong toward the PBRS, but we cannot
say that the [OI] emission anomalously weak toward the PBRS given that the upper limits do not
indicate [OI] luminosities to be significantly lower than other protostars with a similar L$_{bol}$.

In addition to the [OI] 63.18~\micron\ line, we examined the spectra for [OI] emission at 145.5~\micron\ 
in the range scans. As shown in Figure \ref{pacsspectra}, this line is only detected toward 019003. However, we do not
think this is emission from the protostar itself, but extended emission that was not fully subtracted from
the off position as some spaxels have a negative feature, while others have emission.

%***TO FIX****
%The [OI] luminosity can also be used to calculate the outflow rate under the assumption that the [OI] emission is
%the result of a shock, $\dot{M}$ = L([OI]) $\times$ 8.1 $\times$ 10$^{-5}$ M$_{\sun}$ yr$^{-1}$ L$_{\sun}^{-1}$ \citep{hollenbach1985}.  
%The [OI] luminosities, upper limits, and 
%outflow rates inferred from the line luminosities are given in Table 5. 

The [OI] 63 \micron\ luminosity from post-J-shock gas can be used to calculate 
the mass flow rate through the shock \citep{hollenbach1989,hollenbach1985}: 
$\dot{M}$ = L([OI]) $\times$~8.1$\times$ 10$^{-5}$~M$_{\sun}~{\rm yr}^{-1}$~L$_{\sun}^{-1}$.  Since 
our observations encompass, in each case, all the regions in which the outflows from our targets 
drive shocks. Thus, the result is the mass-loss rate from the protostar, averaged over the 
outflow dynamical time. The [OI] luminosities and 
outflow rates inferred from the line luminosities (and their upper limits) are given in Table 6.

\subsubsection{Extended emission}
The high-J CO and water line emission is extended across multiple spaxels in some sources,
the most obvious of which is 135003. We overlay the spectra in each spaxel on the CO ($J=1\rightarrow0$) map in 
Figure \ref{135003-footprint} for the longer and shorter wavelength ends 
of the PACS spectrometer red channel. H$_2$O and CO emission is a detected in all spaxels
that overlap with the blue-shifted side of the CO ($J=1\rightarrow0$) 
outflow, and the line emission is actually brighter than that of the
central spaxel. However, there is not corresponding line emission extended along
the red-shifted side of the outflow, possibly indicating that the southern side
of the outflow is being driven into a less-dense medium.
Similarly, 019003 also has some extended H$_2$O and CO emission on the blue-shifted side of the
CO ($J=1\rightarrow0$) and like 135003 the 
extended emission is also brighter than the central spaxel.

\subsubsection{CO Luminosities and Rotation Temperatures}

We have calculated the high-J CO luminosities and rotation temperatures 
for the 5 PBRS with multiple detected CO transitions. We calculate the column densities of each
CO line and luminosity of each line following \citet{manoj2013}; however, instead of 
fitting Gaussian functions to the unresolved line profiles, we directly sum the spectral
elements around the wavelength of a particular CO line and subtract the background emission 
estimated from line-free continuum regions adjacent to the emission line. We regard this
method as more reliable than fitting Gaussians given the low spectral resolution of the data; similar
results are obtained for the Gaussian method, however (Manoj et al. 2016 submitted).

We show the rotation diagrams for the 5 sources with robust CO detections in multiple lines
in Figure \ref{rotdiagrams}. All sources show the characteristic warm component ($\sim$300~K) of the
CO rotation diagrams \citep[e.g.][]{vankempen2010,karska2013} and only HOPS 373 shows evidence
of another temperature component in CO lines with J$_u$ $\ge$25; all other PBRS have non-detections
for CO lines with J$_u$ $\ge$25. Thus, we fit a linear slope to the rotation diagrams for all 
detected CO lines with J$_u$ $\le$25, finding T$_{rot}$ between 216~K and 282~K. 
HOPS 373 has the highest T$_{rot}$ and 119019 has the lowest
T$_{rot}$.

We plot the PACS CO luminosities (L(CO)) versus L$_{bol}$ and T$_{bol}$ in Figure \ref{LCO-others}. The PBRS
have CO luminosities that are consistent with the observations from the HOPS, WISH, WILL and DIGIT\footnote{WISH
stands for Water In Star forming regions with \textit{Herschel}, WILL stands for WIlliam Herschel Line Legacy,
and DIGIT stands for Dust, Ice, and Gas, In Time. WISH and DIGIT were \textit{Herschel} key programmes
and WILL was an Open Time 2 programme.} samples 
\citep[][Mottram et al. submitted; Karska et al. in prep.]{karska2013,manoj2013,green2013,karska2014}\footnote{
The PACS CO data to be published in Karska et al. is a synthesis and updated analysis of the WISH, WILL, and DIGIT data, while
Mottram et al. focuses on the [OI], HIFI, and ground-based low-J CO observations of the WILL survey only.}. 
However, HOPS 373 has nearly the highest CO line luminosity for all protostars in the 
samples considered here for protostars with L$_{bol}$ $<$ 30 L$_{\sun}$. Looking
at L(CO) vs. T$_{bol}$, also in Figure \ref{LCO-others},
 the PBRS are comparable to other sources with low values of T$_{bol}$.

The comparison of CO T$_{rot}$ to the HOPS/WISH/WILL/DIGIT samples is shown in Figure \ref{Trot-others}; these rotation
temperatures are all measured using CO lines with 14 $\le$ J$_u$ $\le$25. The
PBRS have T$_{rot}$ values that are among the lowest observed for all protostars in the other sample at any luminosity.
However, given the uncertainties in our own measurements and those in the literature, the PBRS are consistent with 
the observed distribution of T$_{rot}$, but on the low-side of the distribution.
We discuss the possible causes for the PBRS have lower T$_{rot}$ values
further in Section 4.4.

\subsubsection{Far Infrared Line Ratios}
We calculated diagnostic line ratios that have been used by \citet{karska2014}
to compare the WISH and WILL observations with various shock models \citep{kaufman1996,flower2010,flower2015}
and list them in Table 7 for the sources with detected lines. For most ratios, the values calculated
for the PBRS are either within the range observed in the WISH/WILL samples \citep{karska2014} or the values are
within 1$\sigma$ of the observed range. The primary line ratio that is systematically
different from the WISH/WILL samples is the CO ($J=16\rightarrow15$)/CO ($J=21\rightarrow20$); 
the ratios are systematically larger for all the PBRS. This likely reflects the colder CO T$_{rot}$
values that are derived for the PBRS, relative to the WISH/WILL sources. We also list
ratios for CO ($J=17\rightarrow16$)/CO ($J=22\rightarrow21$), 
CO ($J=16\rightarrow15$)/CO ($J=17\rightarrow16$), 
and CO ($J=21\rightarrow20$)/CO ($J=22\rightarrow21$) because 
CO ($J=17\rightarrow16$) and CO ($J=22\rightarrow21$) are also accessible from 
SOFIA\footnote{Stratospheric Observatory For Infrared Astronomy https://www.sofia.usra.edu/}.

One source, HOPS 373, also had detections of OH transitions, enabling further comparison to
the WISH/WILL results. Note that one of the OH~84~\micron\ doublet lines is contaminated
by CO ($J=31\rightarrow30$) and to correct for this we measured the flux of the uncontaminated
doublet line and multiplied its flux by two. The ratio of
 OH~84~\micron\ to OH~79~\micron\ is larger than WISH/WILL, but 
within the uncertainties, H$_2$O (4$_{04}$-3$_{13}$) to OH~84~\micron\ 
is consistent with WISH/WILL, and CO ($J=16\rightarrow15$) to OH~84~\micron\ is slightly
in excess of the WISH/WILL results. Thus, for HOPS 373, the H$_2$O line emission relative to OH
is weaker than predicted by the shock models, consistent with the suggested interpretation of 
\citet{karska2014} that UV irradiation of the shocks is needed 
in order to explain the H$_2$O and OH line ratios as suggested by \citet{karska2014}.

%9.0*0.77**2*(5.0e-5*1.99e33/3.14e7)**2/(4.0*8.0*3.14*6.67e-8*0.1*1.99e33)/1.496e13

\section{Discussion}

The PBRS have been demonstrated, through multiple lines of evidence, to be consistent
with being the youngest known Class 0 protostars. Their SEDs indicate that they are 
surrounded by very dense envelopes (ST13) and this was further confirmed by the CARMA
2.9 mm dust continuum luminosities (Paper I). If these sources truly are a sample of the
youngest protostars, the results from the outflow diagnostics presented here can offer
valuable clues to the properties of outflows toward very young protostars. Given the multitude of the data presented,
including the continuum results from Paper I, we have compiled a list of
PBRS properties determined from the follow-up observational data and present a summary
of these data in Table 5.

\subsection{Nature of the PBRS Very Red Colors}

A principal uncertainty in the characterization of the PBRS was if the extremely
red 24~\micron\ to 70~\micron\ colors observed by ST13 were strongly influenced
by source viewing angle. If the PBRS were typical Class 0 sources and observed
in exactly edge-on orientation, then the combined opacity of the envelope and
disk midplane could result in the very red 24~\micron\ to 70~\micron\ colors. However, ST13 showed that
even if the PBRS were all viewed edge-on, the envelope densities would still have to be $>$2$\times$
higher than typically found toward HOPS protostars; the median envelope density for Class 0 protostars
in HOPS at a radius of 1000 AU is found to be 5.9 $\times$ 10$^{-18}$ g cm$^{-3}$ from
SED modeling \citep{furlan2016}.

For the sources with detected CO ($J=1\rightarrow0$) outflows, the clear spatial
separation of the blue and redshifted CO emission clearly shows
that 093005, 090003, 082012, HOPS 372, HOPS 373, 135003, 019004 are \textit{not} observed
with edge-on orientation and must be observed at an intermediate viewing angle 
(neither
edge-on nor face-on). The distribution of inclinations is consistent with being random;
therefore,
the extremely red colors of these protostars are not the result of extreme edge-on viewing
angle, but are due to the high density of the infalling envelope itself. 
We are unable to make a definitive conclusion about 061012 since the outflow is not clearly
detected, but there appear to be separated blue and red lobes.

However, for two sources, 119019 and 302002, only low-velocity
CO emission is found for those outflows. The outflow
toward 302002 (Figure \ref{302002}) had a small velocity gradient from across the source,
and we mentioned
in Section 3.1.3 that the inclination is likely between 50\degr\ and 80\degr, but closer to 80\degr.
In the more extreme case of 119019, this PBRS
had no detectable
velocity gradient and there is roughly equal amounts of emission at both blue
and redshifted velocities (Figure \ref{119019}. 
Thus, these two sources may only have been classified as PBRS because
their of edge-on (or nearly edge-on) orientation. 

In summary, we confirm that the extremely red colors of the PBRS
are not the result of inclination for 7 out of 9 sources 
with detected CO ($J=1\rightarrow0$) outflows. The sources without
detections of CO ($J=1\rightarrow0$) outflows may have low-velocity outflows
that are confused with the cloud emission, or the outflows are still too small in
spatial extent and are not bright enough to detect with the sensitivity of
our current observations.

\subsection{Outflow Properties}

The outflows exhibit a range of masses, momenta, energies, and forces;  HOPS 373 has
outflow properties typical of those in \citet{plunkett2013} and 082012 has outflow properties
in excess (Table~4). In contrast, the two most compact outflows in the sample (090003 and 093005)
have quite low outflow masses, momenta, energies, and forces.
Since the \citet{plunkett2013} sample includes single-dish data to 
measure the total flux, a comparison with \citet{arce2006}, using interferometer-only data, is more appropriate.
The ranges for the observed outflow parameters from \citet{arce2006} and \citet{plunkett2013} are given in Table 4. 
We note, however, that neither of those studies computed F$_{CO}$.
The sources
093005, 090003, and 302002 have values all less than the range from \citet{arce2006}, HOPS~223 is
within the range, and HOPS~373 and 082012 have values in excess of these numbers. 

The outflow toward 082012 is truly exceptional, its high-velocity nature was first reported
by \citet{sandell1999}; it is more energetic and has more 
momentum than the strongest outflows in the \citet{plunkett2013} sample. The increased collimation 
and large velocity extent bears resemblance to NGC 1333 IRAS 4A, L1448C \citep{hirano2011}, 
and IRAS 04166+2706 \citep{santiago2009}. 
This outflow has energies and momenta in excess of all the outflows observed by
\citet{arce2006} and \citet{plunkett2013}, but it is comparable to L1448C \citep{hirano2011}.
The outflow of 082012 is likely even more
powerful than we measure it to be, given that our properties are lower-limits due to lack of $^{13}$CO
observations to determine the optical depth and because we do not cover the full extent of the outflow.
The outflow of 082012 is also likely blended with that of HOPS 372 at low velocities, but 
at higher velocities it appears to only come from 082012.
Even if we are measuring the combined outflow 
properties, it is very strong relative to those observed in the nearby star forming
regions.

The outflows from 090003 and 093005 represent the most compact (i.e., shortest) CO outflows
found in our data.
The outflows of 093005 and 090003 are not observed to 
extend further than their apparent envelope sizes observed at 870 \micron.
This and the compact 4.5 \micron\ emission
may indicate that the outflows are just beginning to break out from 
their dense, natal envelopes.
These outflows are not particularly powerful either, the outflow forces
plotted in Figure \ref{outflow-lbol} are on the low-end for Class 0 sources and
090003 is lower that the linear relationship from \citet{bontemps1996}, above which 
all Class 0s lie in current data \citep[][Mottram et al. submitted]{yildiz2015}.
Furthermore, the well-developed outflow from 135003 is also found to lie below the 
L$_{bol}$ vs. F$_{CO}$ relationship.
Alternatively, the outflows could be more powerful, but since their energies and momenta
are calculated using entrained material, observed CO ($J=1\rightarrow0$), the outflows only appear weak with these
measures. 

The deeply embedded sources without 4.5 \micron\ emission or outflow 
detections (097002, 091015, 091016, and 082005) may have outflows that are
 too weak/faint to detect in our observations.
However, the lack of outflow detections toward these most embedded sources
and the lack of particularly powerful outflows from 093005 and 090003, could indicate that
outflows may be weak during the early Class 0 phase, given the apparent youth of
 the sources and small spatial extent of the outflows. 
Thus, it possible that the outflow momentum/energy/force may be initially small early-on and
are rising early in the Class 0 phase
such that the Class 0 outflows will
be systematically more powerful than Class I outflows \citep[e.g.,][]{bontemps1996,yildiz2015}. Weak initial
outflows from protostars are predicted from simulations of the FHSC phase \citep{tomida2013,price2012}
where the outflows are $<$15~\kms. If the PBRS have recently transitioned out of the FHSC phase,
then they may not have reached their full outflow power as of yet. This
will be further studied using single-dish data by Menenchella et al. (in prep.).

The absence of detected outflow activity in CO ($J=1\rightarrow0$) 
toward the four sources
mentioned above cannot be construed as evidence of outflow absence
because of our finite resolution and sensitivity. 
For example, the outflow toward OMC MMS6N (also known as HOPS 87) 
was only detected when it was observed
at the highest resolutions with the SMA \citep{takahashi2011}, due to its
very small spatial extent. Thus, the non-detected outflows could be very compact and
in the process of breaking out from the envelopes, necessitating higher resolution data.
On the other hand, OMC MMS6N did have strong H$_2$O and CO emission lines observed
in the far-infrared spectrum from \textit{Herschel} \citep{manoj2013} and 091015/091016 had no detected
emission lines in their PACS spectra. In contrast, 091015/091016 are low-luminosity sources (L=0.65~L$_{\sun}$ and 0.81~L$_{\sun}$)
and OMC MMS6N is a higher-luminosity source (L $>$30~L$_{\sun}$), making direct comparisons between the sources 
difficult.

\subsection{Relationship of Outflows and 2.9 mm Continuum Properties}

In Paper I, the 2.9 mm continuum luminosities and visibility amplitude
profiles were analyzed. We found that most PBRS had
2.9 mm continuum luminosities (median of 1.0$\times$10$^{-5}$ L$_{\sun}$) and L$_{2.9mm}$/L$_{bol}$ 
ratios (median of 8.8$\times$10$^{-6}$) greater than most nearby Class 0 protostars, which have
a median L$_{2.9mm}$ = 3.2$\times$10$^{-6}$ L$_{\sun}$ and a median L$_{2.9mm}$/L$_{bol}$ = 8.5$\times$10$^{-7}$.
The nearby Class 0 continuum samples are drawn from \citet{tobin2011}, \citet{looney2000}, and \citet{arce2006}, which are sensitive to comparable spatial scales;
L$_{2.9mm}$ is calculated assuming a 4 GHz bandwidth centered at 2.9 mm.
The PBRS
have a median L$_{2.9mm}$ that is 3$\times$ larger that typical Class 0s and L$_{2.9mm}$/L$_{bol}$ that 
is 10$\times$ larger. This means that the more nearby Class 0 protostars with high L$_{2.9mm}$ also
have a high L$_{bol}$, whereas the PBRS tend to have lower L$_{bol}$. Furthermore, the
highest L$_{2.9mm}$ for nearby Class 0 protostars is 2.9$\times$10$^{-5}$~L$_{\sun}$ 
toward NGC 1333 IRAS 4A, in contrast to the highest L$_{2.9mm}$ 
of 3.4$\times$10$^{-5}$~L$_{\sun}$ for the PBRS 082012;
see Figure 2 from Paper I.
Finally,
6 out of 14 PBRS (093005, 090003, 091016, 091015, 097002, and 082005) had flat 
visibility amplitude profiles (and small 5 k$\lambda$ to 30 k$\lambda$ visibility
amplitude ratios), consistent with most emission being emitted from scales $<$ 2000 AU (Figures 3 and 4 from Paper I).
Thus, the PBRS tend to have more massive envelopes relative to their bolometric luminosities as
compared to other Class 0 sources and the flat visibility amplitude ratios indicate 
high densities in the inner envelopes (Paper I).

Here we more closely examine the two PBRS have apparent
inclination angles that are close to edge-on: 119019, being almost
exactly edge-on, and 302002 being near 80\degr (between 50\degr\ to 80\degr).
The PBRS 119019 has L$_{2.9mm}$/L$_{bol}$ (1.47$\times$10$^{-6}$) and L$_{2.9mm}$ (2.3$\times$10$^{-6}$~L$_{\sun}$)
values consistent with typical Class 0 protostars from the
literature. Thus, in addition to having an nearly edge-on outflow, 
the 2.9~mm continuum emission from 119019 is not consistent with it
having a massive, dense envelope like the rest of the PBRS (Table 5). 
This points to 119019 perhaps being more evolved than the rest of the PBRS
and its very red colors can be attributed to an edge-on inclination. On the other hand, 302002 
has values of L$_{2.9mm}$/L$_{bol}$ (1.2$\times$10$^{-5}$) and L$_{2.9mm}$ (1.0$\times$10$^{-5}$~L$_{\sun}$)
consistent with rest of the PBRS. Both of these
sources also have declining visibility amplitudes (Paper I).

We also find a tendency for the PBRS with flat visibility amplitudes to show either a
compact outflow or have no detectable outflow in the CO ($J=1\rightarrow0$) line and
\textit{Spitzer} 4.5 \micron\ emission. We suggested
in Paper I that the PBRS with flat visibility amplitudes might be less-evolved than 
the PBRS with more rapidly declining visibility amplitudes. The sources with rapidly
declining visibility amplitudes tend to have more extended, well-developed outflows 
(i.e., 082012, HOPS 373, and 119019) than sources with flat visibility amplitudes.
Therefore we suggest that the flat visibility amplitude sources have
outflows that are only beginning to break out of their envelopes.
Thus, the PBRS with flat visibility amplitudes may indeed be the 
initial stages of the Class 0 protostellar
phase.

The change in visibility amplitude profile could be related to the outflows
carving out cavities and lowering the overall mass of the inner envelope. On the other hand, if
the inner envelope mass is rapidly accreted onto the protostar, then the visibility amplitude profiles 
would also dramatically decline. Using the example from Paper I, the
free-fall time of 2 $M_{\sun}$ confined to a constant density 
sphere with R = 1500~AU is only the $\sim$10,000 yr, quite short on the timescale of protostellar collapse.
For the case of inside-out collapse \citep{shu1977}, the rarefaction wave would take $\sim$36,000~yr to
propagate out 1500~AU (assuming a sound speed of 0.2~\kms), the boundary of the rarefaction wave is where the
density profile changes from r$^{-2}$ to r$^{-1.5}$, reflecting free-fall collapse. Moreover, in the case of strong rotation,
a portion of the density profile inside of the rarefaction wave can have a density profile of
r$^{-0.5}$ \citep{cassen1981,tsc1984}. Thus, in either case, the density structure of the
inner envelopes can be significantly altered on a timescale shorter than the 
Class 0 phase \citep[$\sim$150,000 yr,][]{dunham2014}.
Thus, the outflow detection and extents may simply correlate with the decrease 
in the visibility amplitude profiles and not cause it.

Lastly, the only flat visibility amplitude source with detected far-infrared line emission
is 093005; only continuum emission was detected toward 091015 and 091016. The remaining sources
with line emission had declining or uncertain visibility amplitude profiles.

\subsection{Far-Infrared Diagnostics in the Context of the PBRS}

A key finding of our study is that 
in the absence of other outflow indicators (CO ($J=1\rightarrow0$), 
\textit{Spitzer} 4.5~\micron\ scattered light/H$_2$), the PACS line emission (CO, H$2$O, or [OI])
does not independently show evidence for outflows in the form of shocks from the inner envelopes 
of the protostars.
Thus, we only find far-infrared line emission toward
sources that have detected CO ($J=1\rightarrow0$) outflows. This hints at a strong
link between the mechanisms that produce the cold CO outflows and the warm/hot component
observed in the far-infrared. Furthermore, the [OI] 63~\micron\ transition is only
convincingly detected toward 1 PBRS (HOPS~373) out of the 6 PBRS 
for which we could reliably subtract the background [OI] emission
from the edge spaxels.
We do not consider the detections and non-detections
of 135003 and 019004 meaningful because of the strong, extended, and
spatially variable [OI] emission in the OMC2/3 region. 
HOPS 373 has one of the more well-developed
outflows, has an H$_2$O maser \citep{haschick1983}, and has the brightest line spectrum
of all the PBRS.

\citet{hollenbach1989} predict strong far-infrared CO and [OI]~63~\micron\ emission for
densities $>$ 10$^3$ cm$^{-3}$ 
for fast, dissociative J-shocks with velocities $>$30~\kms.
The [OI] luminosity detected toward HOPS 373 is comparable to other protostars with 
similar luminosity \citep{green2013}. While the tentative detections and non-detections toward the remaining PBRS
do not point to anomalously weak [OI], we can confirm that the PBRS do not have exceptionally strong
[OI] emission. Thus, we conclude that the outflows from the PBRS that give rise to the 
[OI] and high-J CO luminosities appear comparable in those tracers of other Class 0 protostars.

If PBRS are typical of the youngest protostars, early Class 0 protostars, 
then we posit that outflows may be very weak initially. 
At a minimum, 
the PACS [OI] and CO observations, in addition to CO ($J=1\rightarrow0$),
demonstrate that the PBRS are not accompanied by significantly 
stronger outflows than typical Class 0 protostars.
While the PBRS are inconsistent
with the expected properties of first hydrostatic cores (FHSC) due to their luminosities
and colors (ST13), the outflows predicted from FHSCs are quite 
weak $<$ 15~\kms\ \citep{tomida2013,price2012}. The outflows are expected to increase
in velocity as the source evolves, though the simulations did not follow the longer
term evolution. Such slow outflows from the PBRS would be consistent with them
having recently transitioned out of a FHSC phase. If the outflow power is
directly linked to the mass accretion rate, then the time in which protostars have
very low outflow power is likely quite short $<$ 10000 yr, consistent with the
apparent youth of the PBRS.

Alternatively, at 63~\micron\ the opacity from the infalling envelopes may be obscuring the
[OI] emission. Following \citet{kch1993}, the optical depth through an envelope with a density profile consistent with free-fall (r$^{-1.5}$)
density profile \citep{ulrich1976} is given by 
\begin{equation}
\tau_{\lambda} = \frac{\kappa_{\lambda}\dot{M}}{2\pi(2GM_*)^{1/2}}r^{-1/2}
\end{equation}
where $\kappa_{\lambda}$ is the wavelength dependent dust opacity,
$G$ is the gravitational constant, $\dot{M}$ is the mass infall rate, $M_{*}$ is
the protostar mass, and $r$ is the inner radius for which the optical depth
is being calculated. M$_{*}$ is taken to be 0.5 M$_{\sun}$, which is adopted to set the envelope density
for a given infall rate; the absolute value for the mass is not important, only the envelope density.
Under the assumption of free-fall collapse, the
infall rate is directly proportional to the envelope density 
\begin{equation}
\rho_{1000} = 2.378 \times 10^{-18} \left(\frac{\dot{M}_{env}}{10^{-5} M_{\sun} yr^{-1}}\right) \left(\frac{M_*}{0.5 M_{\sun}} \right) g\ cm^{-3}
\end{equation}
which is the volume density at a radius of 1000 AU, following the notation of \citet{furlan2016}. 
From spectral energy distribution model fitting to the Orion protostars \citep{furlan2016}, 
the Class 0 protostars in Orion had median $\rho_{1000}$ of 5.9$\times$10$^{-18}$~g~cm$^{-3}$ with 
upper and lower quartiles of 1.8$\times$10$^{-18}$~g~cm$^{-3}$ and 1.8$\times$10$^{-17}$~g~cm$^{-3}$. The PBRS considered here
are modeled by \citet{furlan2016} to have a median $\rho_{1000}$ of 1.8$\times$10$^{-17}$~g~cm$^{-3}$, 
and the SED fits tend to prefer densities of 3$\times$ to 10$\times$ higher than the typical and lowest 
density Class 0 protostars, respectively.

This difference in density translates to significantly more opacity at 63~\micron\ for the PBRS, a factor
of 4$\times$ to 13$\times$ higher than the median Class 0 density and lower quartile; this results in a transmission
of only 0.09 for a typical PBRS, versus 0.55 and 0.84 for the Class 0 
median and lower Class 0 quartile, respectively. High opacity may be a particularly 
important consideration for 093005 which has a clear outflow
in CO ($J=1\rightarrow0$), PACS CO, and H$_2$O emission but without [OI] emission.

The high envelope opacities can also influence the CO rotation temperatures because the increasing
optical depth at shorter wavelengths would cause the rotation temperatures to decrease due
to flux attenuation of the line emission. To characterize the
magnitude of this effect, we examined the difference in transmission for the PACS CO lines
 down to a radius of 1000 AU 
\citep[where much of the PACS CO emission appears to be emitted,][]{green2013,manoj2013}.
For typical Class 0 envelope densities ($\rho_{1000}$ = 5.9$\times$10$^{-18}$~g~cm$^{-3}$),
the typical density of the PBRS envelopes $\rho_{1000}$ = 1.8$\times$10$^{-17}$~g~cm$^{-3}$, and assuming
dust opacities from \citet[][Table 1, column 5]{ossenkopf1994}, we found that 
the 3$\times$ higher envelope density could decrease the CO rotation temperatures by $\sim$20~K. Thus, 
the CO rotation temperatures of 220~K - 230~K would be higher if corrected for optical depth, making them even more consistent
with the WISH/WILL/DIGIT/HOPS samples

\section{Summary and Conclusions}

We have presented an observational study of both the cold and warm/hot molecular gas in outflows
from the youngest known protostars in the Orion molecular clouds, the PACS Bright Red Sources (PBRS).
The cold gas was probed toward 14 out of 19 PBRS 
using observations of the CO ($J=1\rightarrow0$) transition from CARMA, and
the warm/hot gas was examined  for 8 out of the 19 PBRS using full spectral scans (55~\micron\ to 200~\micron) from the
\textit{Herschel} PACS far-infrared spectrometer. Finally, we also examined \textit{Spitzer} 4.5 \micron\ imaging
to look for evidence of both compact and extended outflow activity from both scattered light and shocked H$_2$ emission.
The results from the follow-up work done in this study and Paper I demonstrate the 
critical need for complementary data in the determining the nature of protostellar
sources that are otherwise only characterized by their SEDs. 
Our main conclusions are as follows.

1. We detect clear outflows toward 8 out of 14 PBRS (119019, 090003, 093005, 
135003, HOPS~373, 082012, and 019003) in the CO ($J=1\rightarrow0$) molecular 
transition. There is tentative evidence for outflows toward an additional three PBRS 
(HOPS~372, 302002, and 061012). 
We also detect outflows from two non-PBRS HOPS 223, a FU Ori-like
outbursting protostar \citep{fischer2012} and HOPS 68 \citep{poteet2011};
the HOPS 68 outflow also appears to be quadrupolar. No detectable
outflow activity is found toward the PBRS 097002, 082005, 091015, and 091016 in
CO ($J=1\rightarrow0$), 4.5 \micron\ emission, or far-infrared spectroscopy (only 091015 and 091016).

2. The outflows toward 090003 and 093005 are the most compact, subtending less 
than 20\arcsec\ (8400 AU) in total extent, having dynamical ages 
$\le$2,500 yr. These outflows are also found to have
momenta, energies, and forces 
that are at the low end for Class 0 protostars. 
This observation, in addition to the lack of detectable
outflows toward several other PBRS, leads us to suggest that outflows may start out 
weak in protostellar sources and become more energetic with time. These sources are also the only ones
with flat visibility amplitudes to have detected outflows 
and we find a tentative tendency for the sources with flat visibility amplitudes in the
2.9~mm continuum (see Paper I) to either have no detected outflow activity or 
the most spatially compact outflows. This is further
evidence for the sources with flat visibility amplitude being among the youngest protostars
and the youngest PBRS.

3. The outflow from 082012 is extremely powerful, with red-shifted 
emission detected out to $+$40 \kms\ from line center
and extent greater than the CARMA primary beam. Its total energy is in 
excess of any individual outflow in the NGC 1333 
star forming region \citep{plunkett2013}
and comparable to some of the most powerful
known outflows from Class 0 protostars \citep[e.g.,][L1448C]{hirano2011}.

4. We detect far-infrared CO emission lines toward 6 out of the 8 PBRS observed. H$_2$O lines
are detected toward 5 out of 8 PBRS, and OH and [OI] are detected toward 1 PBRS. 
The far-infrared
CO, H$_2$O, and [OI] lines do not reveal outflows in the absence of outflow detections from other
diagnostics. 
The CO luminosities and [OI] detections/upper limits are consistent with the results from larger samples
of Class 0 protostars. However, the CO rotation temperatures tend to be 
lower than the typically observed 300~K CO rotation temperature
for protostars; however, given the uncertainties the PBRS are consistent with the larger samples.
Nevertheless, with a simple calculation of envelope
opacity to a radius of 1000~AU, we find that the observed rotation temperatures of the PBRS
could appear $\sim$20~K lower due to envelope opacity, given that the PBRS seem to have
denser envelopes than typical Class 0 protostars.

We wish to thank the anonymous referee for excellent suggestions which
have significantly improved the quality of the manuscript.
The authors also wish to acknowledge fruitful 
discussions with M. Dunham, L. Kristensen, and J. Mottram regarding this work.
J.J.T. is currently supported by grant 639.041.439 from the Netherlands
Organisation for Scientific Research (NWO).
J.J.T acknowledges past support provided by NASA through Hubble Fellowship 
grant \#HST-HF-51300.01-A awarded by the Space Telescope Science Institute, which is 
operated by the Association of Universities for Research in Astronomy, 
Inc., for NASA, under contract NAS 5-26555. The work of A.M.S. was supported by the Deutsche
Forschungsgemeinschaft priority program 1573 ('Physics of the
Interstellar Medium'). 
AK acknowledges support from the Foundation for Polish 
Science (FNP) and the Polish National Science Center grant 2013/11/N/ST9/00400. 
This work is based in part on observations 
made with Herschel, a European Space Agency Cornerstone Mission with significant 
participation by NASA. Support for this work was provided by NASA through an award 
issued by JPL/Caltech.  We are very grateful to have had the opportunity
to conduct these follow-up observations with the CARMA array in California. The discontinuation
of support for this productive facility is a loss that will continue 
to be felt into the future. Support for CARMA construction 
was derived from the states of Illinois, California, and Maryland, the 
James S. McDonnell Foundation, the Gordon and Betty Moore Foundation, the Kenneth 
T. and Eileen L. Norris Foundation, the University of Chicago, the Associates of the 
California Institute of Technology, and the National Science Foundation. Ongoing 
CARMA development and operations are supported by the National Science Foundation 
under a cooperative agreement, and by the CARMA partner universities. 

{\it Facilities:}  \facility{CARMA}, \facility{\textit{Herschel}}, \facility{\textit{Spitzer}}, \facility{Magellan}
\appendix

\section{Individual Sources}

\subsection{HOPS 373}

HOPS~373 is the close neighbor of 093005, located 110\arcsec\ 
to the south. The dust continuum emission observed in D-configuration
only showed some asymmetry and the combined D and C configuration
data resolved a second component, separated by 4\arcsec\ (Paper I). An outflow was
previously detected in CO ($J=3\rightarrow2$) observations 
with the JCMT \citep{gibb2000} and an associated water maser
by \citet{haschick1983}. Our observations of CO ($J=1\rightarrow0$) 
in Figure \ref{HOPS373} show that the outflow has quite a wide
angle and is extended beyond the primary beam. We also tentatively detected an
outflow originating from the secondary source that has blue and red-shifted lobes
opposite of the main outflow. The wide separation of the blue and red-shifted 
lobes indicates that the source is viewed at an inclination angle between 50\degr\ and 80\degr.
There is higher-velocity redshifted emission observed away from the
continuum source toward the edge of the primary beam. 

The far-infrared line emission from this source is quite intense, detecting 
[OI], OH, CO, and H$_2$O. The line emission from this source is the third brightest
of all HOPS protostars and the only PBRS in our sample with confidently detected [OI] and OH emission.

\subsection{093005}
The reddest PBRS, 093005, is located at the intersection of 
three filaments and $\sim$110\arcsec\ north of HOPS 373 (ST13).
At wavelengths shorter than 70~\micron, 093005 was only 
detected in \textit{Spitzer} 3.6~\micron\ and 4.5~\micron\ imaging (Figure \ref{spitzer-a}). 
The 4.5~\micron\ emission could be indicative of shocked H$_2$ emission 
and/or scattered light in an outflow cavity. Thus, a detection at 4.5~\micron\
is indicative of possible outflow activity
toward this source. We clearly detect the CO outflow 
originating from 093005, as shown in Figure \ref{093005}. The outflow appears 
compact with an offset between the red and blueshifted lobes of only $\sim$3\arcsec.
The position-velocity diagram of the outflow simply 
shows high-velocity emission offset from the protostar
position, not the typical increasing velocity with distance 
as typical for many protostellar outflows \citep{arce2007}. The features could result
from a compact bow shock component as the outflow begins to break out from its envelope.
However, the resolution of our observations 
was only $\sim$3\arcsec\ (1200 AU), making clear determinations as to the nature
of the high-velocity features difficult. The relative 
velocities of the red and blue-shifted lobes and their
close spatial location indicate that the source is not 
oriented edge-on and is at an inclination angle of $\sim$30\degr. 
Compact bow-shocks viewed at an intermediate inclination could 
show observed morphology \citep{arce2007}. Far-infrared CO and H$_2$O 
line emission is also clearly detected toward this source.

\subsection{090003}

The PBRS 090003 \citep[also called Orion B9 SMM 3; ][]{miettinen2012a} is 
located in a loose filamentary complex north of NGC 2024 with several protostars and 
starless cores over a 0.5 pc region \citep{miettinen2012a}.
Much like 093005, the only detection shortward of 24 \micron\ for this source is at 4.5 \micron, where there
is a small feature offset from the location of the protostar. This may be
indicative of a knot of shocked H$_2$ emission \citep{miettinen2012a,stutz2013}. 
The CO ($J=1\rightarrow0$) outflow from this source appears 
similar to that of 093005 and is indicative of $\sim$30\degr\ inclination, as
shown in Figure \ref{090003}; however, in contrast, there is a more spatially extended, 
low-velocity component. The high-velocities near the 
source and low velocities extended away from the source could be 
indicative of a wide-angle wind driving this outflow. 
Moreover, only $\pm$1 \kms\ around the systemic 
velocity is corrupted by $^{12}$CO emission from the cloud, 
so we are able to see lower-velocity features than in 093005. 
\citet{miettinen2012a} observed $^{13}$CO ($J=2\rightarrow1$) 
with APEX ( $\sim$ 30\arcsec\ resolution) and did not detect any indication of outflow
emission from 090003, suggesting that the outflow is quite compact.

\subsection{082012 and HOPS 372}

The outflow from 082012 is the brightest and one of the two most spatially extended outflows in 
the sample. Moreover, the outflow is visible over the largest 
velocity range (aside from HOPS 68) as shown by the 3 panels integrated at low, moderate, and
high velocities in Figure \ref{082012}. \citet{sandell1999} previously reported
single-dish CO ($J=3\rightarrow2$) and continuum maps at 450~\micron\ and 850~\micron toward this 
region. They resolved the dust emission around both protostars, and found a high-velocity outflow, 
consistent with our data, but mapped over a larger region, $\pm$150\arcsec\ from the source.

The clear separation of the
blue and redshifted lobes indicates an intermediate orientation of the source(s).
 The driving source of the collimated,
high-velocity emission seems to be 082012; however, at lower velocities the red-shifted
lobe extends back to HOPS 372 and there is blue-shifted emission that
appears associated with HOPS 372 as well. Thus, the two outflows are nearly parallel
and are perhaps interacting, but at a minimum their emission is clearly overlapping at lower velocities.

The highest observed outflow velocities toward 082012 are in excess 
of $\pm$40 \kms\ with multiple components being evident in the PV diagram and
we can see the characteristic `Hubble-flow' in the PV diagram.
Furthermore, there are also red and blue-shifted
CO emission clumps nearly orthogonal to the main outflow 
of 082012 which could be yet another outflow in the region. Furthermore, 
there are extended H$_2$ knots along the position angle of the outflow from
082012 as shown in Figure \ref{spitzer-a}.

\subsection{135003}

The PBRS 135003 is located in the OMC2/3 region of the Orion A cloud and located near OMC2-FIR6.
The outflow from 135003 is well-collimated on the blue-shifted side, another source with a characteristic
`Hubble-flow' in the PV diagram, see Figure \ref{135003}. The red-shifted, however,
 side does not appear as well-collimated near the 
source, but there is another red-shifted feature along the position angle, but outside the primary beam.
An initial outflow detection was reported for this source by \citet{shimajiri2009}, consistent with our
measured position angle. Moreover, their single-dish CO ($J=3\rightarrow2$) data show that the outflow
does extend outside our primary beam.
The \textit{Spitzer} 4.5 \micron\ map in Figure \ref{spitzer-b} shows a few knots of emission extending 
along the blue-shifted side of the outflow. H$_2$ imaging from \citet{stanke2002} (SMZ 1-38) shows emission along
both the northern (blue-shifted) and southern (red-shifted sides of the outflow). This source also shows
bright far-infrared CO and H$_2$O features along its outflow, 
coinciding with the blue-shifted side of the outflow as shown in Figure \ref{135003-footprint}.
We do not detect an outflow from its neighbor HOPS 59 within our sensitivity limits in low-J CO or PACS 
far-infrared lines.

\subsection{019003}

The PBRS source 019003 is also located in the OMC 2/3 region, northward of 135003. 
In Paper I, we detected 2 continuum sources toward the location of 019003 that were 
separated by $\sim$10\arcsec; the source associated with the PBRS is 019003-A and the other appears
starless and is denoted 019003-B (Paper I). We detect an apparent outflow from 019003-A as shown in Figure \ref{019003} and 
the 4.5 \micron\ emission is also offset from the
main outflow axis, similar to 090003, HOPS 373, and 302002. The surface 
brightness of the outflow is low, thus its detection is not as definitive
as some of the others due to the crowded, confused region. Finally, there was no complementary
detection in H$_2$ from \citet{stanke2002}.

\subsection{HOPS 68}

The Class I protostar HOPS 68 is detected at the edge of the primary beam in the 019003 field. An outflow
is well-detected from this source; the red-shifted lobe falls within the half-power point of the primary beam, while the blue-shifted lobe is
located just outside the half-power point. The velocity distribution of the outflow indicates that it is located
at an inclination angle of 50\degr\ from comparison to the models of \citet{cabrit1986}.
 An intermediate outflow inclination was necessary for a model by \citep{poteet2011} to explain
the relatively flat SED between 3.6 \micron\ to 24 \micron, deep silicate absorption
feature, and crystalline silicate features observed with \textit{Spitzer}; the crystalline dust is
postulated to have been formed by shocks in the outflow cavity walls. 
Furthermore, there are apparently two outflows from this source, a lower 
velocity outflow that is more east-west in orientation (PA $\sim$ 230\degr), 
but at higher velocities there is an apparent shift and the outflow is more
north-south in orientation (PA $\sim$ 160\degr). However, the blue-shifted 
side has both components out to the highest velocities
we can measure. We overlay the integrated intensity maps of the 
high-velocity emission on a Ks-band image
in Figure \ref{HOPS68}, and there are apparent outflow features 
associated with both the northwest-southeast component and the
northeast-southwest component. H$_2$ imaging by \citet{stanke2002} confirms that
these knots are H$_2$ emission (SMZ 1-28) and they also suggested that the driving source was
FIR2 from \citet{chini1997} (coincident with HOPS 68).

\subsection{302002}

The PBRS 302002 is located at the end of an isolated filamentary structure in NGC 2068, it is located 
$\sim$20\arcsec\ to the south of a Class I source (HOPS 331). 302002 is undetected at 24 \micron\, but does seem to 
have emission at 4.5 \micron, indicative of an outflow cavity or shocked H$_2$ emission
in the outflow, see Figure \ref{spitzer-b}.
We show in Figure \ref{302002} that there appears to be outflow emission associated with this source; however, the
blue and red-shifted emission are not located on the same position angle from the source. The blue-shifted emission is narrow and
extends out to the edge of the primary beam. The red-shifted emission on the other hand is quite compact and located in
a single clump offset west of the protostar. The CO outflow direction is marginally consistent with the apparent
orientation in the 4.5 \micron\ imaging. The poorly detected outflow and low-velocity of the emission may indicate
that this source is close to edge-on.
From the comparison to \citet{cabrit1986}, the outflow could be between
50\degr\ and 80\degr\, but likely closer to 80\degr.

\subsection{061012 and HOPS 223}

The PBRS 061012 is located very near the outbursting protostar V2775 Ori (HOPS~223) in the L1641 region \citep{fischer2012}.
The outflow toward 061012 cannot be unambiguously disentangled from that of HOPS 223 in the integrated velocity
map shown in Figure \ref{061012}. However, looking at the PV diagram of the $^{12}$CO emission centered on the
continuum source of 061012, we do see evidence of higher velocity emission near the protostar. The position
angle of the outflow is estimated from the resolved 4.5 \micron\ emission shown in \citet{stutz2013} and there
are H$_2$ emission knots at 4.5 \micron\ extending almost 0.3 pc from the source (Figure \ref{spitzer-b}). Thus, 
we may be detecting an inner, compact outflow toward this protostar. The outflow from HOPS 223 appears quite wide, bright
and clumpy in the integrated intensity map and PV diagram in Figure \ref{061012}.
The clumpiness could in part result from spatial filtering and that the source is toward the edge of
the primary beam with increased noise. However, episodic mass ejection episodes could contribute
to the clumpiness of the outflow emission, which has been seen in outflow data toward HH 46/47 \citep{arce2013}.

\subsection{091015 and 091016}

The PBRS 091015 and 091016 are close neighbors in NGC 2068, 091016 being $\sim$40\arcsec\ east of 091015;
these sources are completely undetected at wavelengths shortward of 70 \micron.
We do not detect evidence of outflow emission from these sources
at any wavelength.
Given that a substantial amount of cloud emission is resolved-out at line center, there could be 
lower-velocity outflow emission associated with these sources that
we simply cannot detect with the current data. Observations of 
higher-excitation CO transitions at higher resolution may better
distinguish potential outflow emission from these sources. 
However, we also did not detect any far-infrared line emission from
these sources, a further indication that any outflows from these sources may be weak,
or completely buried within their the optically thick envelopes.

\subsection{082005}

The PBRS 082005 is located about 4\arcmin\ south of 082012, and these sources are connected by a 
filamentary structure detected at 870~\micron\ and 160~\micron. This source 
is also undetected at wavelengths shorter than 70~\micron. 
No CO outflow emission was detected in our CARMA observations toward this source and we 
see no evidence for outflow emission from
the \textit{Spitzer} 4.5~\micron\ maps in Figure \ref{spitzer-c}.

\subsection{097002}

The PBRS 097002 is found near a bright 4.5 \micron\ and 24 \micron\ source as seen in \textit{Spitzer} data shown
by \citet{stutz2013}; however, this short wavelength emission is not from 097002, which is only detected
at 70 \micron\ and longer wavelengths. We do not detect an outflow from this source in our CO ($J=1\rightarrow0$) maps, but
the continuum is quite bright (Paper I). However, there is some emission detected
near line-center at the source position.

\subsection{119019}

\begin{figure}
\figurenum{12b}
\begin{center}

\includegraphics[scale=0.35]{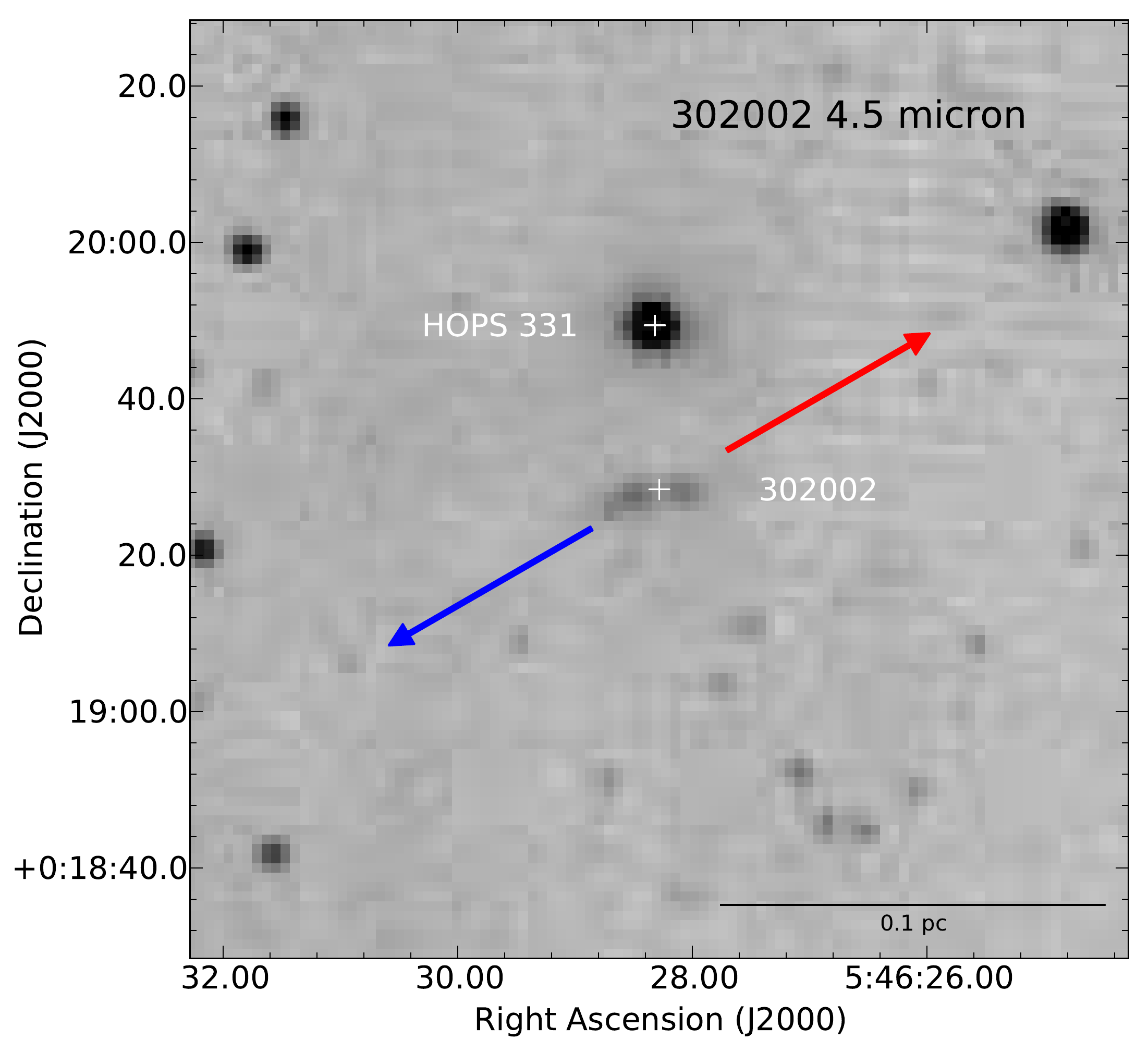}
\includegraphics[scale=0.35]{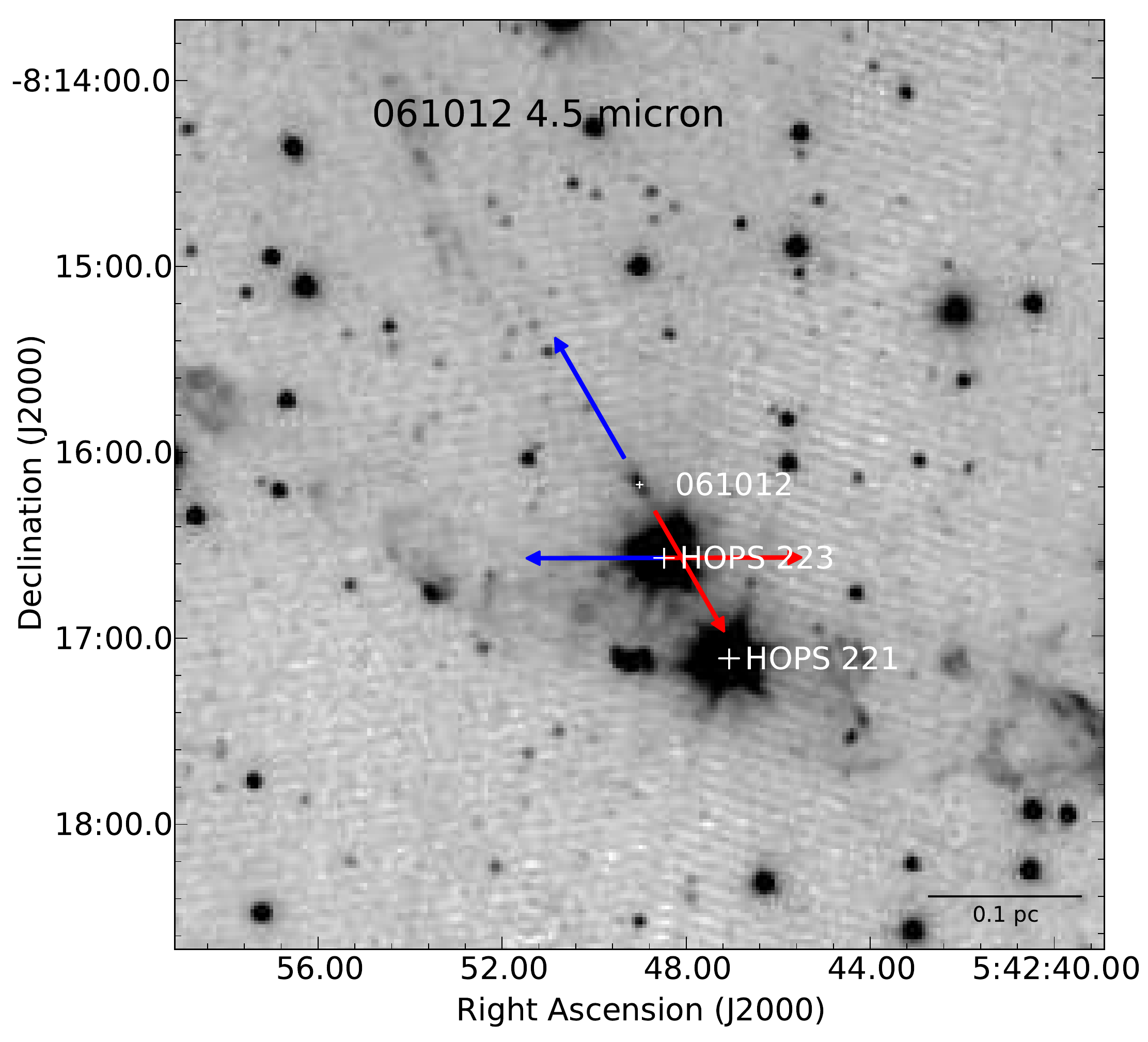}
\includegraphics[scale=0.35]{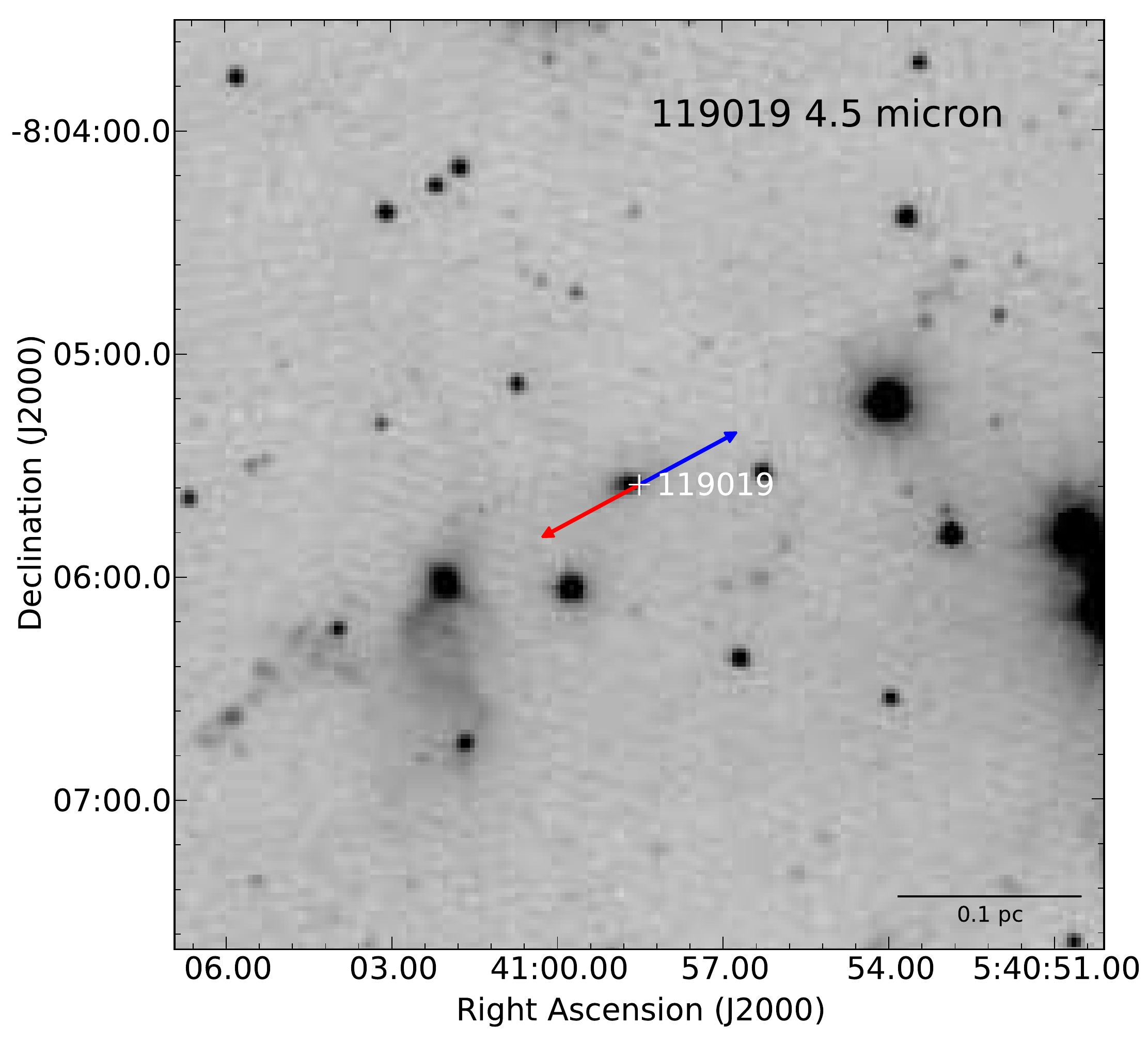}
\includegraphics[scale=0.35]{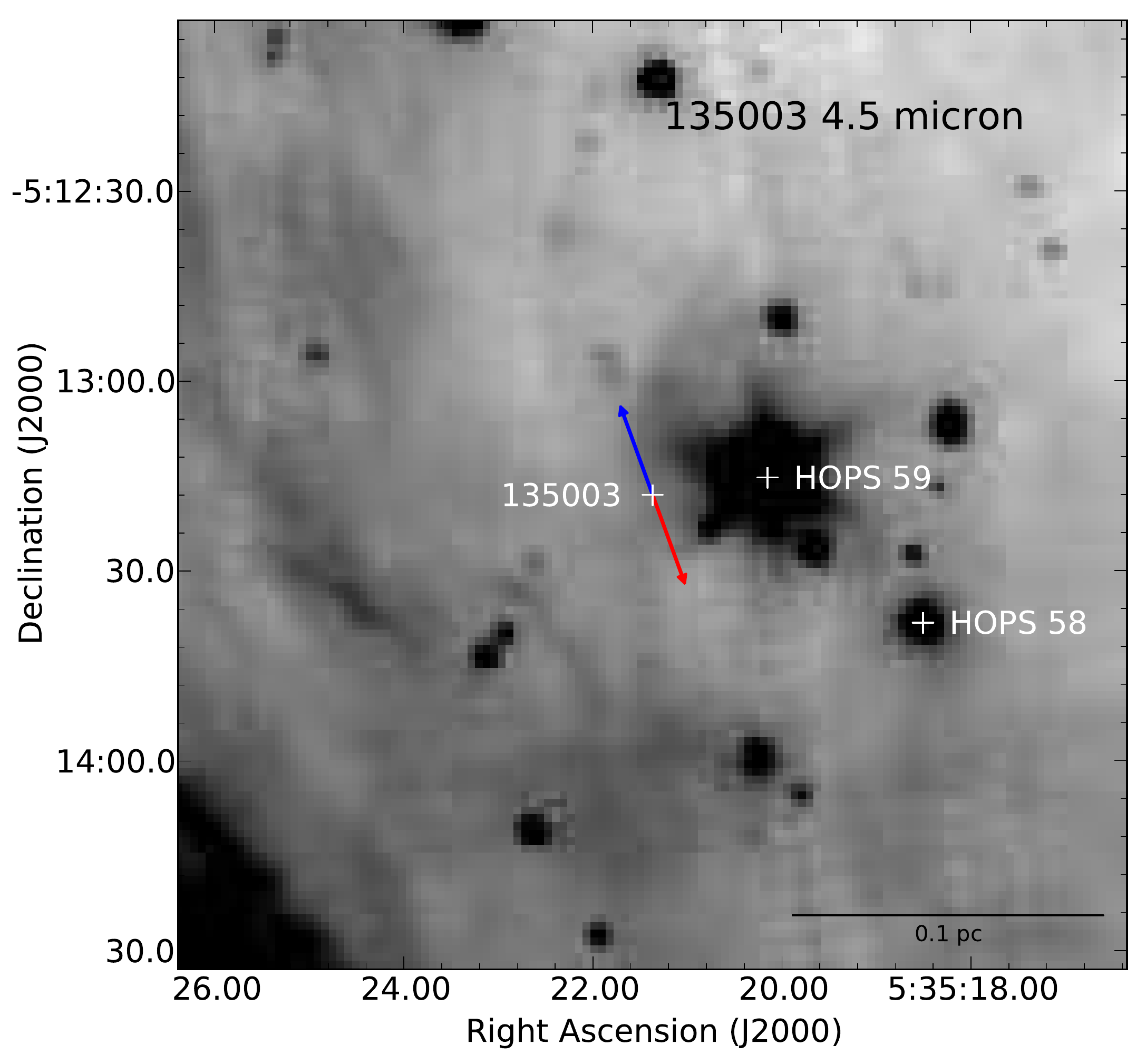}
\includegraphics[scale=0.35]{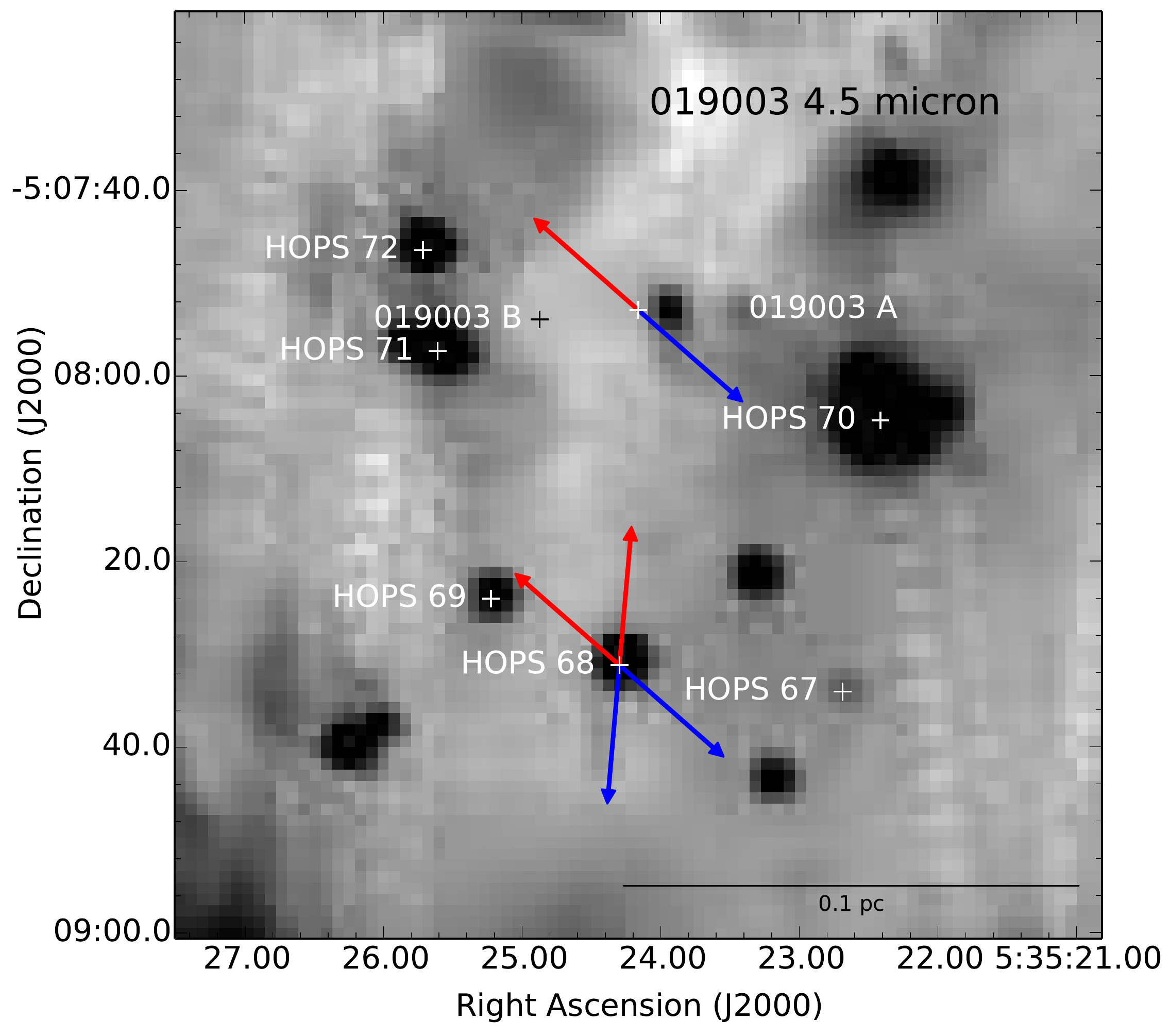}
\includegraphics[scale=0.35]{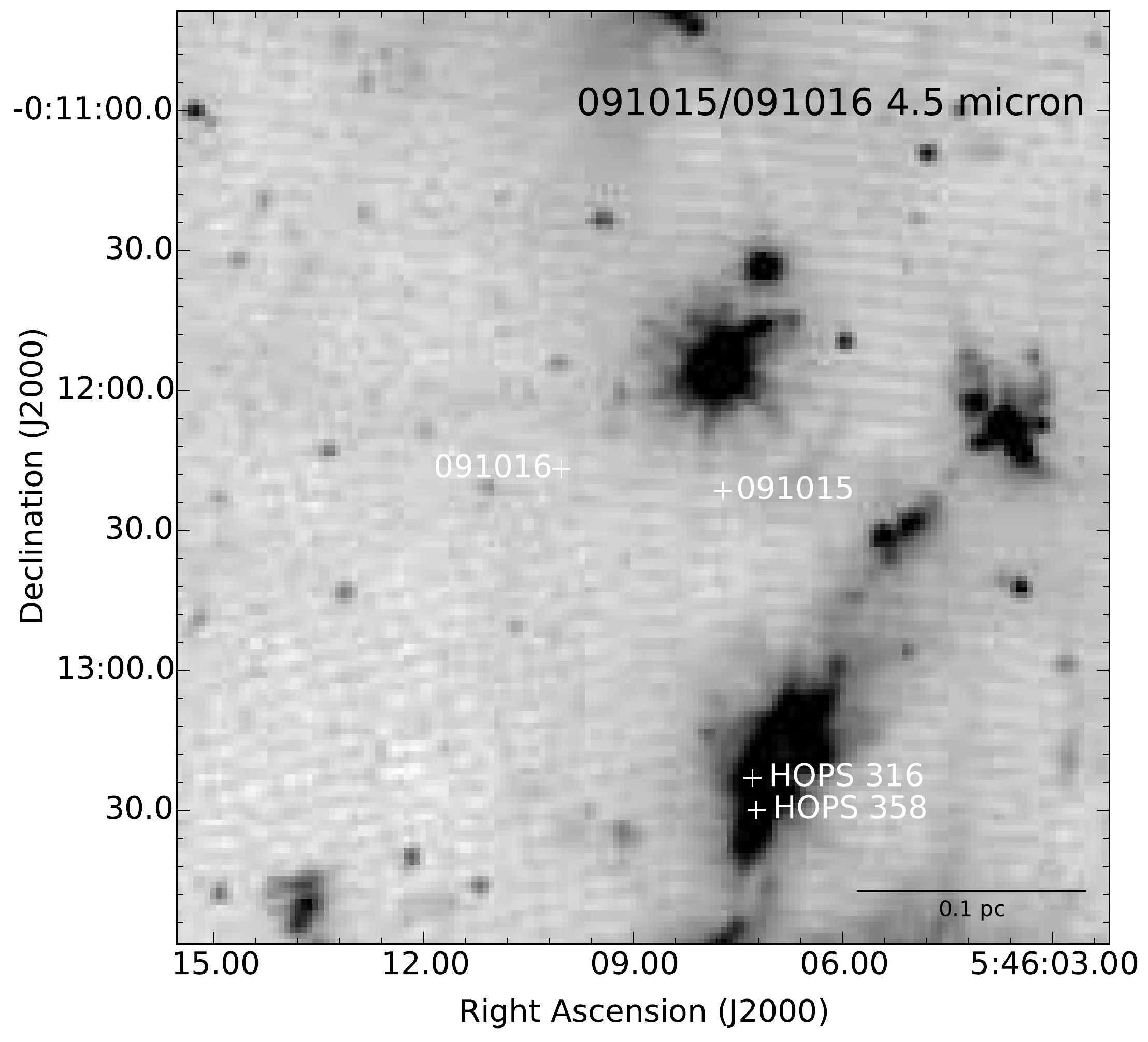}
\end{center}
\caption{}
\label{spitzer-b}
\end{figure}

\begin{figure}
\figurenum{12c}
\begin{center}

\includegraphics[scale=0.35]{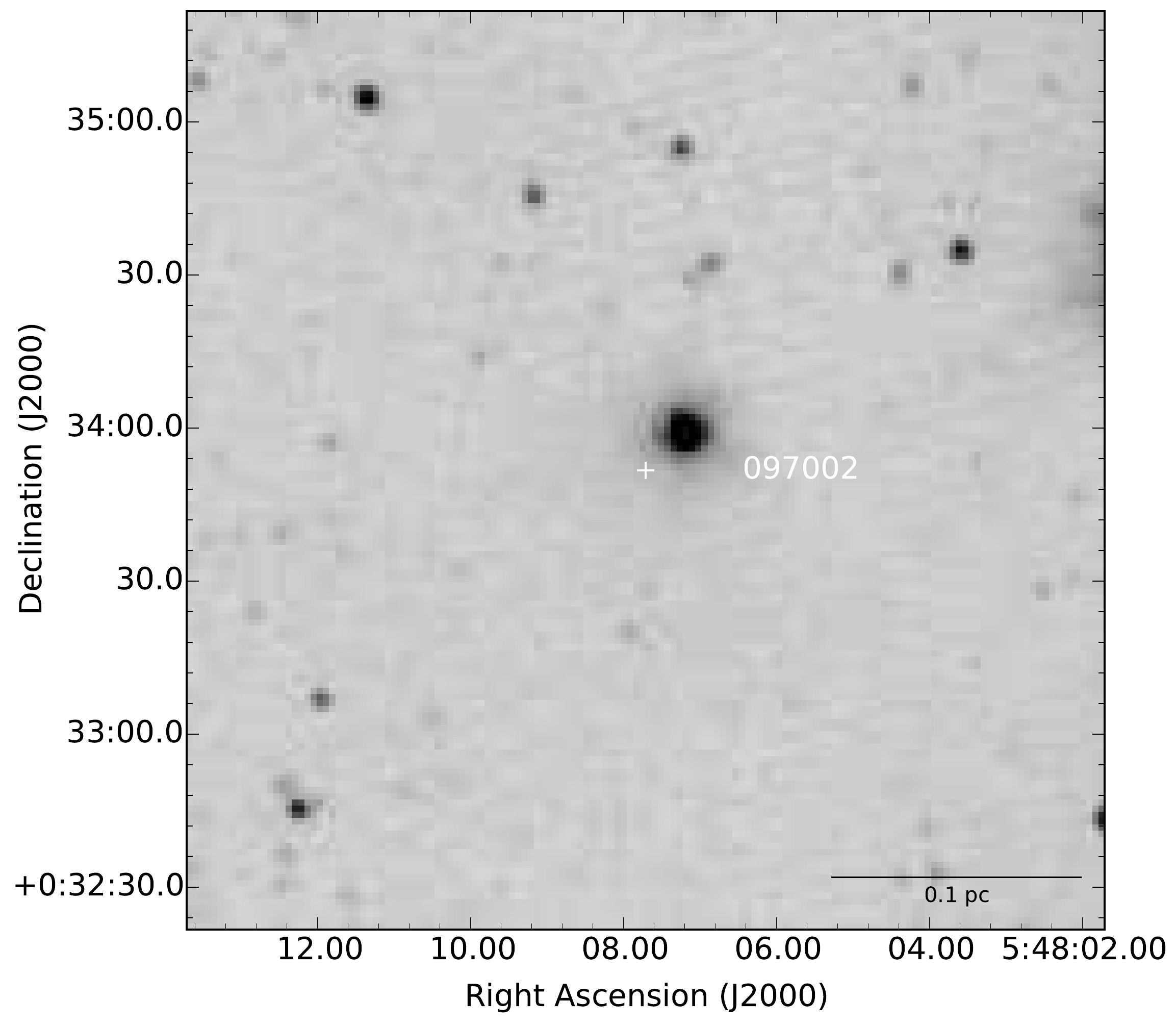}

\end{center}
\caption{}
\label{spitzer-c}
\end{figure}

\begin{figure}
\figurenum{13}

\begin{center}
\includegraphics[scale=0.65]{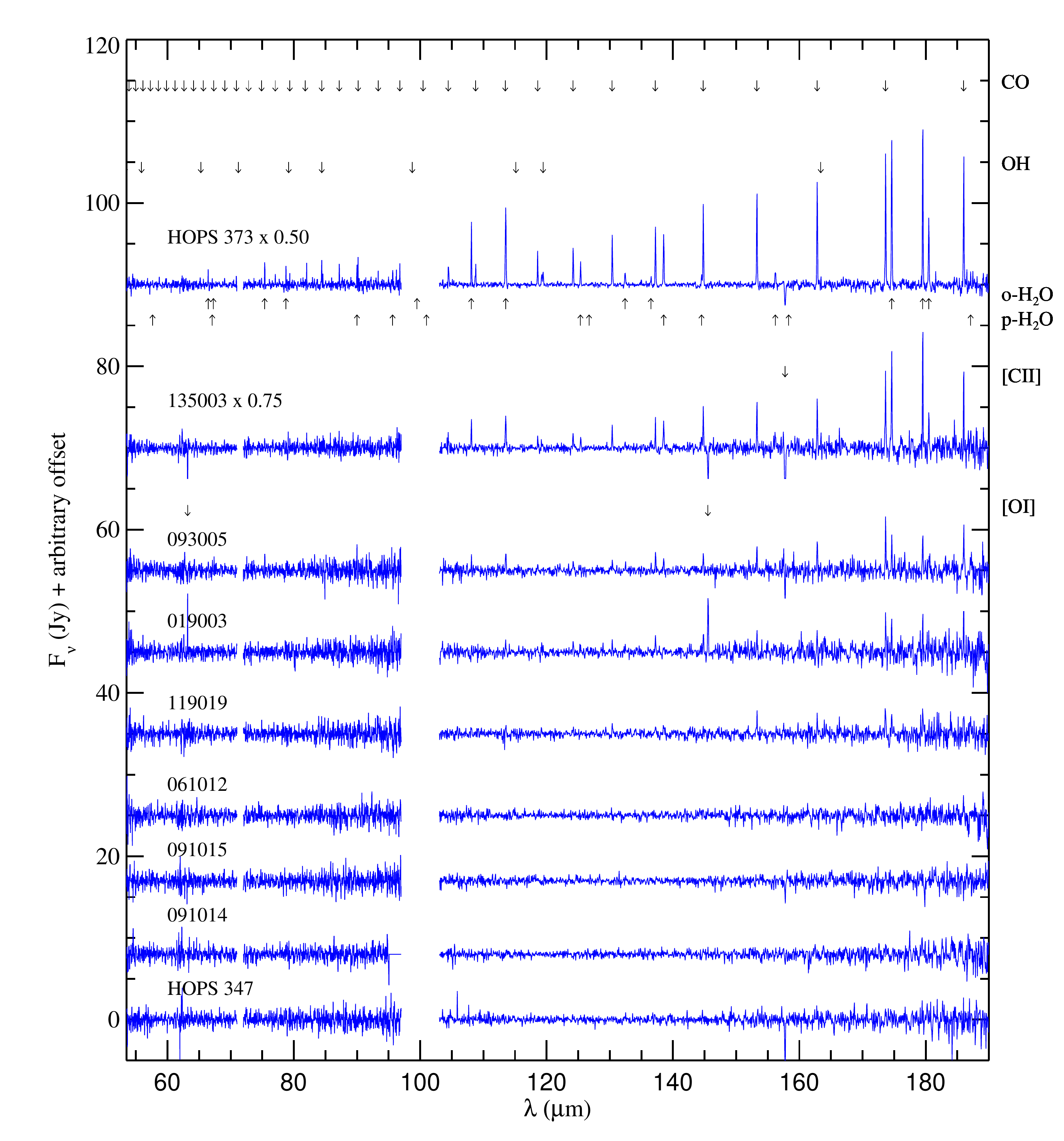}

\end{center}
\caption{Continuum subtracted PACS Spectra for the observed sample;
the sources are plotted in descending order of line brightness.
The wavelengths of common spectral lines are marked with arrows with labels located to the
right of the plot. Negative features are not absorption, but reflect line contamination
in the off position. Only [OI] (63 \micron\ and 143 \micron) and [CII] were 
found to be contaminated by the off positions for some sources.
}
\label{pacsspectra}
\end{figure}

\begin{figure}
\figurenum{14}
\begin{center}
\includegraphics[scale=0.25]{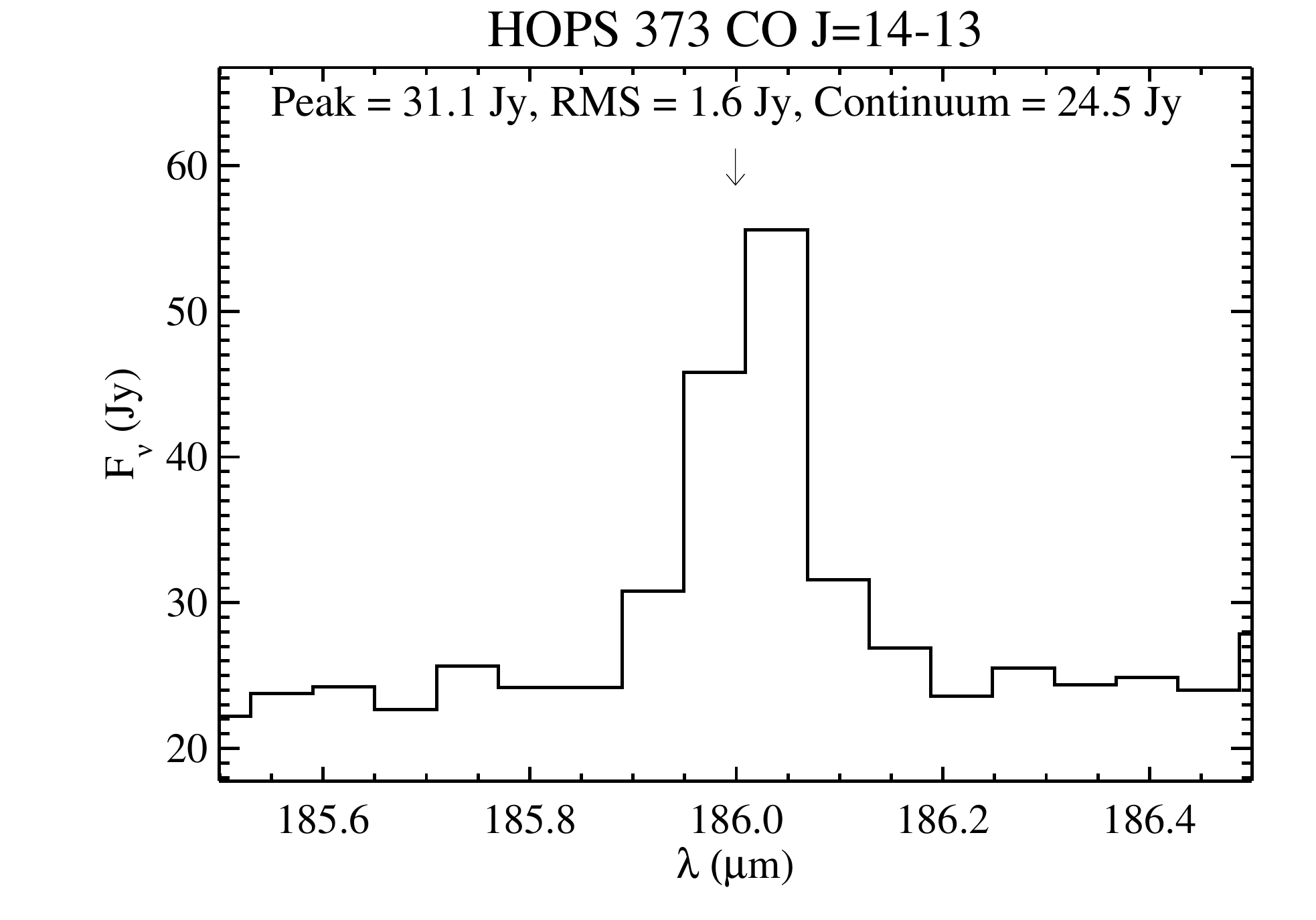}
\includegraphics[scale=0.25]{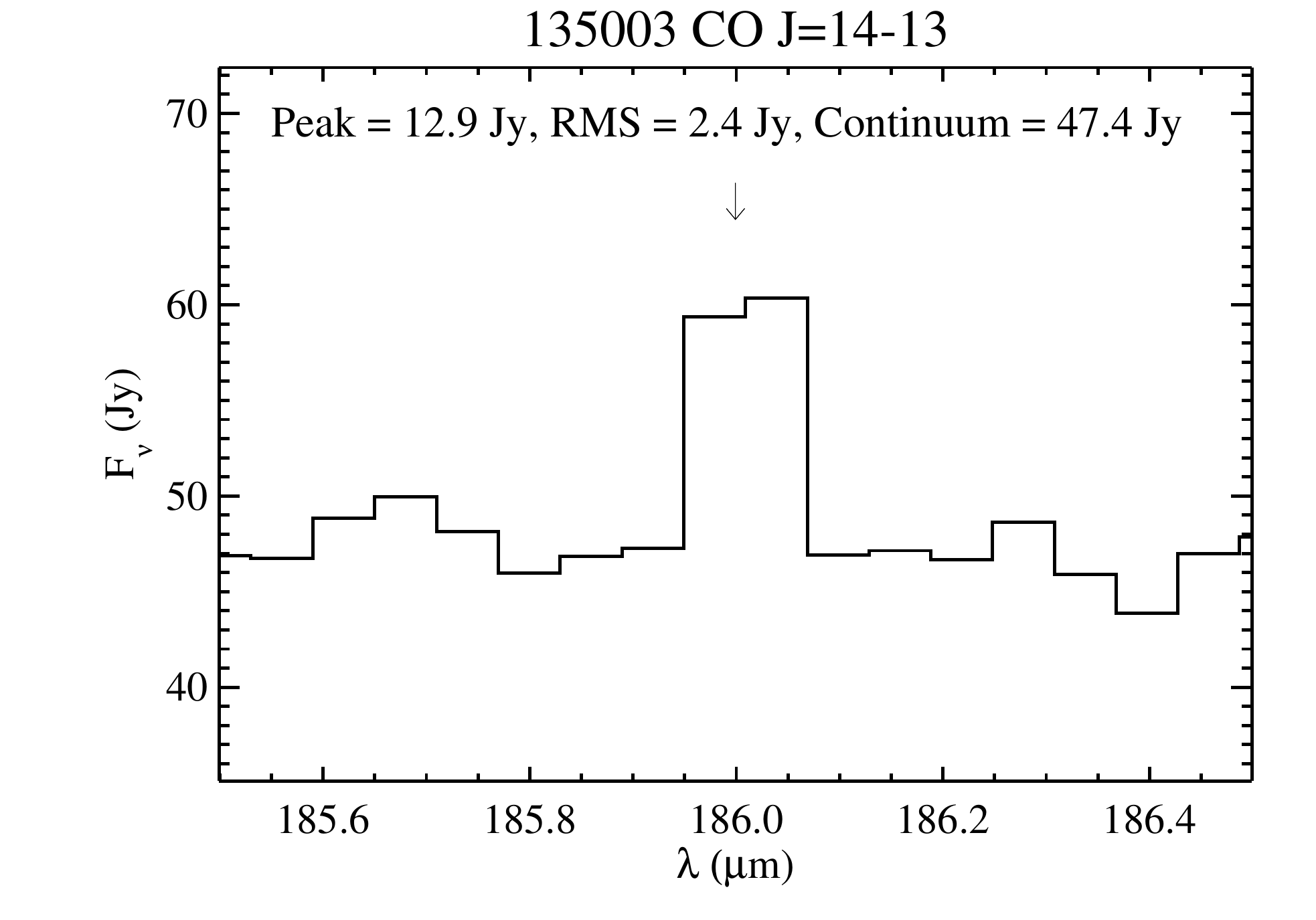}
\includegraphics[scale=0.25]{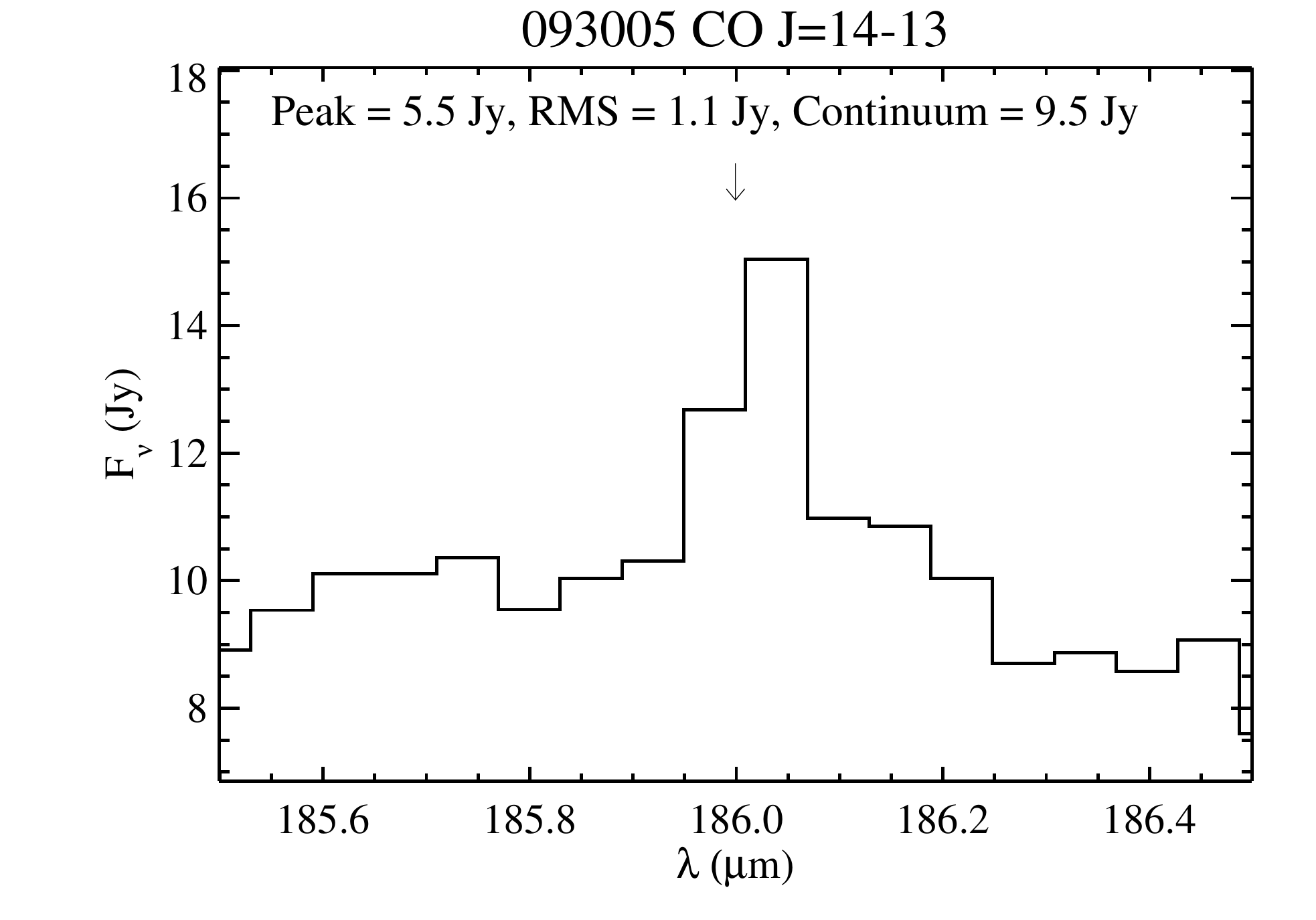}
\includegraphics[scale=0.25]{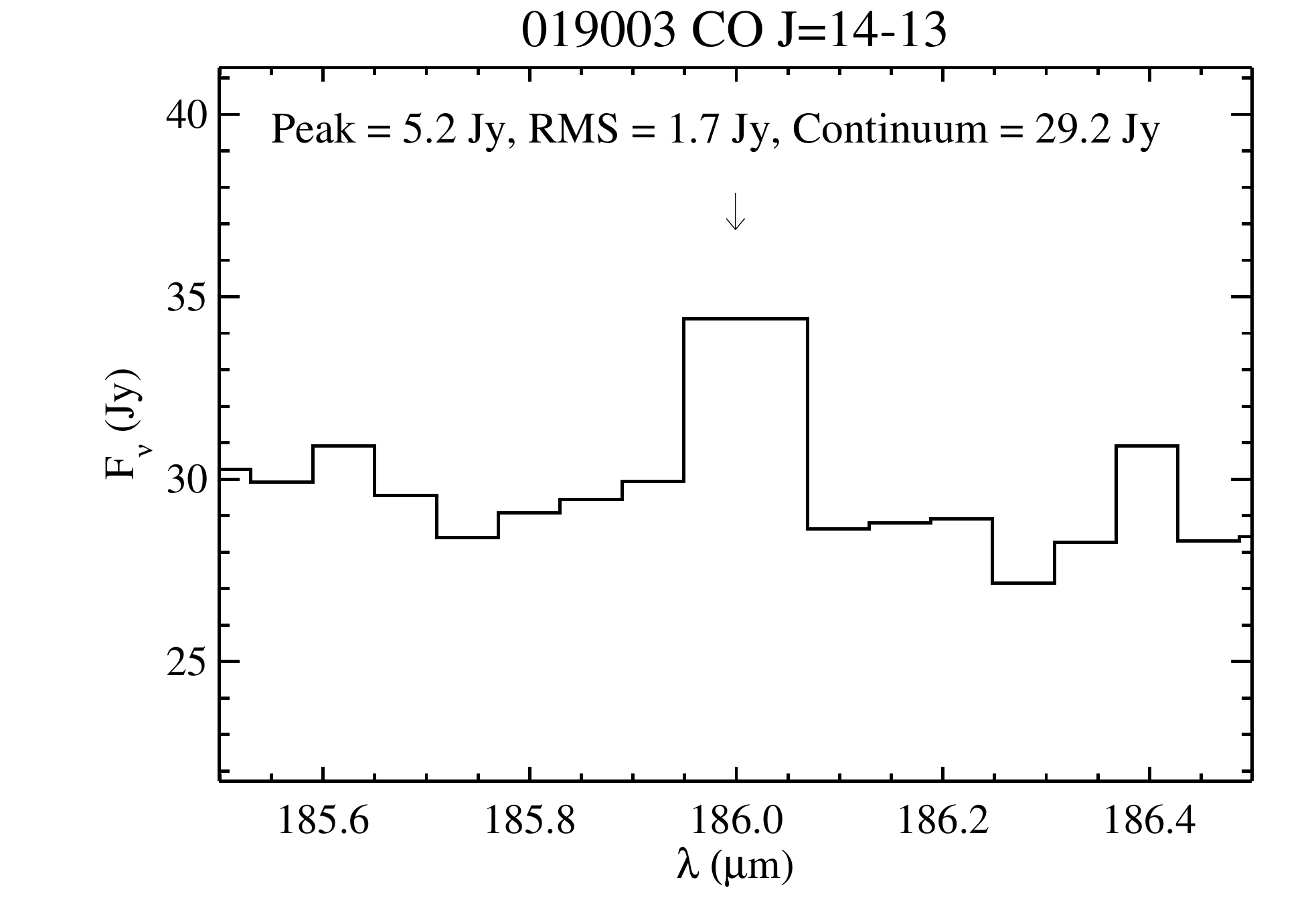}
\includegraphics[scale=0.25]{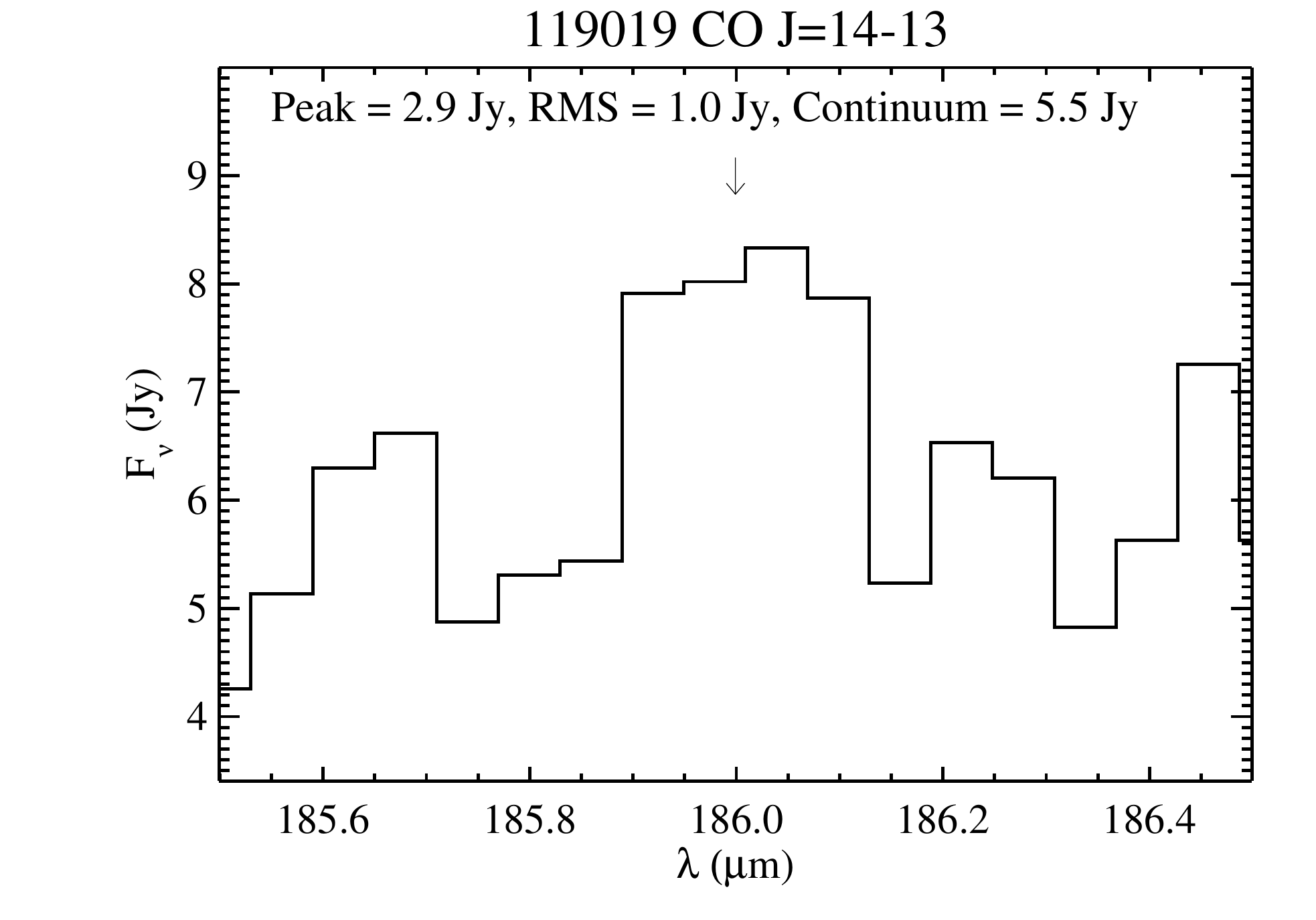}
\includegraphics[scale=0.25]{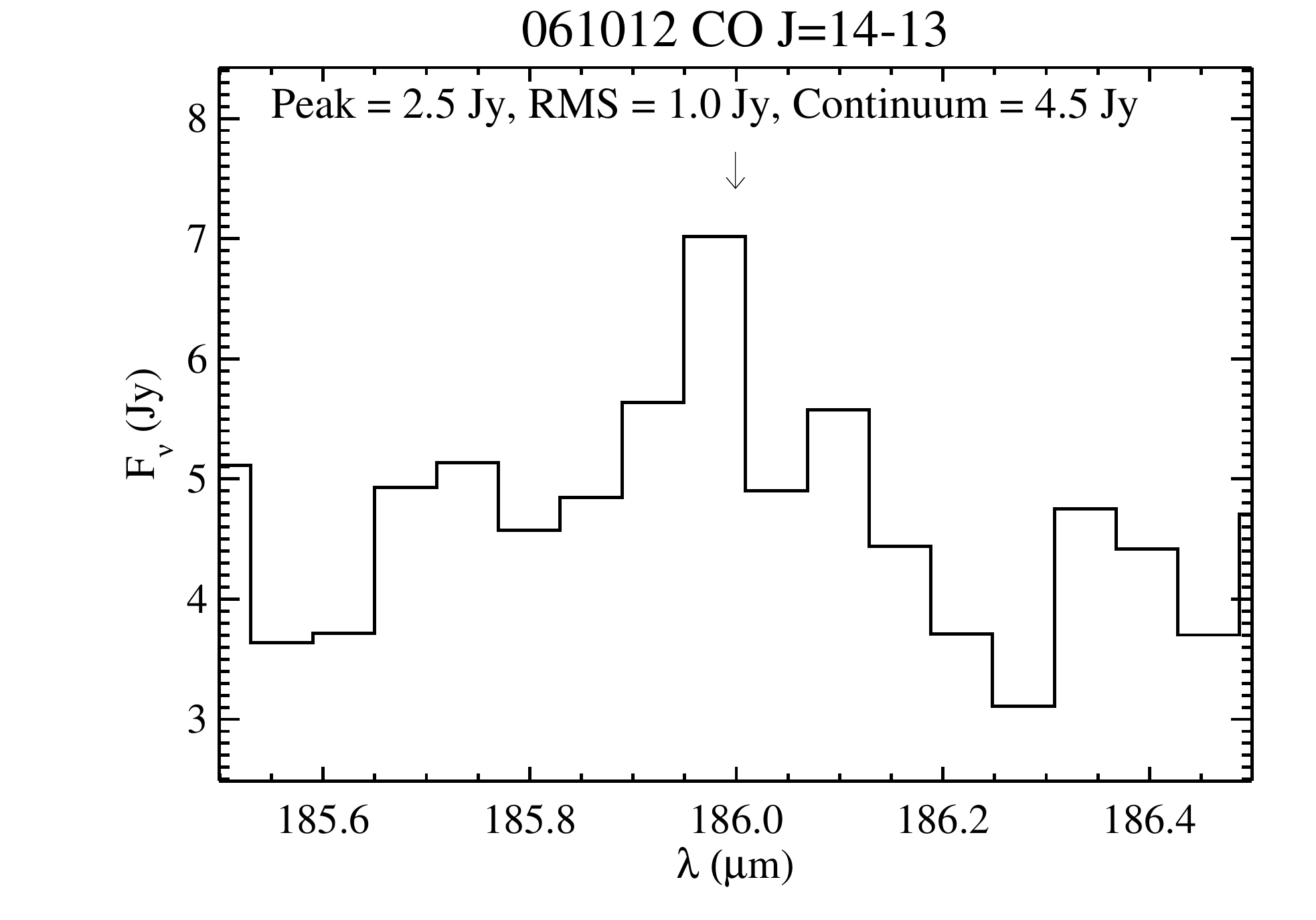}
\includegraphics[scale=0.25]{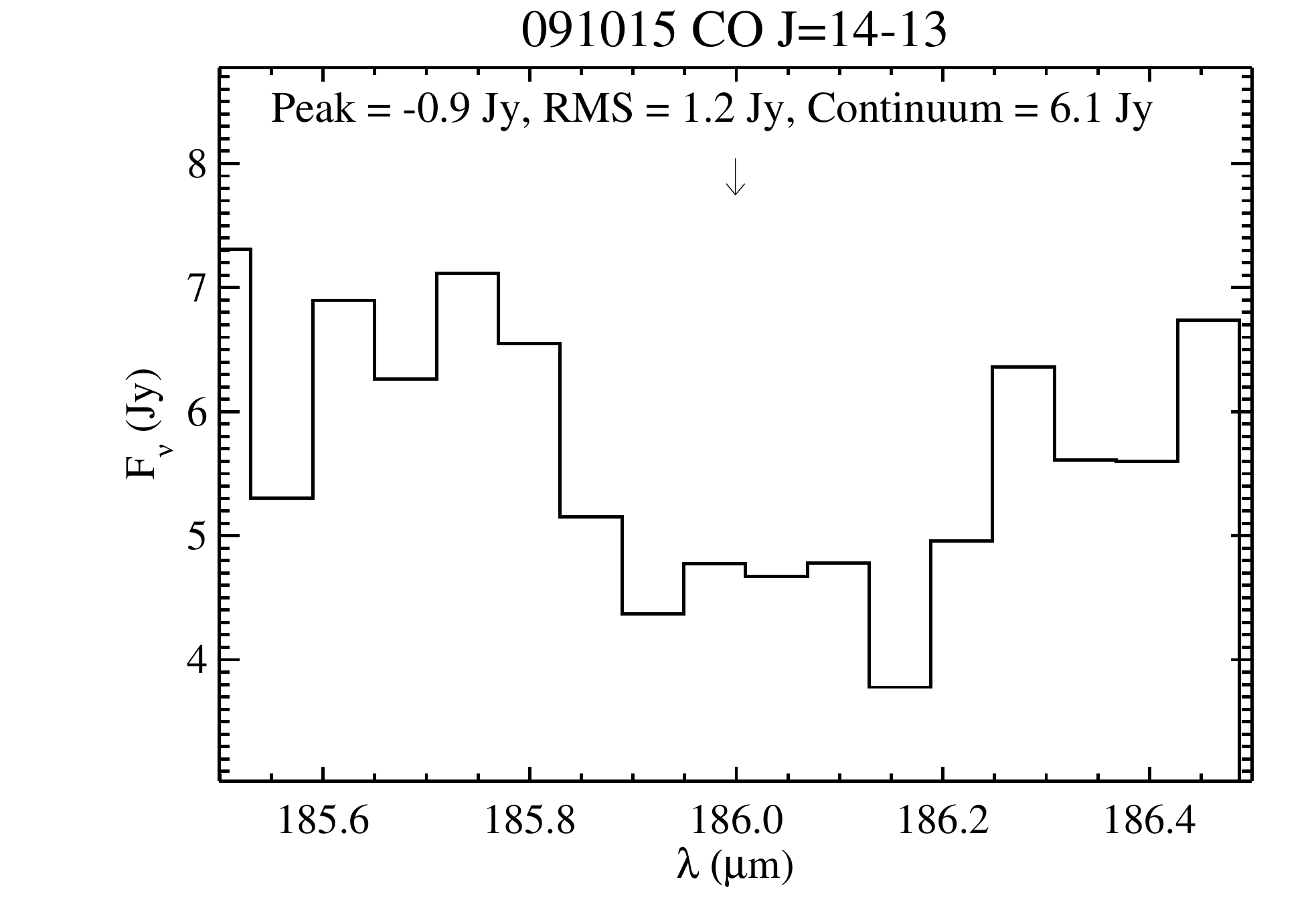}
\includegraphics[scale=0.25]{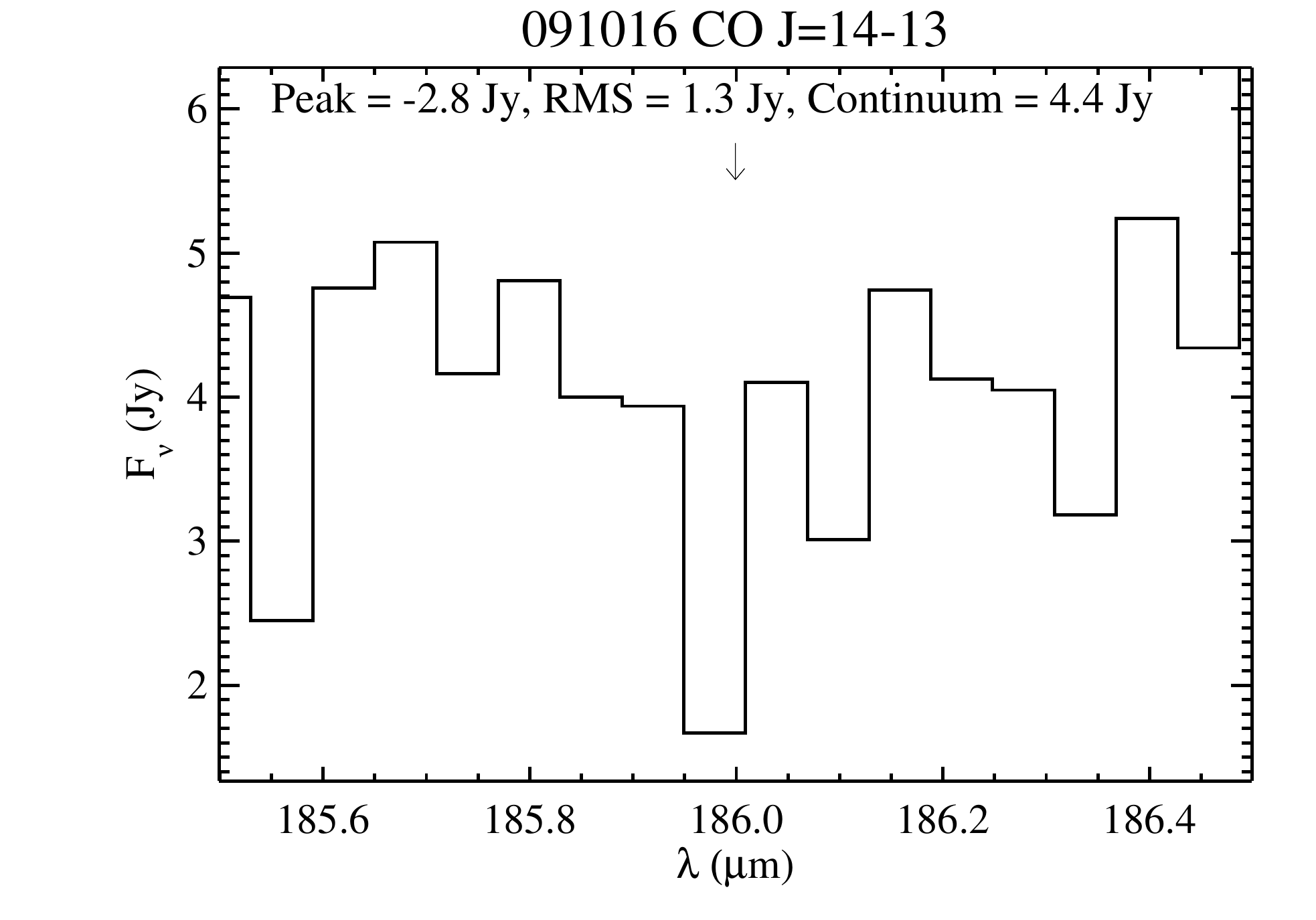}
\includegraphics[scale=0.25]{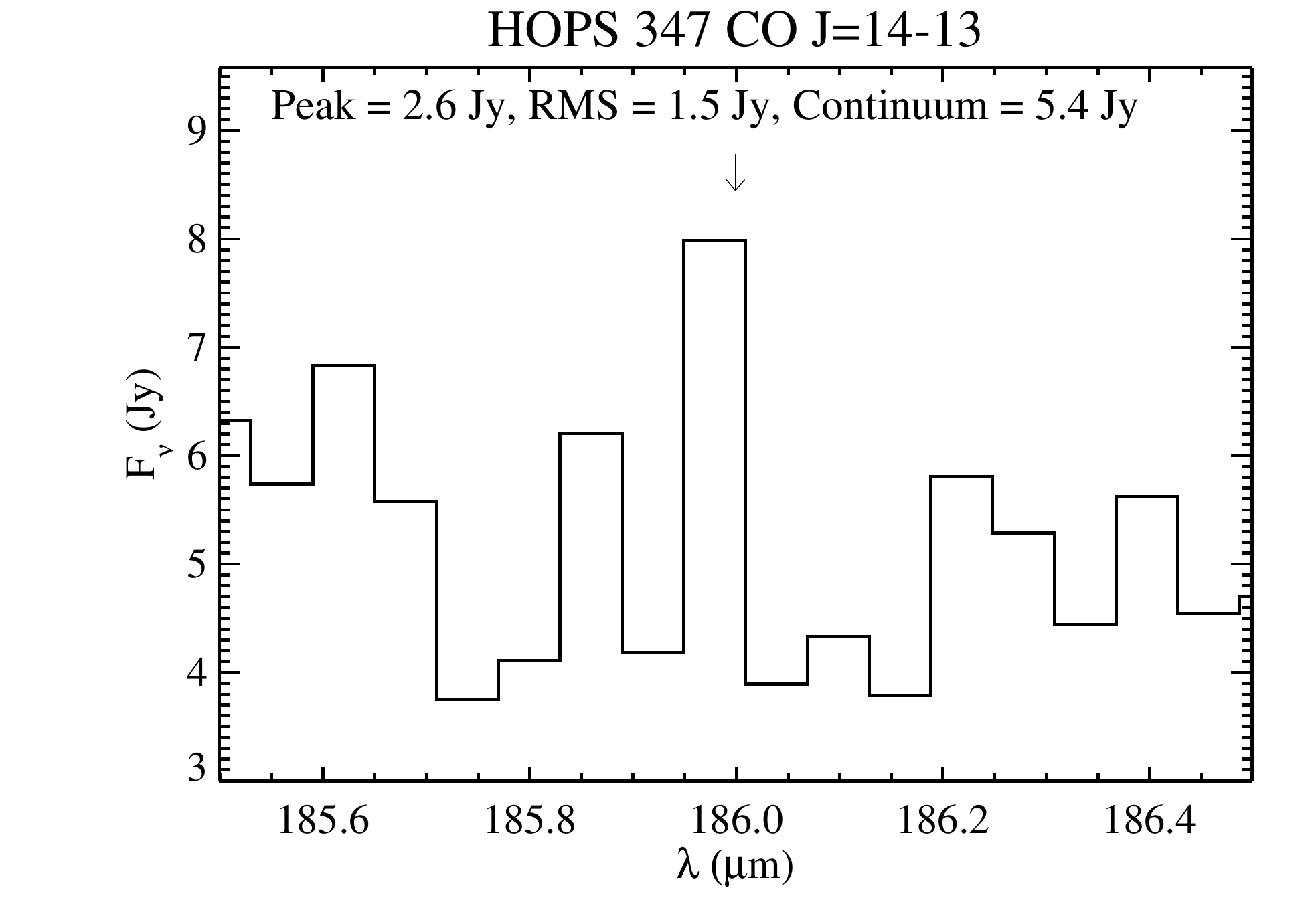}
\end{center}
\caption{PACS spectra centered on the CO ($J=14\rightarrow13$) transition without
continuum subtraction. The downward pointing arrow marks
the wavelength of CO ($J=14\rightarrow13$). We also note the peak line flux density, rms of the continuum, and continuum level 
in each plot; the peak line flux density is relative to the continuum level.
PBRS 119019 only has a 2.9$\sigma$ detection of the CO ($J=14\rightarrow13$), but other
CO transitions are detected with higher significance, thus we regard this as a robust detection.
On the other hand the PBRS 061012 has only a tentative (2.5$\sigma$) detection
of CO ($J=14\rightarrow13$) and no other CO transitions detected; 091015 and 091016 do not have detections. HOPS 347 has a peak at
the expected wavelength of CO ($J=14\rightarrow13$) but it is not significant given 
the noise around the line. The peak line flux density, RMS, and continuum level are denoted in each plot.
}
\label{co14-13spectra}
\end{figure}

\begin{figure}
\figurenum{15}

\begin{center}
\includegraphics[scale=0.33,trim=3cm 12cm 5cm 5cm, clip=true]{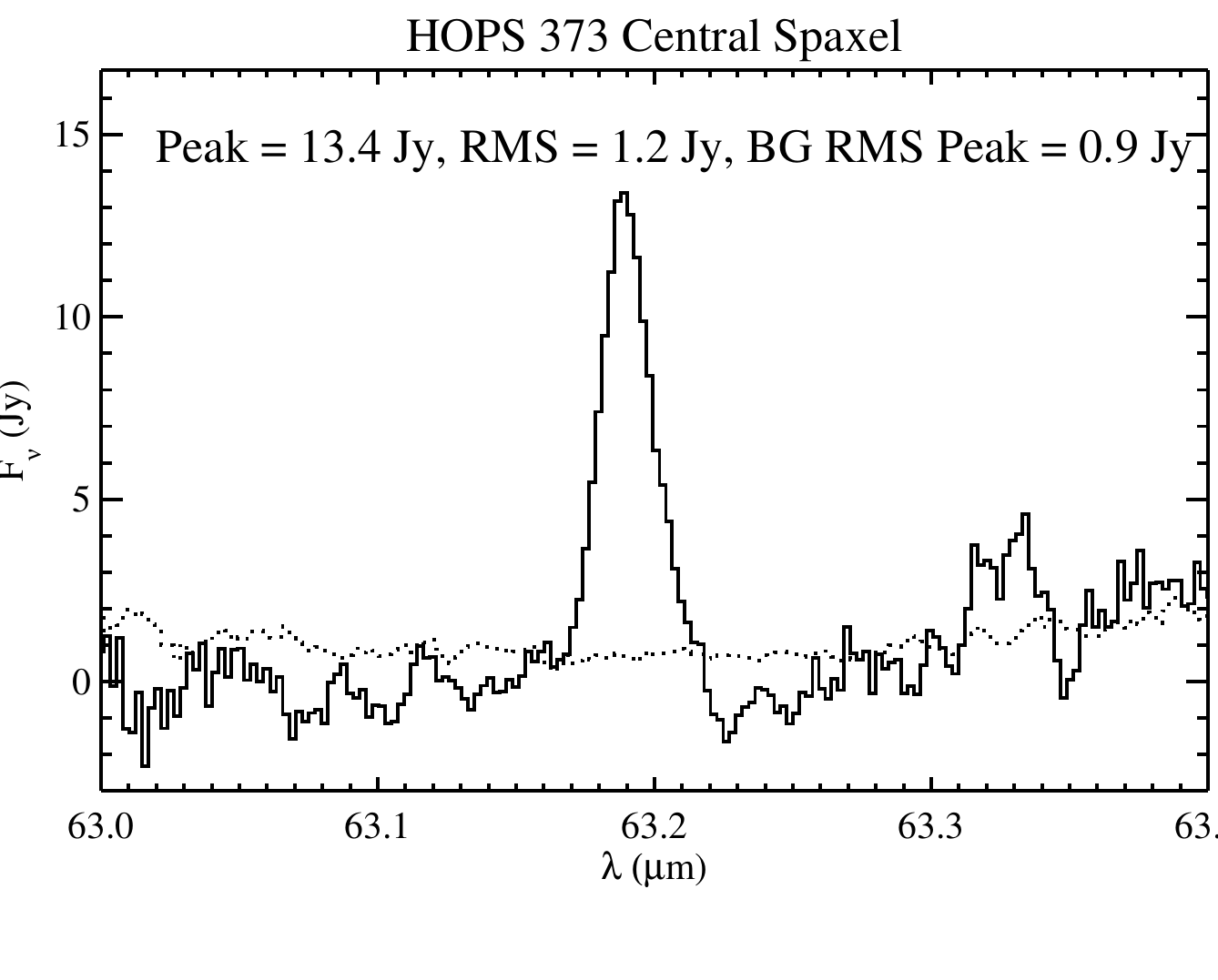}
\includegraphics[scale=0.33,trim=3cm 12cm 5cm 5cm, clip=true]{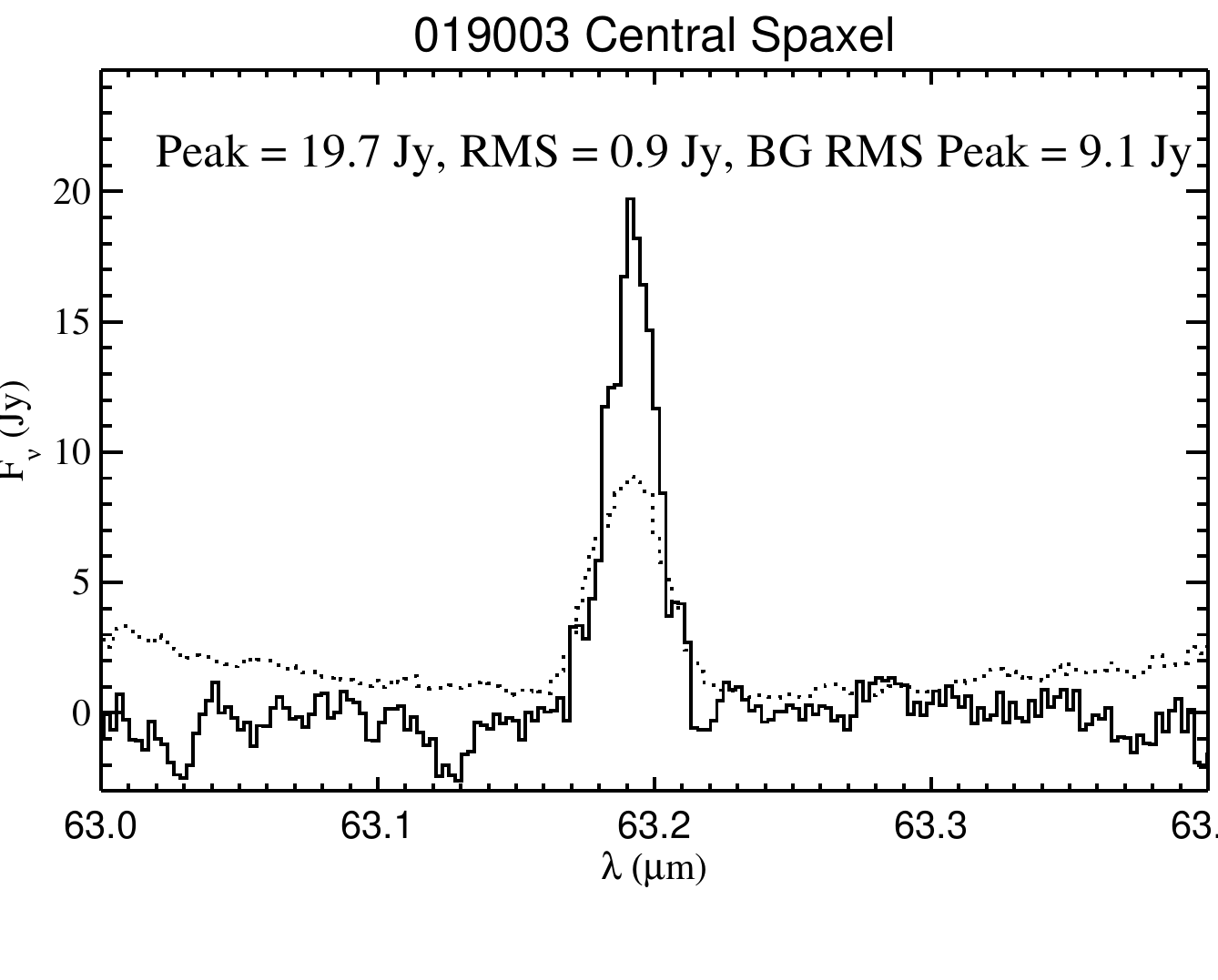}
\includegraphics[scale=0.33,trim=3cm 12cm 5cm 5cm, clip=true]{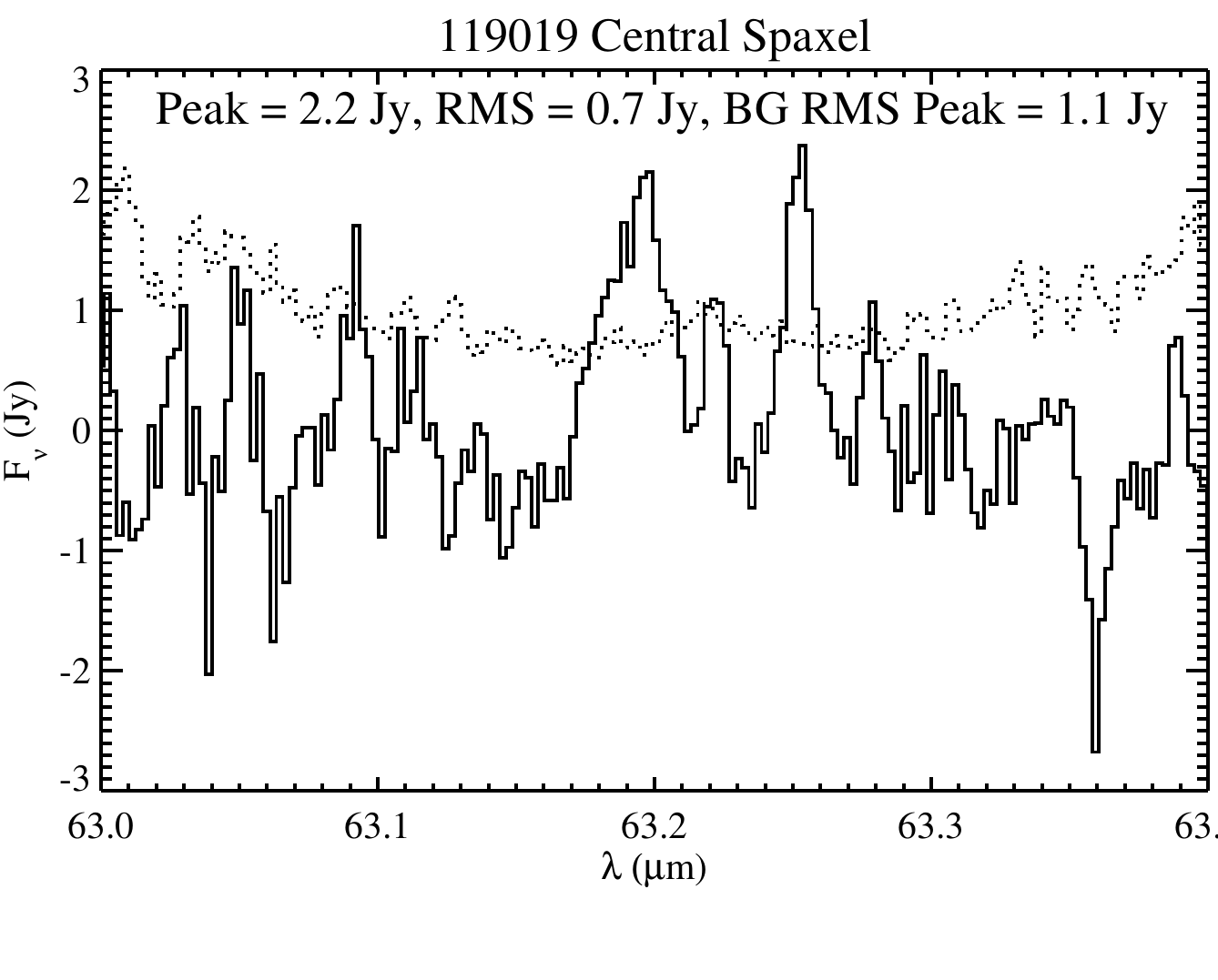}
\includegraphics[scale=0.33,trim=3cm 12cm 5cm 5cm, clip=true]{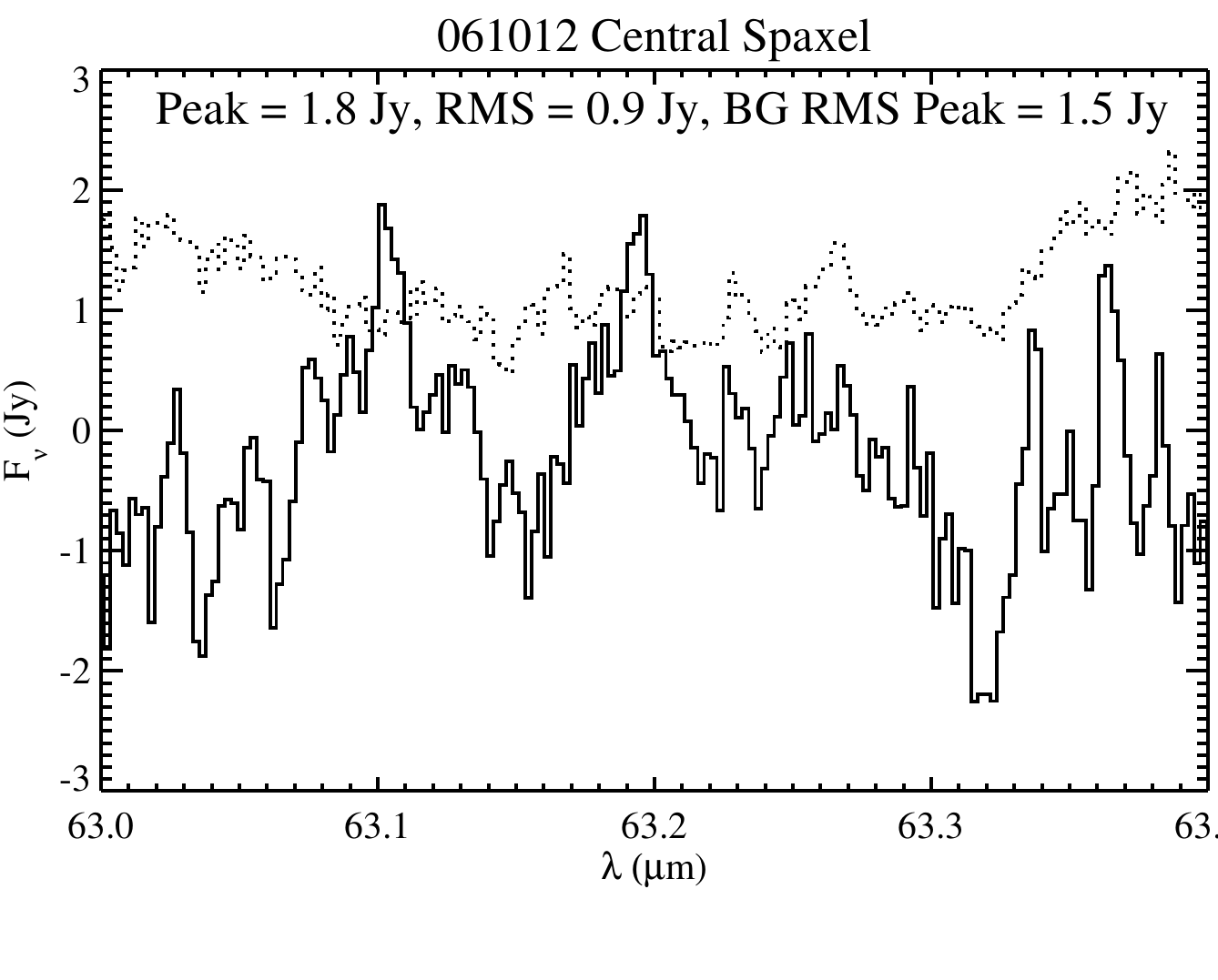}
\includegraphics[scale=0.33,trim=3cm 12cm 5cm 5cm, clip=true]{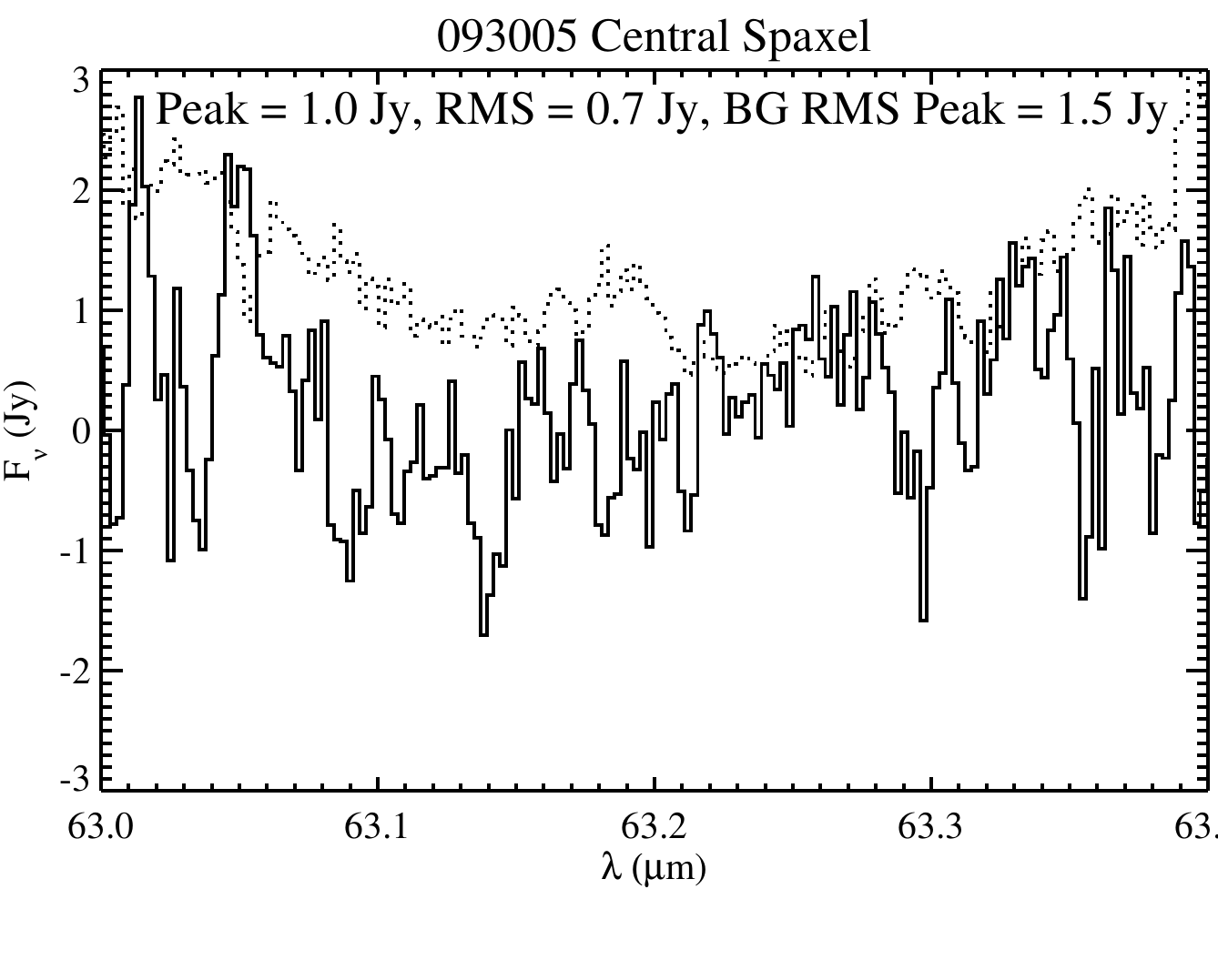}
\includegraphics[scale=0.33,trim=3cm 12cm 5cm 5cm, clip=true]{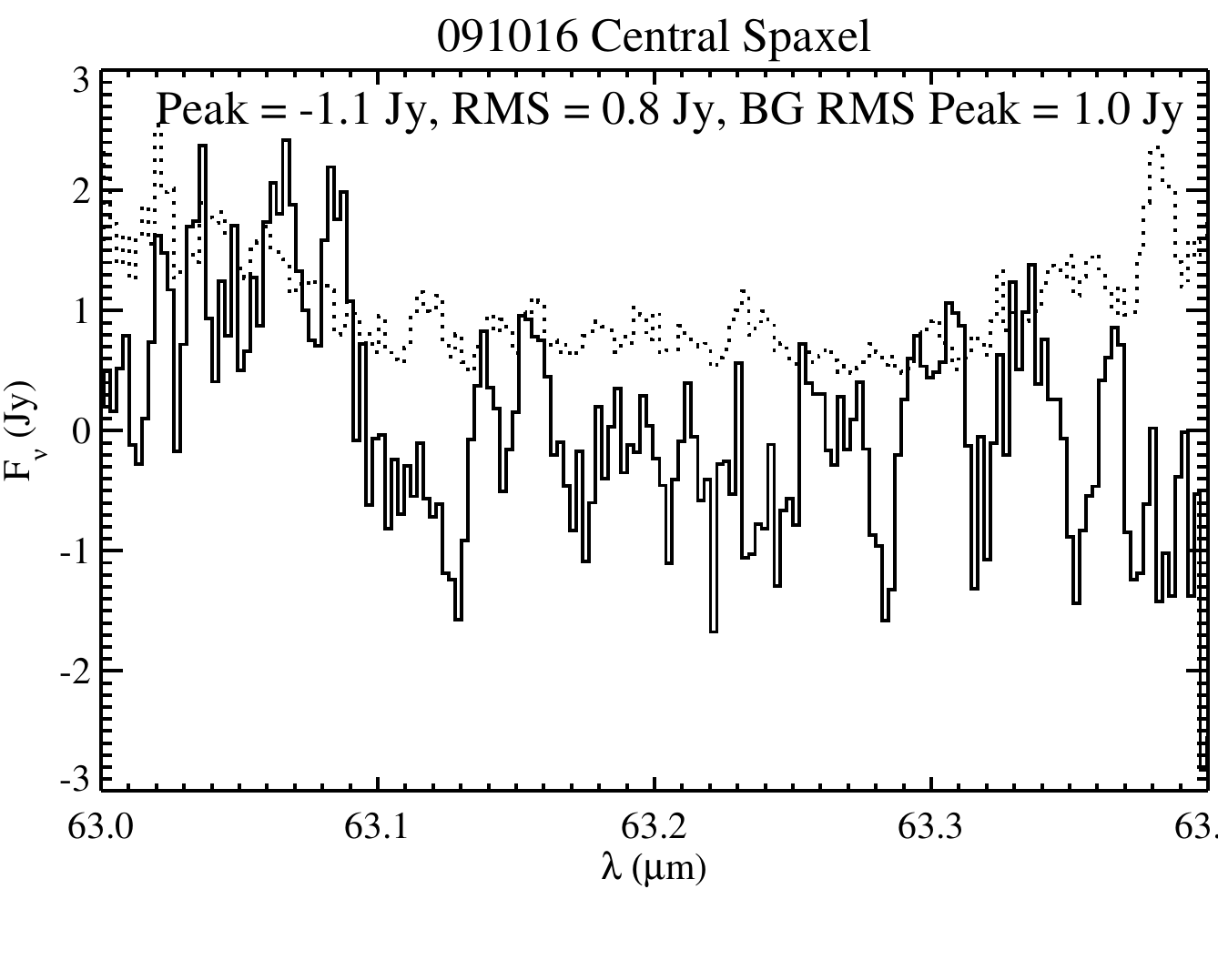}
\includegraphics[scale=0.33,trim=3cm 12cm 5cm 5cm, clip=true]{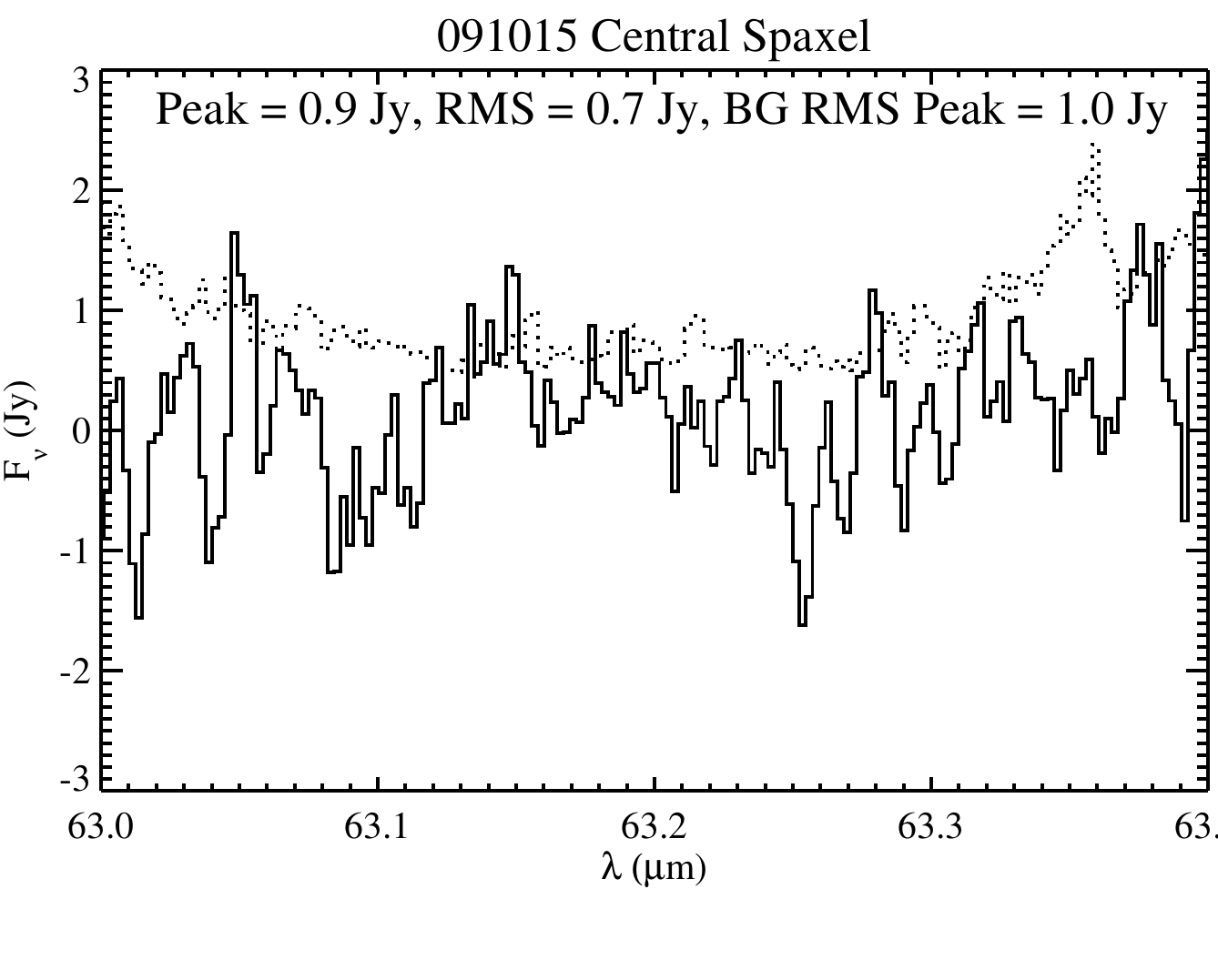}
\includegraphics[scale=0.33,trim=3cm 12cm 5cm 5cm, clip=true]{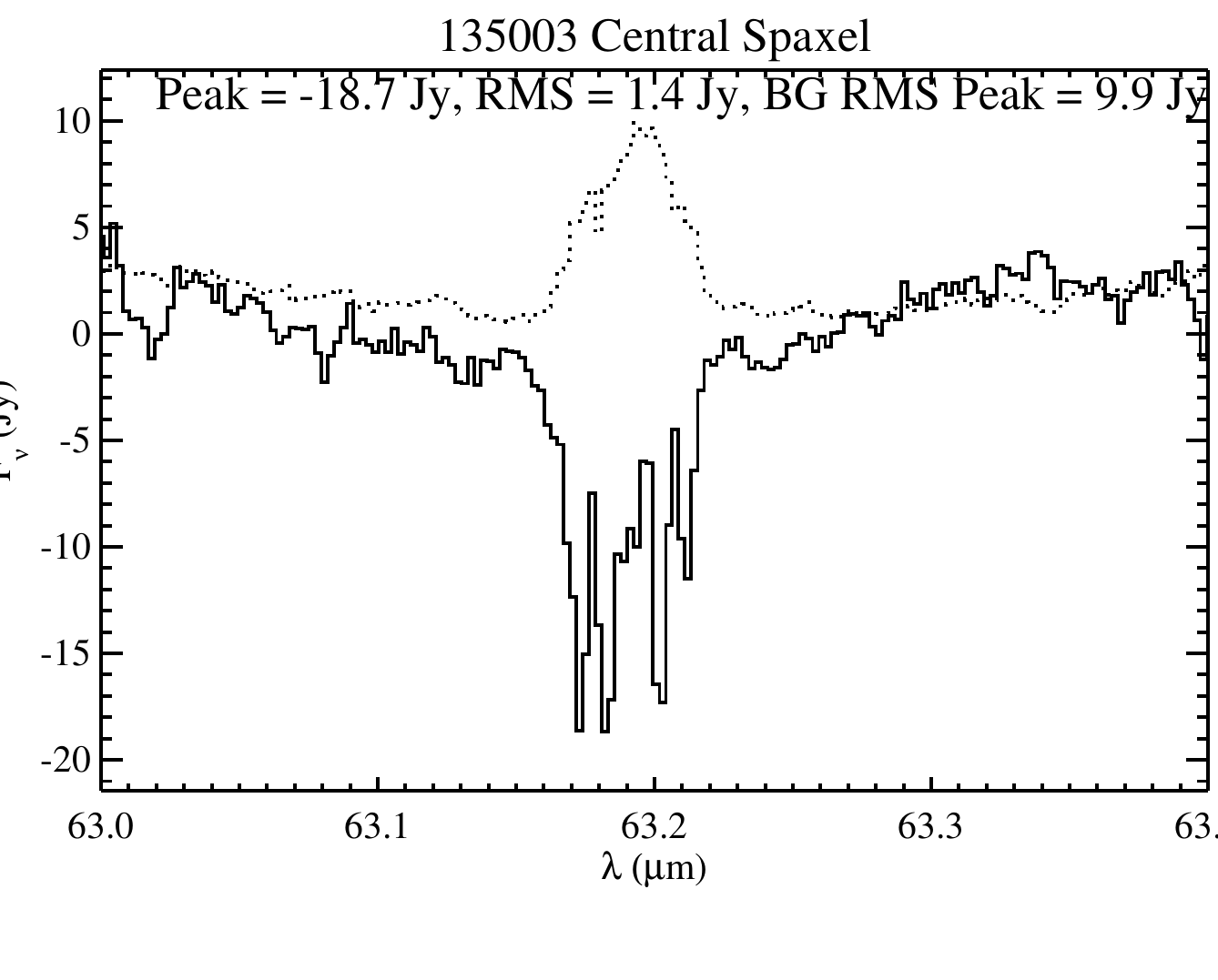}
\includegraphics[scale=0.33,trim=3cm 12cm 5cm 5cm, clip=true]{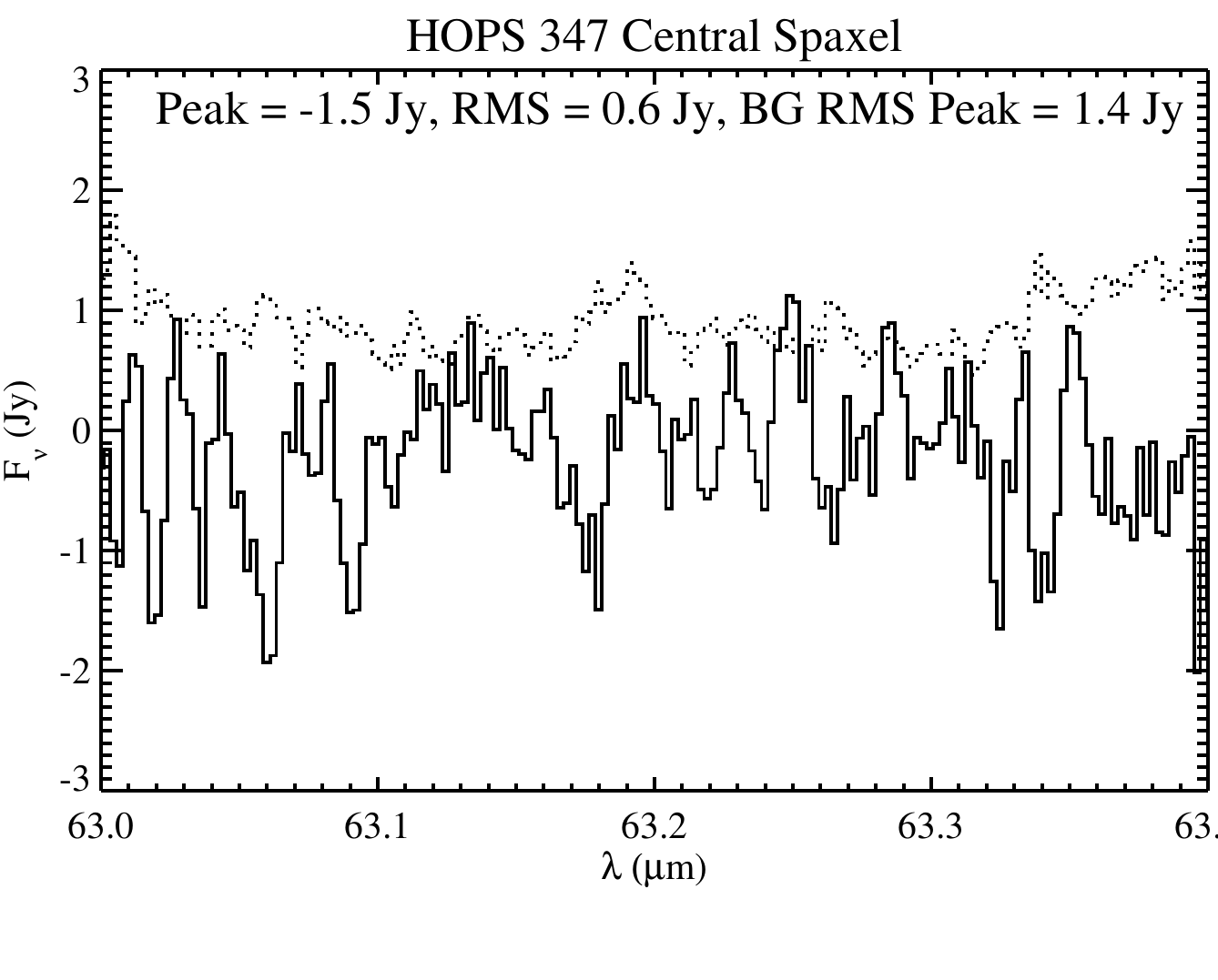}
\end{center}
\caption{PACS spectra around the [OI] 63.18 \micron\ transition. The
solid-line is the foreground/background subtracted spectrum.
The foreground/background is measured from the edge spaxels and the 
fine dashed line is the standard deviation of the foreground/background spectrum. This is 
a good representation of the noise level in the spectral band since the [OI] variations
will dominate the noise. Only HOPS~373 has a convincing detection in the [OI] line,
the detection of 019003 is tentative (2.2$\sigma$) given the large variations in the foreground/background spectrum. 
There are features at the expected wavelength of [OI] toward 061012 and 119019, 
but there are other features that have the same level of peak intensity that do
not correspond to an expected spectral line. The foreground/background [OI] near 135003 is highly variable, 
resulting in the negative spectrum. The peak line flux density, spectrum RMS, and the RMS of the background (BG RMS Peak)
emission at the wavelength of the [OI] line are noted in each panel.
}
\label{OI-spectra}
\end{figure}

\begin{figure}
\figurenum{16a}
\begin{center}
\includegraphics[scale=0.66]{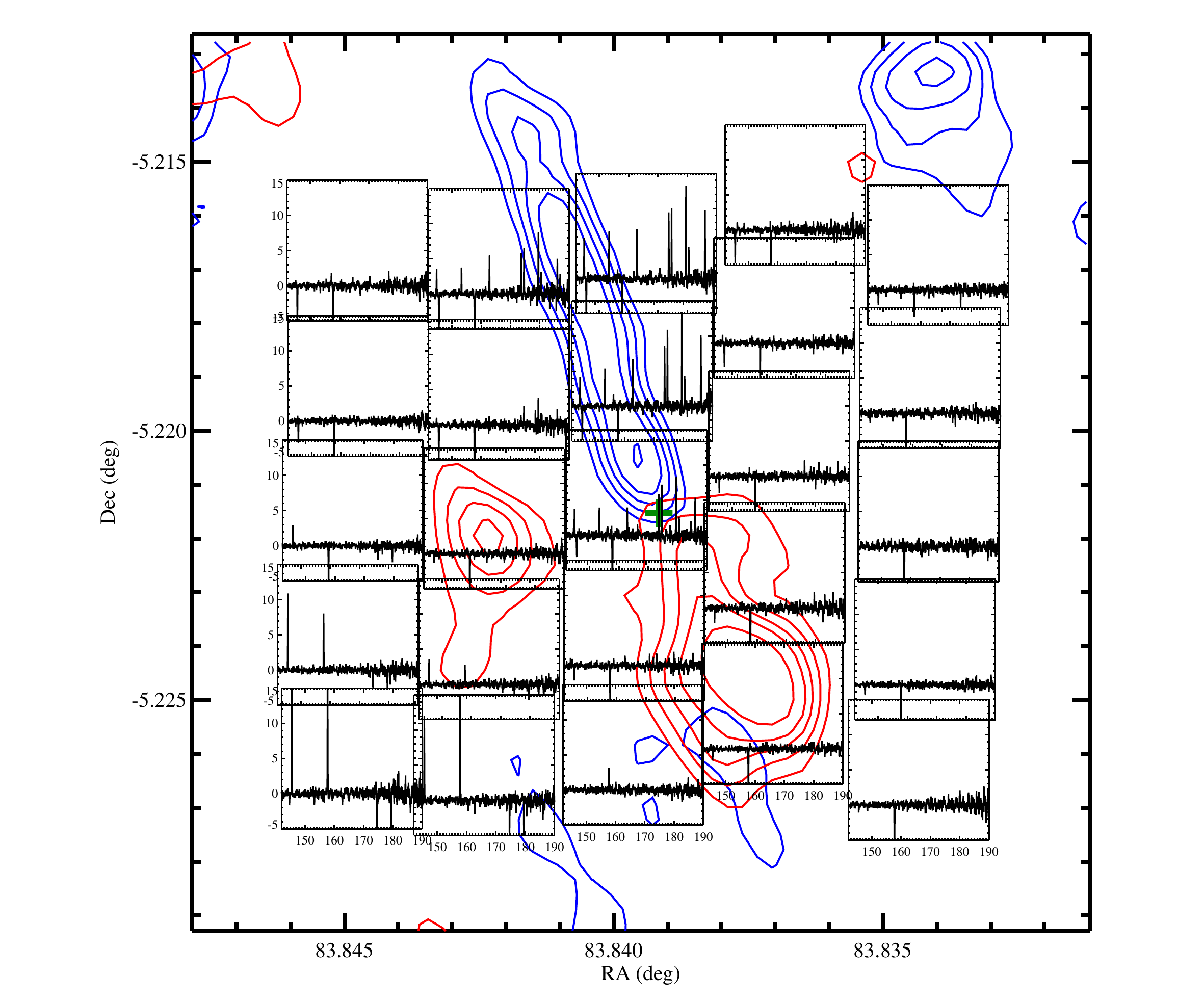}
\end{center}
\caption{PACS Spectrometer footprint observed toward PBRS 135005 ans
 overlaid on the CO ($J=1\rightarrow0$) 
contours from Figure \ref{135003}, plotting the positive contours only. 
The plots of the two long wavelength
channels of the spectrometer show remarkable correspondence between the 
blue-shifted (north) side of the CO outflow and high-J water and CO
line emission. The wavelength range from 140 \micron\ to 190 \micron\
is shown in (a) and 100 \micron\ to 140 \micron\ is shown in (b).  There is an apparent lack of similar high-J CO and water 
emission on the red-shifted side of the outflow (south); however,
maps of the far-infrared and submillimeter continuum show that there is 
extended cold dust emission north of 135003 but not south. Therefore,
the blue-shifted outflow is likely impacting ambient material causing 
shocks, while the red-shifted outflow is being driven into a less dense 
medium. The green cross in the central spaxel marks the location of the 
2.9~mm continuum source, where the red and blue-shifted contours meet.
}
\label{135003-footprint}
\end{figure}

The outflow toward 119019 has complete spatial overlap between the red and blue-shifted emission
meaning that this source is viewed almost exactly edge-on. This source was also one of the fainter
continuum sources detected in (Paper I). Therefore, unlike the rest of the sample,
this source may only have been identified as a PBRS due to an edge-on orientation.
The outflow extends outside the CARMA primary beam
and the velocity width of the outflow is quite narrow, only $\pm$3 \kms; however,
the outflow may have greater speeds given that we are viewing it in the plane of the sky. 
Some diffuse emission is detected at 4.5~\micron\ near the protostar location and along the outflow
in figure \ref{spitzer-b}; \citet{davis2009} also detects H$_2$ knots that appear to be part 
of this outflow (DFS 136).
This source also has the faintest far-infrared line emission for which we have a confident detection.

\begin{small}
\bibliographystyle{apj}
\bibliography{ms}
\end{small}

\clearpage

\begin{figure}
\figurenum{16b}
\begin{center}

\includegraphics[scale=0.66]{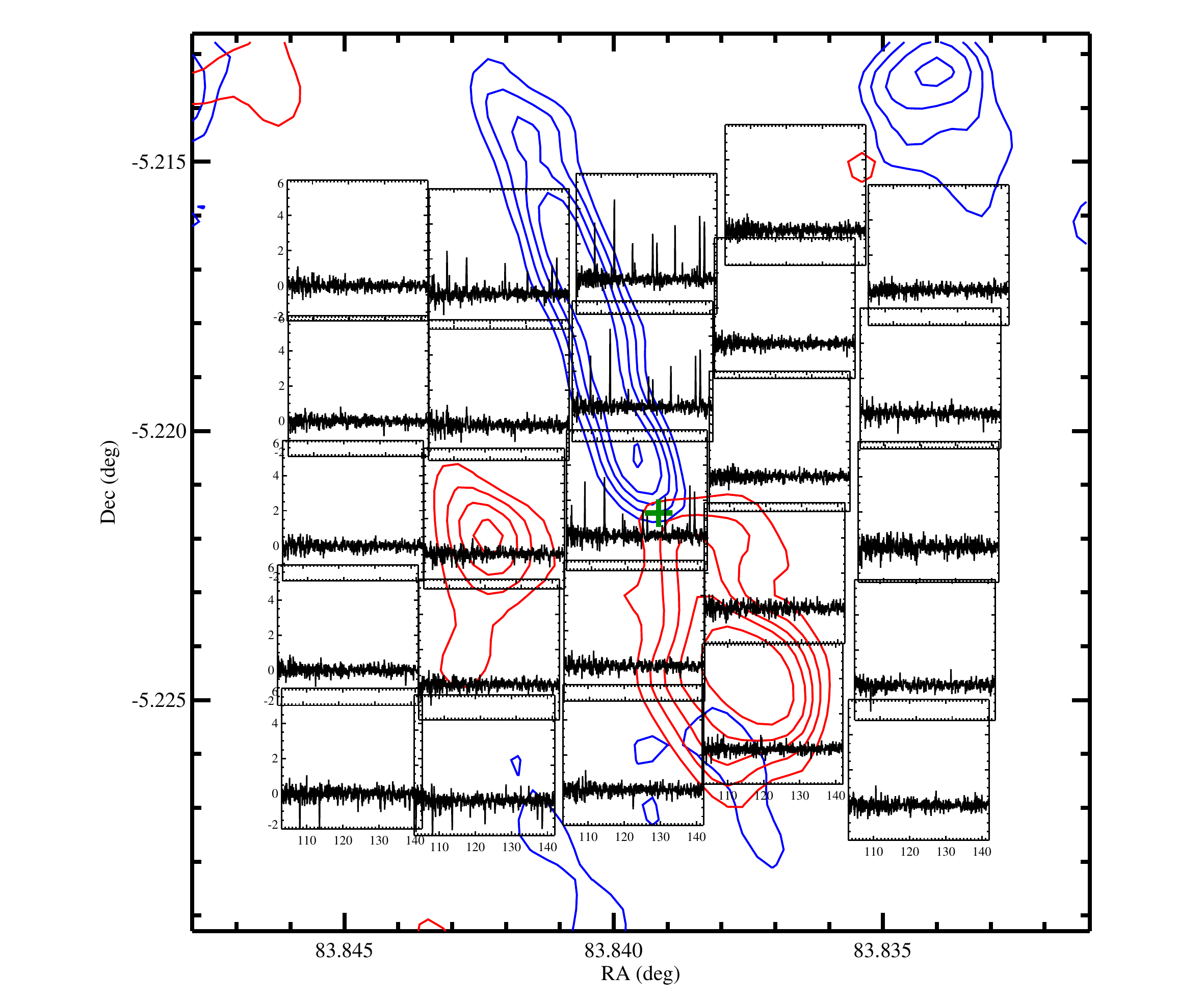}

\end{center}
\caption{}
\end{figure}

\begin{figure}
\figurenum{17}

\begin{center}
\includegraphics[scale=0.4]{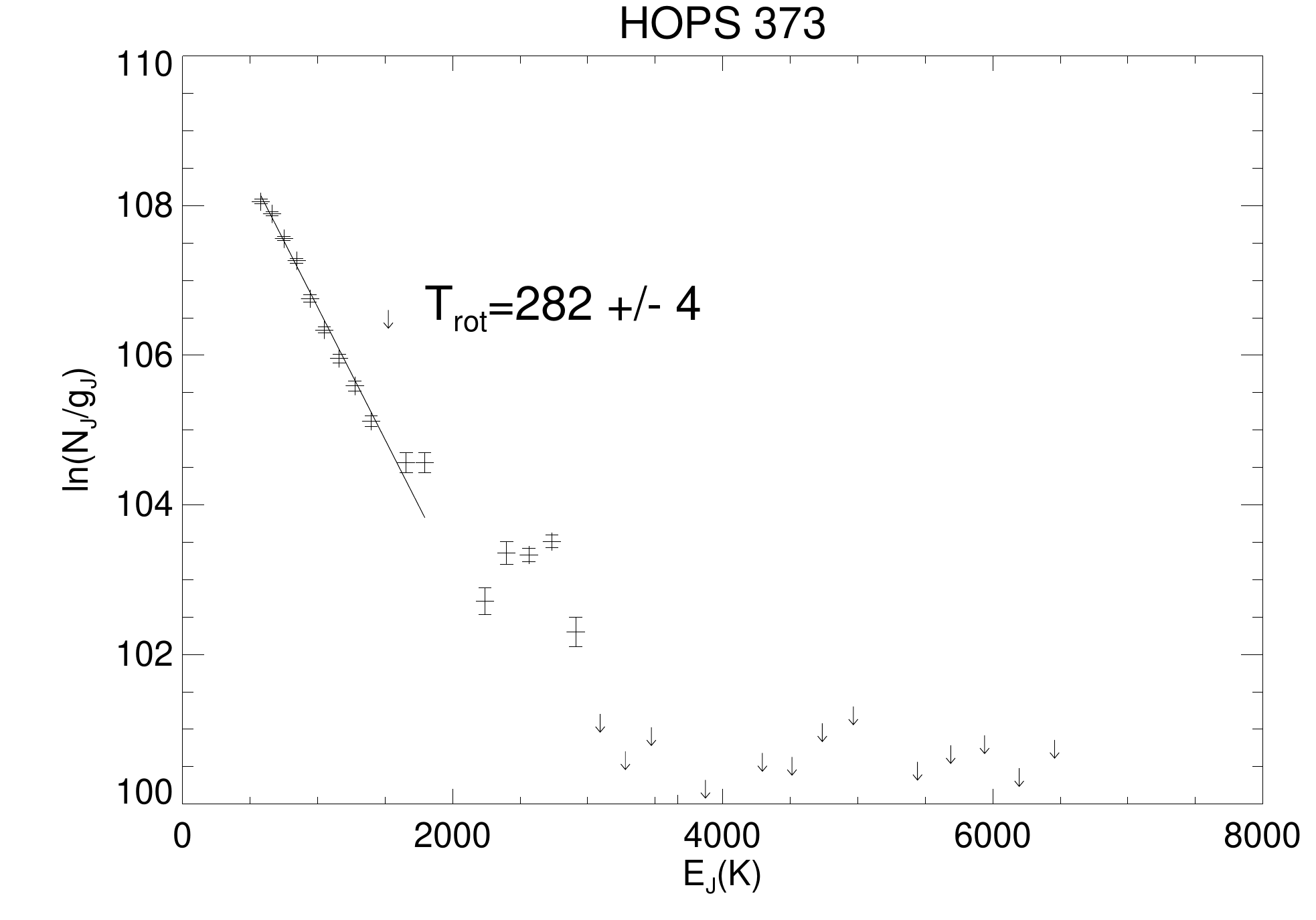}
\includegraphics[scale=0.4]{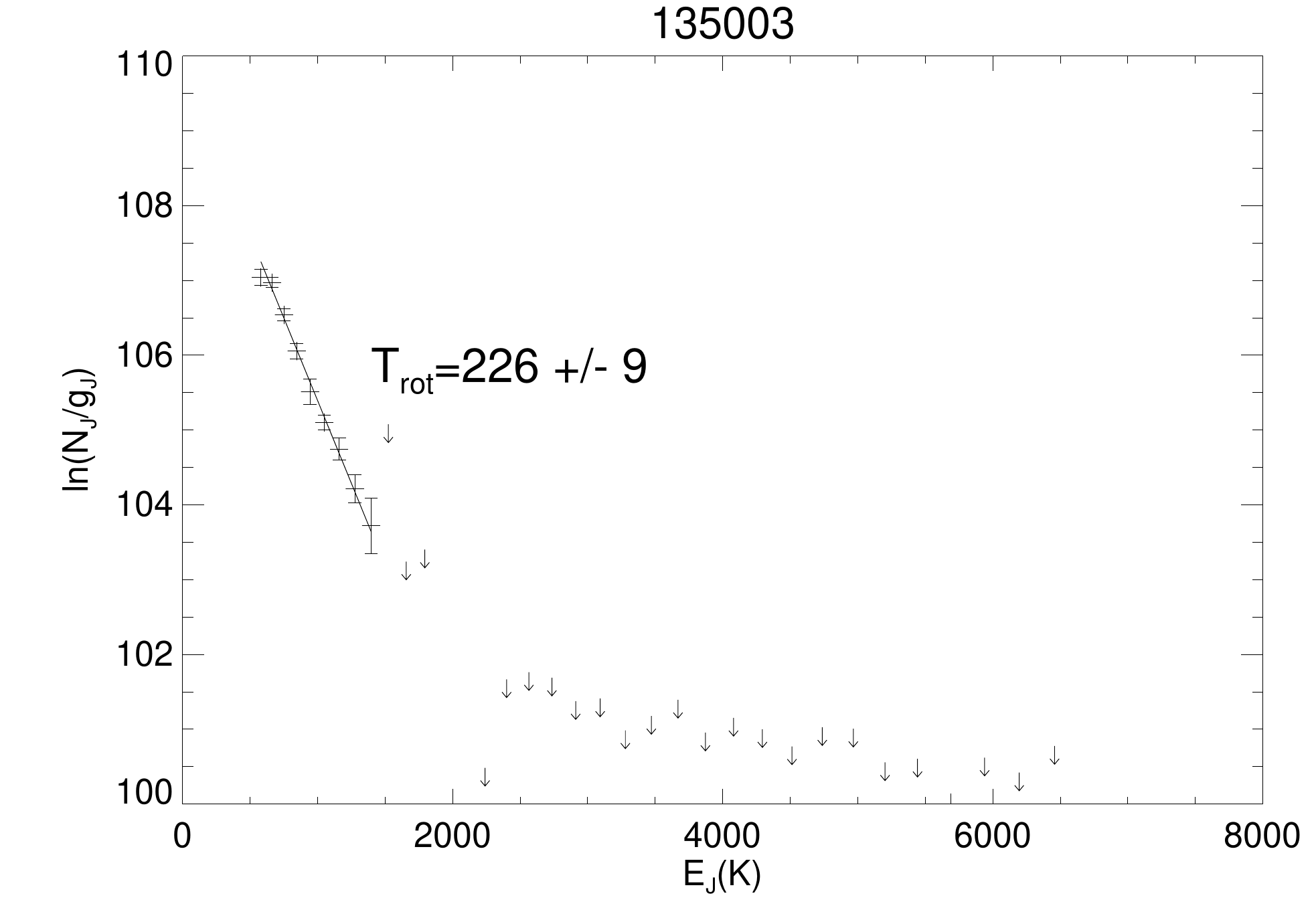}
\includegraphics[scale=0.4]{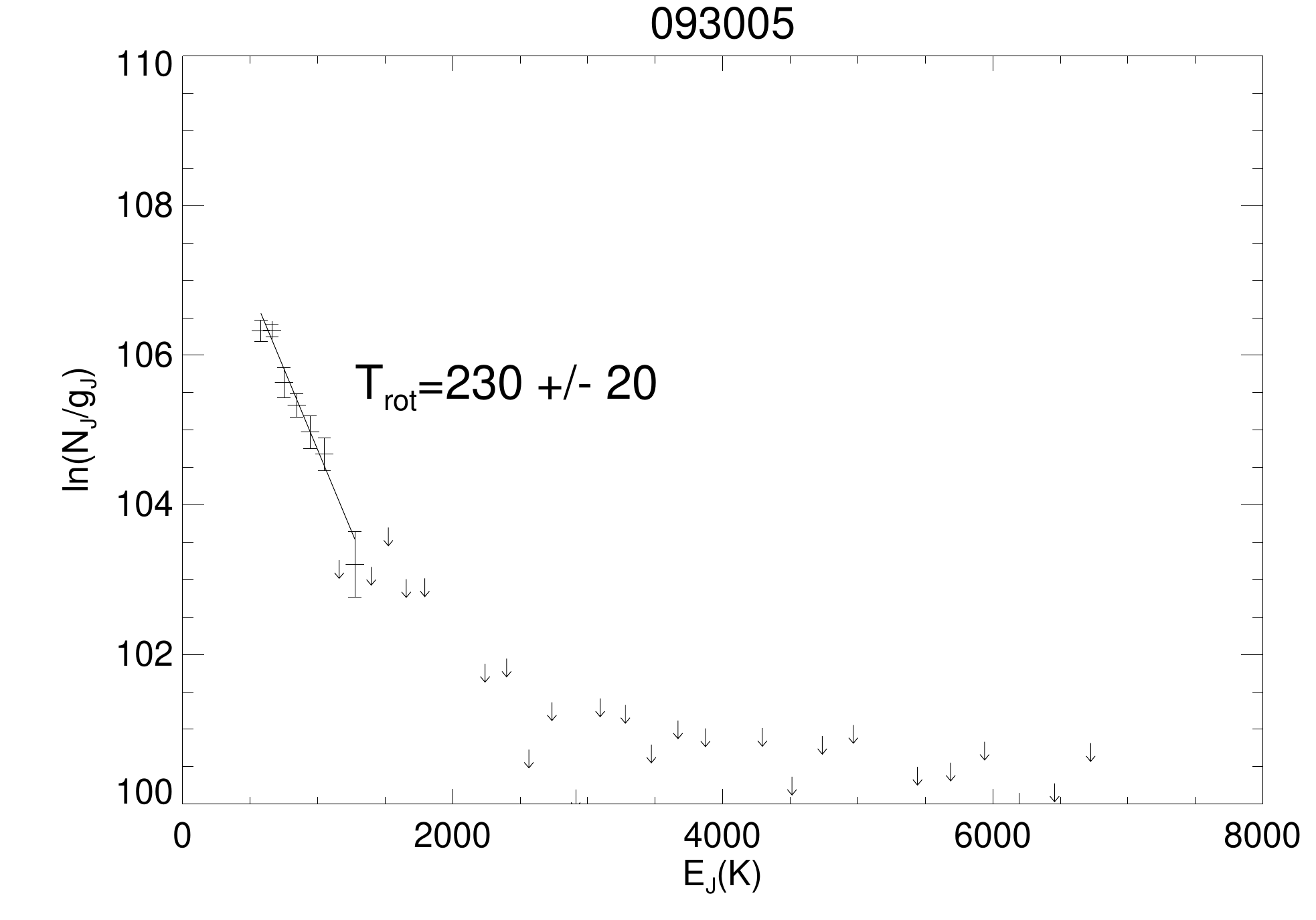}
\includegraphics[scale=0.4]{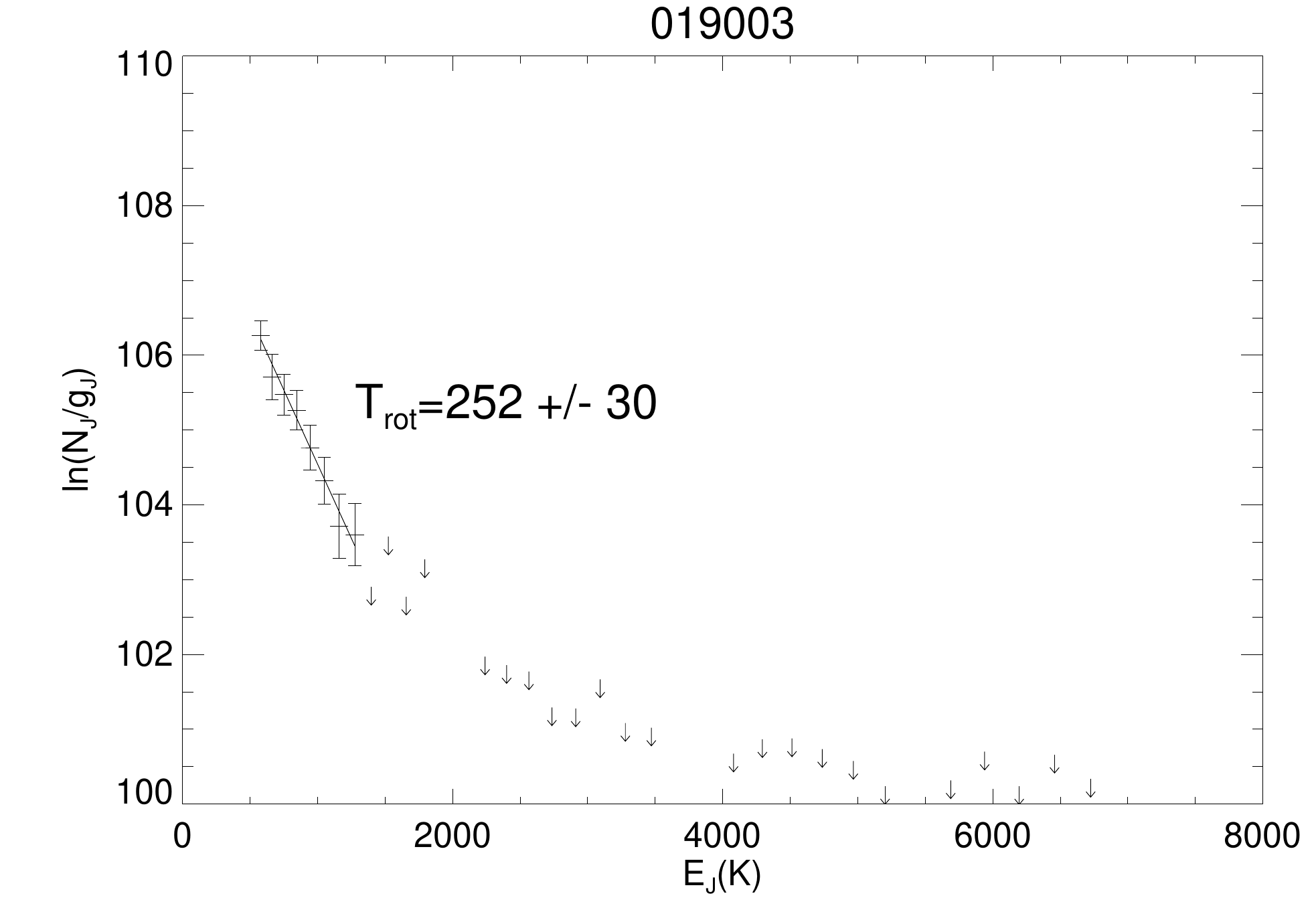}
\includegraphics[scale=0.4]{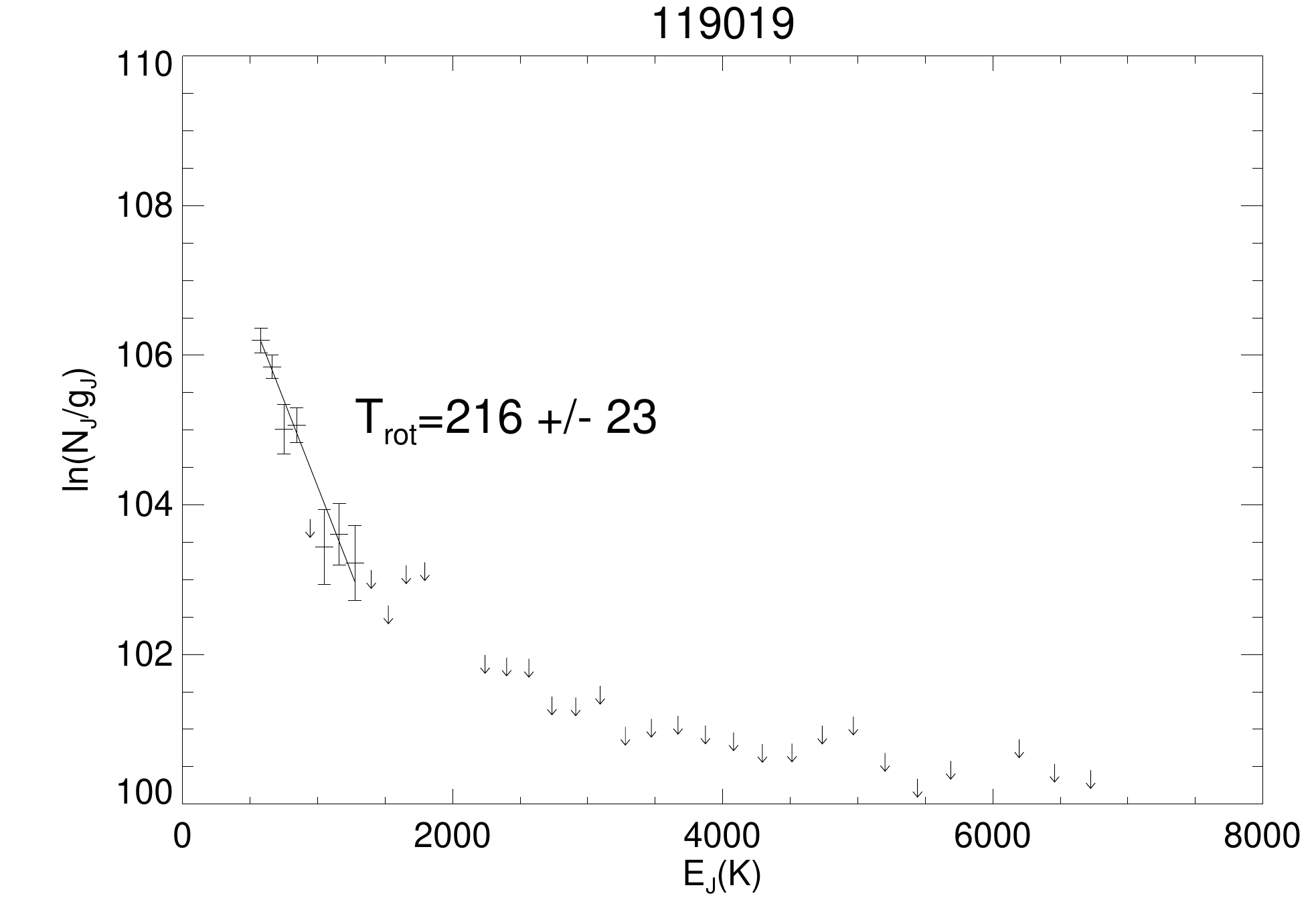}

\end{center}
\caption{Rotation diagrams of PACS CO emission, assuming optically thin emission. The quantity plotted
on the y-axis is the natural logarithm of the total number of CO molecules in the J$th$ state divided
by the degeneracy of that state.
}
\label{rotdiagrams}
\end{figure}

%\begin{figure}
%\figurenum{17}
%\begin{center}
%\includegraphics[scale=0.4]{f17.pdf}
%\end{center}
%\caption{CO ($J=1\rightarrow0$) outflow energy versus, CO luminosity observed in the PACS spectral range. This plots shows that higher
%outflow energies \textit{may} be related to increased high-J CO luminosities. However, three sources have essentially the same outflow
%energies but the high-J CO luminosity is increasing; 119019 does have the lowest L$_{bol}$, but 019003 has a higher L$_{bol}$ than 093005. 
%Note that the outflow energies are lower limits due to our lack of sensitive CO data and 119019 in particular is observed with an edge-on
%geometry, thus the outflow energy is particularly uncertain.
%}
%\label{outflowLCO}
%\end{figure}

\clearpage

\begin{figure}
\figurenum{18}

\begin{center}
\includegraphics[scale=0.4]{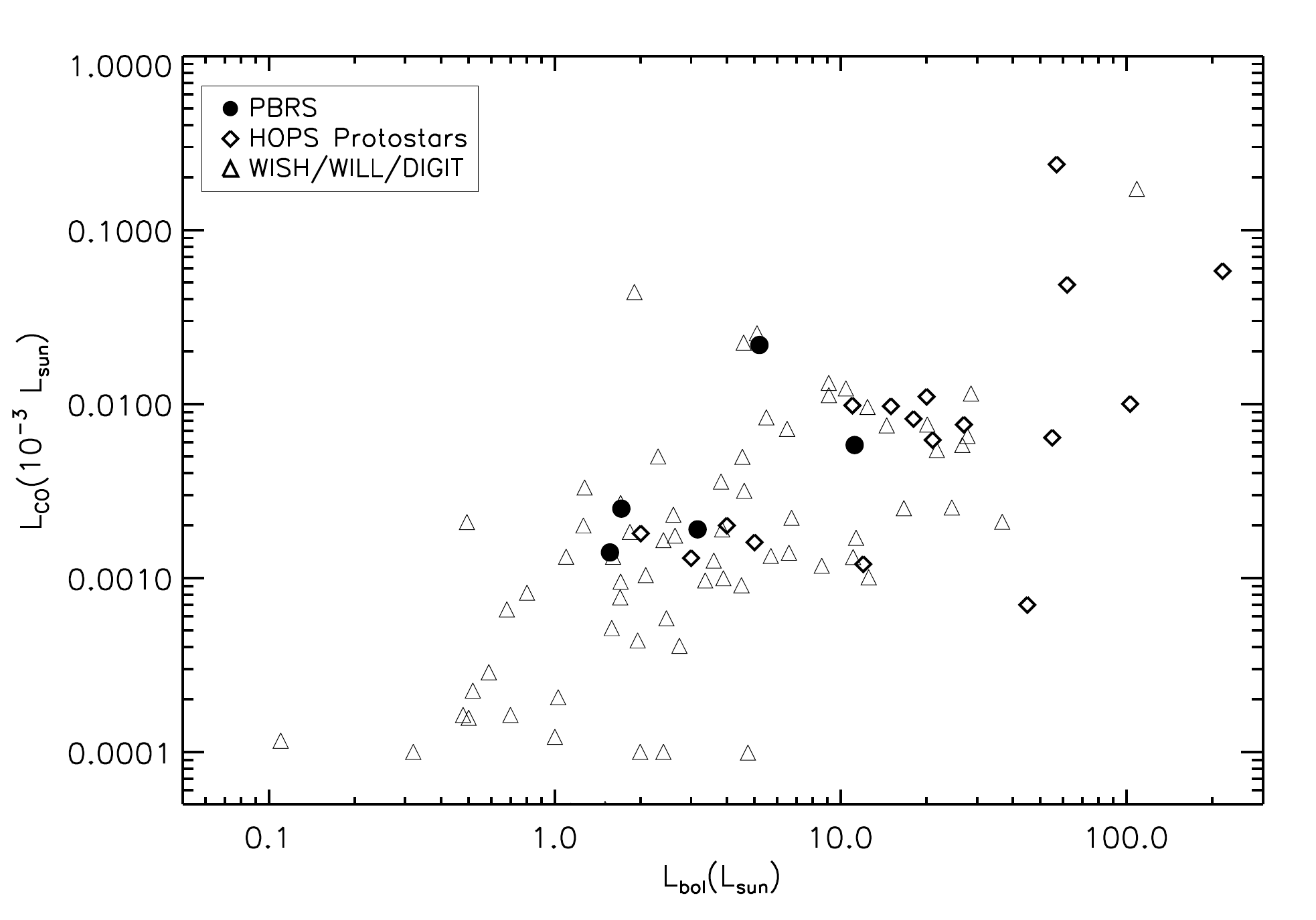}
\includegraphics[scale=0.4]{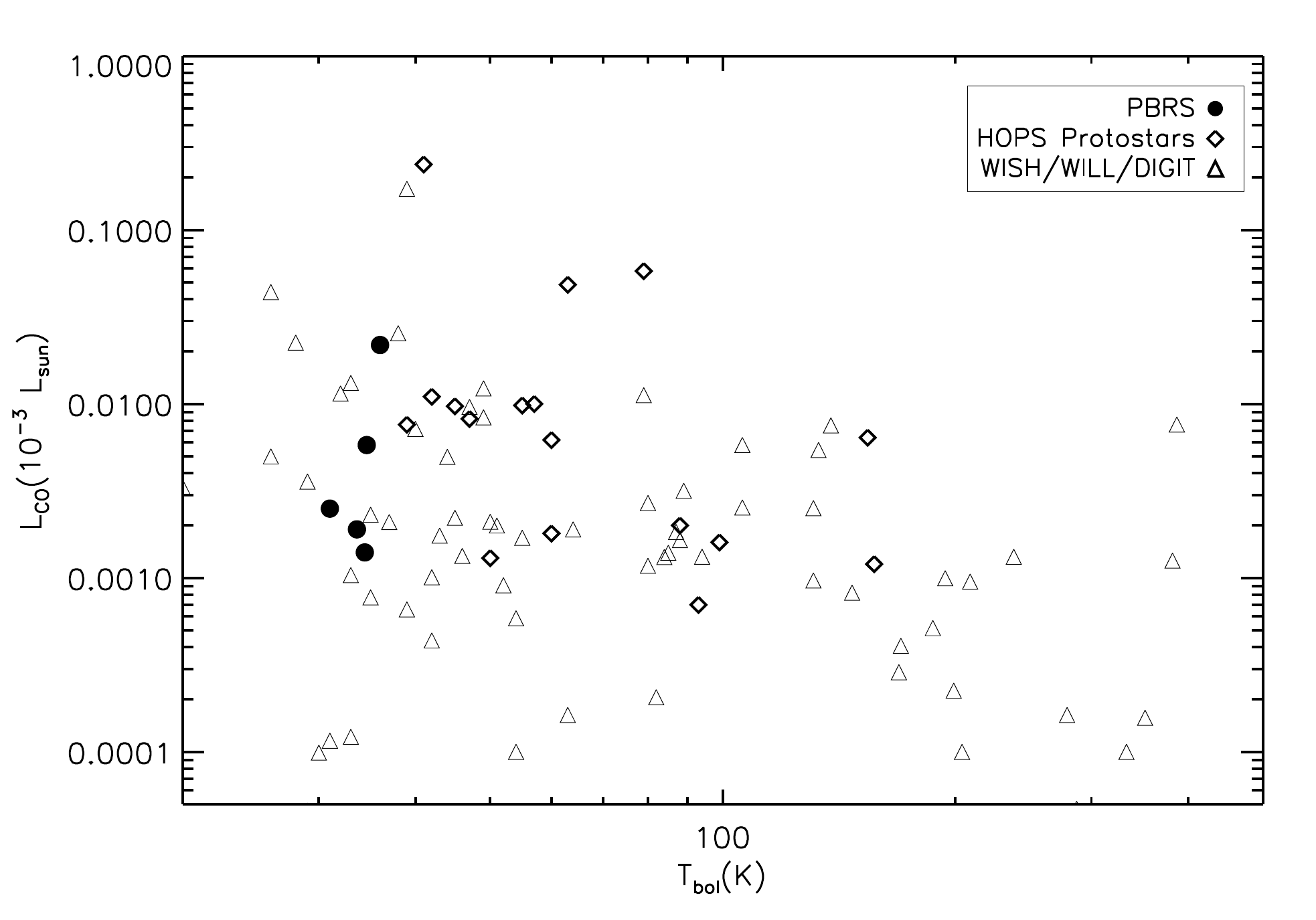}
\end{center}
\caption{CO luminosity versus L$_{bol}$ (left) and T$_{bol}$ (right) for the PBRS, and WISH/WILL/DIGIT/HOPS samples. 
The CO luminosity for the PBRS and HOPS sources is a summation of all detected CO lines
in the PACS spectral range for the PBRS, and WISH/WILL/DIGIT/HOPS samples.
The WISH and WILL CO luminosities are calculated by extrapolation of 
the CO ladder given that not all CO lines were observed.
}
\label{LCO-others}
\end{figure}
\begin{figure}
\figurenum{19}
\begin{center}
\includegraphics[scale=0.4]{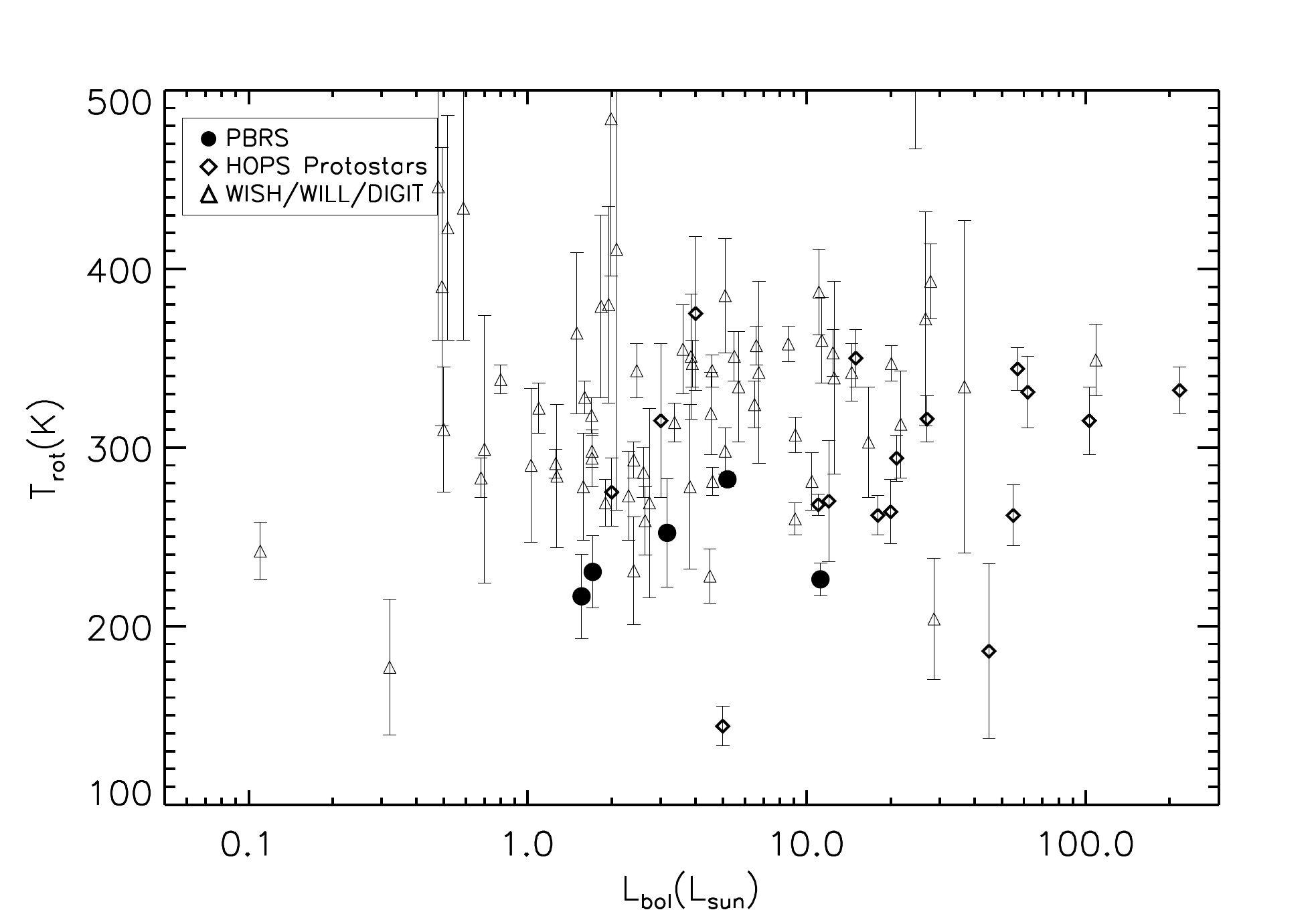}
\includegraphics[scale=0.4]{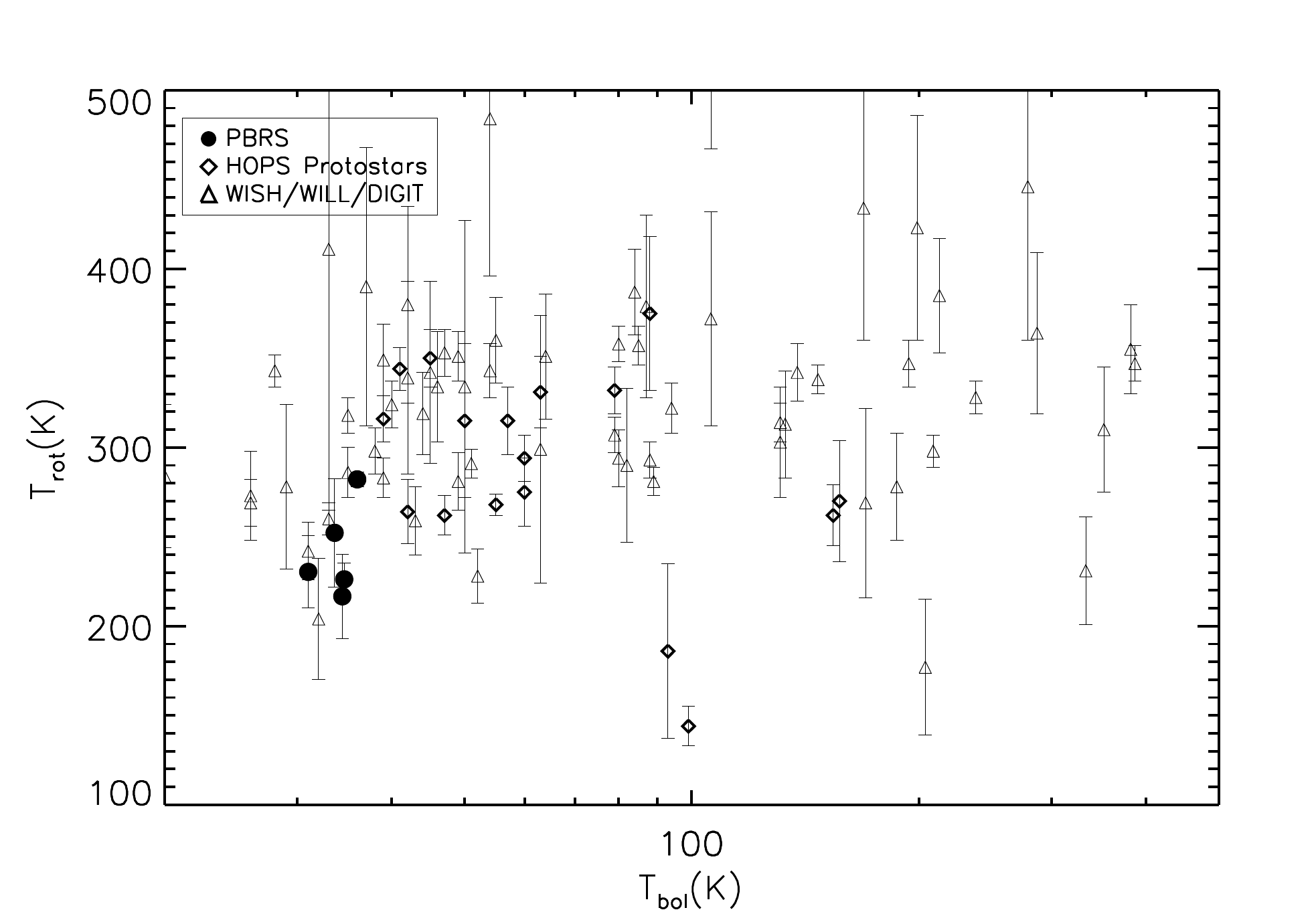}

\end{center}
\caption{CO rotation temperatures (T$_{rot}$) of the PBRS relative 
to the WISH/WILL/DIGIT/HOPS samples. The T$_{rot}$ values for the PBRS are 
among the lowest measured for the luminosity range sampled and are
lower that most protostars with similar T$_{bol}$ measurements. 
The source with the lowest L$_{bol}$ is IRAM 04191 from the DIGIT sample \citep{green2013}.
}
\label{Trot-others}
\end{figure}

\clearpage
\input{tab1}
\input{tab2}
\input{tab3}

\input{tab4}
\input{tab5}

\input{tab6}

\input{tab7}

\end{document}

%% file: tab1.tex
\begin{deluxetable}{lllll}
\tablewidth{0pc}
\tabletypesize{\scriptsize}
\tablecaption{CARMA CO ($J=1\rightarrow0$) Observations}
\tablehead{
  \colhead{Source}   & \colhead{HOPS \#}  & \colhead{Configuration(s)} & \colhead{Beam}      & \colhead{Line RMS}\\
                     &                   &                            & \colhead{(\arcsec)} & \colhead{(Jy beam$^{-1}$ chan$^{-1}$)}   
}
\startdata
097002    &  404 & D   & 6.8 $\times$ 5.0 & 0.1\\
HOPS 373\tablenotemark{a} &  373 & D+C & 3.3 $\times$ 3.0 & 0.15\\
302002   &  407 & D   & 4.9 $\times$ 3.7 & 0.13\\
093005   &  403 & D+C & 2.9 $\times$ 2.7 & 0.11\\
091016   &  402 & D+C & 2.3 $\times$ 2.2 & 0.1\\
091015   &  401 & D+C & 2.3 $\times$ 2.2 & 0.1\\
061012/HOPS 223   &  397 & D   & 5.4 $\times$ 4.1 & 0.19\\
090003   &  400 & D+C & 3.2 $\times$ 2.9 & 0.11\\
082005   &  398 & D+C & 3.1 $\times$ 2.9 & 0.1\\
082012/HOPS 372\tablenotemark{a}   &  399 & D+C & 4.5 $\times$ 4.1 & 0.15\\
119019   &  405 & D   & 6.3 $\times$ 3.4 & 0.12\\
019003 A/HOPS 68 &  394 & D   & 5.4 $\times$ 3.7 & 0.15 \\
135003   &  409 & D   & 5.4 $\times$ 3.7 & 0.12 \\

%\\
\enddata
\tablecomments{The RMS values given are for 0.5 \kms\ channel widths.} 
\tablenotetext{a}{UV tapering applied, increasing the beam size and sensitivity to the extended outflow structures.}
\end{deluxetable}

%% file: tab2.tex
\begin{deluxetable}{lll}
\tablewidth{0pt}
\tabletypesize{\scriptsize}
\tablecaption{\textit{Herschel} Observation Log}
\tablehead{
  \colhead{Source}   & \colhead{Date} & \colhead{Obs. ID} \\
               &    \colhead{(UT)}      &       \\
}
\startdata
PACS Full Range Scans\\
\hline
135003   & 2012-09-14 & 1342250990, 1342250991\\
HOPS 347  & 2012-09-24 & 1342251350, 1342251351\\
093005   & 2012-09-24 & 1342251352, 1342251353\\
019014   & 2012-09-24 & 1342251355, 1342251356 \\
019015   & 2012-09-25 & 1342251357, 1342251358 \\
019003   & 2012-09-25 & 1342251359, 1342251360 \\
119019   & 2012-09-25 & 1342251361, 1342251361 \\
HOPS 373 & 2012-10-02 & 1342252083, 1342252084 \\
061012   & 2012-10-02 & 1342252085, 1342252086 \\
\hline
[OI] 63 \micron\ Unchopped Line Spectroscopy\\
\hline
019003   & 2012-08-14 & 1342249505\\
HOPS 347 & 2012-09-12 & 1342250913\\
093005   & 2012-09-12 & 1342250914\\
091015   & 2012-09-12 & 1342250988\\
091014   & 2012-09-14 & 1342250915\\
061012   & 2012-09-14 & 1342250915\\
119019   & 2012-09-14 & 1342250992\\
135003   & 2012-09-14 & 1342250993\\
HOPS 373 & 2012-09-24 & 1342251354\\

%\\
\enddata
\end{deluxetable}

%% file: tab3.tex
\begin{deluxetable}{llllll}
\tablewidth{0pt}
\tabletypesize{\scriptsize}
\tablecaption{PBRS Observational Summary}
\tablehead{
  \colhead{Source} & \colhead{HOPS ID} & \colhead{RA} & \colhead{Dec}    & \colhead{CARMA Data} & \colhead{\textit{Herschel} PACS Spectra}\\
             &            & \colhead{(J2000)} &  \colhead{(J2000)}           \\
}
\startdata
HOPS 354 &  354 & 05:54:24.1 & 01:44:20.2  & \nodata & \nodata \\
097002   &  404 & 05:48:07.71 & 00:33:51.7 & x & \nodata  \\
HOPS 359 &  359 & 05:47:24.8 & 00:20:58.2  & \nodata & \nodata \\
HOPS 341 &  341 & 05:47:00.9 & 00:26:20.8  & \nodata & \nodata \\
HOPS 373 &  373 & 05:46:30.99 & -00:02:33.9 & x & x     \\
302002   &  407 & 05:46:28.28 & 00:19:28.1 & x & \nodata  \\
093005   &  403 & 05:46:27.90 & -00:00:52.1 & x & x       \\
091016   &  402 & 05:46:10.01 & -00:12:17.3 & x & x      \\
091015   &  401 & 05:46:07.72 & -00:12:21.3 & x & x     \\
HOPS 358 &  358 & 05:46:07.2 & -00:13:30.9 & \nodata & \nodata \\
061012   &  397 & 05:42:49.03 & -08:16:11.8 & x & x        \\
090003   &  400 & 05:42:45.26 & -01:16:13.9 & x & \nodata  \\
082005   &  398 & 05:41:29.40 & -02:21:16.5 & x & \nodata  \\
HOPS 372 &  372 & 05:41:26.34 & -02:18:21.6 & x & \nodata  \\
082012   &  399 & 05:41:24.92 & -02:18:07.0 & x & \nodata   \\
119019   &  405 & 05:40:58.56 & -08:05:35.0 & x & x    \\
HOPS 169 &  169 & 05:36:36.0  & -06:38:54.0 & \nodata & \nodata \\
019003 A &  394 & 05:35:24.23 & -05:07:53.9 & x & x     \\
135003   &  409 & 05:35:21.40 & -05:13:17.5 & x & x      \\
\enddata
\tablecomments{Summary of observational follow-up data for the PBRS. All PBRS have imaging data from
\textit{Spitzer} and \textit{Herschel}. The PBRS 135003 was not included in ST13 because its full width
at half-maximum was slightly larger than the cutoff value adopted by ST13 to filter out extended structures
that were not protostellar sources. However, further examination revealed that it was a robustly
detected PBRS and it was included in subsequent follow-up observations.} 
\end{deluxetable}

%% file: tab4.tex
\clearpage
\hspace*{-4em}
\begin{deluxetable}{lllllllllllllll}
\tablewidth{0pt}
\setlength{\tabcolsep}{4pt}
\rotate
\tabletypesize{\tiny}
\tablecaption{CO ($J=1\rightarrow0$) Outflow Properties}
\tablehead{        &                  &                     &         &   Blue Lobe        &                  &                &  Red Lobe          &                  &                &  Total             &                &    \\
  \colhead{Source} &\colhead{PA} & \colhead{Length} & \colhead{$\delta$v$_{max}$} & \colhead{Mass} & \colhead{Momentum} & \colhead{Energy} & \colhead{Mass} & \colhead{Momentum} & \colhead{Energy} & \colhead{Mass} & \colhead{Momentum} & \colhead{Energy} &\colhead{Force} & \colhead{T$_{dyn}$}\\
                   &\colhead{(\degr)}& \colhead{(AU)}     & \colhead{(\kms)}   & \colhead{(M$_{\sun}$)} &  \colhead{(M$_{\sun}$ \kms)}  & \colhead{(10$^{40}$ erg)}& \colhead{(M$_{\sun}$)} &  \colhead{(M$_{\sun}$ \kms)}  & \colhead{(10$^{40}$ erg)}& \colhead{(M$_{\sun}$)} &  \colhead{(M$_{\sun}$ \kms)}  & \colhead{(10$^{40}$ erg)} &  \colhead{(M$_{\sun}$ \kms yr$^{-1}$)} &  \colhead{(yr)}\\

}
\startdata
P2013 min\tablenotemark{a} & \nodata & 5000 & \nodata & \nodata & \nodata & \nodata & \nodata & \nodata & \nodata & 0.1 & 0.6 & 700 & \nodata & \nodata \\
P2013 max & \nodata & 90000 & \nodata & \nodata & \nodata & \nodata & \nodata & \nodata & \nodata & 1.8 & 5.0 & 21800 & \nodata & \nodata \\
AS2006 min\tablenotemark{b} & \nodata & \nodata & \nodata & \nodata & \nodata & \nodata & \nodata & \nodata & \nodata & 0.015 & 0.044 & 142 & \nodata & \nodata \\
AS2006 max& \nodata & \nodata & \nodata & \nodata & \nodata & \nodata & \nodata & \nodata & \nodata & 0.15 & 0.119 & 484 & \nodata & \nodata \\
\hline  
HOPS 373 & 123 & $\ge$31000 & 18 & 0.03 & 0.14 & 760 & 0.13 & 0.5 & 2100 & 0.16 & 0.64 & 2860 & 78.0$\times10^{-6}$ & 8200\\
082012\tablenotemark{c}  & 151 & $\ge$42000 & 70 & 0.08 & 0.7 & 11000 & 0.08  & 1.2  & 27000 & 0.16 & 2.0 & 38000 & 700$\times10^{-6}$ & 2900\\
093005  & 238 & 4200 & 22 & 0.001 & 0.005 & 36.0 & 0.0008 & 0.005 & 34.0 & 0.0018 & 0.01 & 70.0 & 11.0$\times10^{-6}$ & 900\\
090003  & 86 & 7300 & 14 & 0.001 & 0.004 & 12.0 & 0.0004 & 0.001 & 3.6 & 0.005 & 0.005 & 15.6 & 2.0$\times10^{-6}$ & 2500\\
135003  & 21 & $\ge$42000 & 40 &  0.003    &  0.015   &   87.0    &   0.003    &   0.016   &   91.0    &   0.006   & 0.031    &  178.0 & 6.2$\times10^{-6}$ & 5000  \\
119019  & 114 & $\ge$42000 & 8 & 0.02 &  0.04  &  61.0   &   0.01    &  0.02     &   29.0   &    0.03   &  0.06   &  90.0  & 2.4$\times10^{-6}$ & 25000 \\
302002  & 122 & $\ge$31000 & 4 & 0.004 & 0.007 & 14.8 & 0.002 & 0.002 & 1.9 & 0.006 & 0.01 & 16.7 & 0.3$\times10^{-6}$ & 37000\\
019003 A & 230 & $\ge$21000 & 16 & 0.003   &  0.02   &  121     &   0.002    &   0.01   &   61.0    &  0.005    &  0.03   &  82 & 4.8$\times10^{-6}$ & 6300   \\
HOPS 68\tablenotemark{c} & 230, 166 & $\ge$15000 & 70 &   0.01   &  0.06   &  380     &   0.006    &   0.03   &   199    &  0.04    &  0.09   &  579 & 88.0$\times10^{-6}$ & 1020   \\
HOPS 223 & 90 & $\ge$31000 & 20  & 0.001 & 0.03 & 120 & 0.005 & 0.02 & 90 & 0.006 & 0.05 & 210 & 5.8$\times10^{-6}$ & 8600\\
061012 &  58 & $\ge$ 4200 & $\sim$15 \\
HOPS 372\tablenotemark{d} & 141 & $\ge$10000 & $\sim$10\\
\enddata
\tablecomments{These quantities are derived following the analysis in \citet{plunkett2013} which is based on \citet{bally1999}. These
quantities are at best lower limits given that substantial emission is resolved-out near line center and the outflows were not
detected in our $^{13}$CO data, the inclusion of $^{13}$CO would enable us to more accurately calculate the optical depth of $^{12}$CO. The last column, T$_{dyn}$
is derived from dividing the outflow length by $\delta$v$_{max}$.
}
\tablenotetext{a}{Range of parameters listed for all outflows identified in \citet{plunkett2013}.}
\tablenotetext{b}{Range of paremters listed for the Class 0 outflows listed in \citet{arce2006}.} 
\tablenotetext{c}{Source likely has two blended outflows, properties listed are for both outflows combined.}
\tablenotetext{d}{Outflow extent, position angle, and velocity width uncertain due to blending with another outflow.}
\end{deluxetable}

%% file: tab5.tex
\begin{deluxetable}{lllllllllll}
\tablewidth{0pt}
\rotate
\tabletypesize{\scriptsize}
\tablecaption{PBRS Summary Table}
\tablehead{
  \colhead{Source} & \colhead{HOPS ID} & \colhead{M$_{env}$}         & \colhead{L$_{bol}$}    & \colhead{T$_{bol}$} & \colhead{Visibilities} & \colhead{CO ($J=1\rightarrow0$)} & Inclination & \colhead{4.5 \micron} & \colhead{PACS lines} & \colhead{[OI]}\\
             &         &    & \colhead{($M_{\sun}$)}      & \colhead{(L$_{\sun}$)} & \colhead{(K)}       &                        &  \colhead{outflow}               \\
}
\startdata
097002   &  404  &   2.8 $\pm$ 0.3 & 1.14 & 33.4 & flat     & no    & \nodata   & no    & \nodata  & \nodata  \\
HOPS 373 &  373  &   3.1 $\pm$ 0.4 & 5.2  & 36.0 & decline  & ext   & $\sim$50\degr & comp  & yes      & yes      \\
302002   &  407  &   2.9 $\pm$ 0.3 & 0.85 & 28.6 & decline  & ext (low S/N)   & $\sim$80\degr   & comp  & \nodata  & \nodata  \\
093005   &  403 &   5.4 $\pm$ 0.6 & 1.7  & 30.8 & flat     & comp  & $\sim$30\degr & comp  & yes      & no       \\
091016   &  402  &   2.8 $\pm$ 0.3 & 0.65 & 29.1 & flat     & no    & \nodata   & no    & no       & no       \\
091015   &  401  &   1.9 $\pm$ 0.3 & 0.81 & 30.9 & flat     & no    & \nodata   & no    & no       & no       \\
061012   &  397  &   1.0 $\pm$ 0.2 & 0.75 & 32.1 & decline? & comp (low S/N) & \nodata & ext   & marginal & no       \\
090003   &  400  &   7.0 $\pm$ 0.7 & 2.71 & 36.0 & flat     & comp  & $\sim$30\degr & comp  & \nodata  & \nodata  \\
082005   &  398 &   2.0 $\pm$ 0.3 & 1.02 & 29.3 & flat     & no    & \nodata   & no    & \nodata  & \nodata  \\
HOPS 372 &  372  &   2.2 $\pm$ 0.4 & 4.9  & 36.9 & decline? & ext (blend)  & \nodata & ext   & \nodata  & \nodata  \\
082012   &  399  &   9.4 $\pm$ 1.0 & 6.3  & 32.2 & decline  & ext   & $\sim$50\degr & ext   & \nodata  & \nodata  \\
119019   &  405  &   0.6 $\pm$ 0.1 & 1.56 & 34.4 & decline  & ext   & $\sim$90\degr   & ext   & yes      & no       \\
019003 A &  394  &   2.4 $\pm$ 0.3 & 3.16 & 33.6 & decline  & ext   & $\sim$30\degr & comp  & yes      & no       \\
135003   &  409  &   3.0 $\pm$ 0.4 & 12.0 & 30.0 & decline  & ext   & $\sim$50\degr & ext   & yes      & no       \\

\enddata
\tablecomments{Summary table for PBRS properties. L$_{bol}$ and T$_{bol}$ are from ST13 and \citet{tobin2015}; the visibility amplitude
profiles, envelope masses, and 2.9 mm flux densitis are from \citet{tobin2015}. With regard to the CO outflows, `ext' refers to extended emission,
while `comp' refers to compact emission; compact being less than 12\arcsec\ (5000 AU) in extent. Also, the tentative CO outflows are denoted with 
either `low S/N' or `blend' in the table.  The 4.5 \micron\ emission is used to trace shocked H$_2$ emission
in the outflows, `comp' means there is emission located within 10\arcsec\ and a `ext' means that there are apparent H$_2$ knots more than 10\arcsec\ from the
source but in the outflow direction. } 
\end{deluxetable}

%% file: tab6.tex
\begin{deluxetable}{lllllllllllll}
\tablewidth{0pt}
\tabletypesize{\scriptsize}
\tablecaption{Far-Infrared Spectral Properties}
\tablehead{
  \colhead{Source} & \colhead{HOPS ID}    & \colhead{L(CO)}                   & \colhead{CO T$_{rot}$} & \colhead{L([OI])} & \colhead{$\dot M_{wind}$ ([OI])}\\
                   &                      & \colhead{(10$^{-3}$ L$_{\sun}$)}  & \colhead{(K)}          & \colhead{(10$^{-3}$ L$_{\sun}$)} & \colhead{(M$_{\sun}$ yr$^{-1}$)}\\
}
\startdata
HOPS 373 &  373 & 21.8 $\pm$ 0.3    & 282 $\pm$ 4  & 1.3$\pm$0.05      & 1.1$\times10^{-7}$\\
093005   &  403 & 2.5 $\pm$ 0.2     & 230 $\pm$ 20 & $<$0.1  & $<$2.8$\times10^{-9}$\\
091016   &  402 & $<$0.5            & \nodata      & $<$0.11  & $<$3.1$\times10^{-9}$\\
091015   &  401 & $<$0.5            & \nodata      & $<$0.15  & $<$4.7$\times10^{-9}$\\
061012   &  397 & 0.2 $\pm$ 0.1     & \nodata      & $<$0.12  & $<$9.5$\times10^{-9}$\\
119019   &  405 & 1.4 $\pm$ 0.2     & 217 $\pm$ 24 & $<$0.15   & $<$1.6$\times10^{-8}$\\
019003 A &  394 & 1.9 $\pm$ 0.25     & 252 $\pm$ 30 & $<$1.6     & $<$1.4$\times10^{-7}$\\
135003   &  409 & 5.8 $\pm$ 0.25    & 226 $\pm$ 9  & $<$3.5  & $<$9.5$\times10^{-8}$\\
HOPS 347 &  347 & $<$0.5            & \nodata      & $<$0.13  & $<$3.6$\times10^{-9}$\\
\enddata
\tablecomments{The far-infrared CO luminosities are calculated by summing all the detected line flux densities and
converted to luminosity, assuming a distance of 420 pc. The L([OI]) only considers emission from the 63.18 \micron\
line and the $\dot M_{wind}$ ([OI]) is calculated by multiplying L([OI]) by 8.1 $\times$ 10$^{-5}$ M$_{\sun}$yr$^{-1}$ L$_{\sun}^{-1}$ \citep{hollenbach1985}.
The upper limits given for the line luminosities are 3$\sigma$ upper limits.} 
\end{deluxetable}

%% file: tab7.tex
\begin{deluxetable}{lllllllllll}
\tablewidth{0pt}
\rotate
\tabletypesize{\scriptsize}
\tablecaption{PACS Line Ratios}
\tablehead{
  \colhead{Source} & \colhead{CO 16-15} & \colhead{CO 17-16} & \colhead{CO 16-15} & 
  \colhead{CO 21-20} & \colhead{H$_2$O 2$_{12}$-1$_{01}$} & \colhead{H$_2$O 4$_{04}$-3$_{13}$} & 
  \colhead{H$_2$O 4$_{04}$-3$_{13}$} & \colhead{OH 84} & \colhead{CO 16-15} &\colhead{H$_2$O 4$_{04}$-3$_{13}$}\\
  \colhead{} & \colhead{/CO 21-20} & \colhead{/CO 22-21} & \colhead{/CO 17-16} & 
  \colhead{/CO 22-21} & \colhead{/H$_2$O 4$_{04}$-3$_{13}$} & \colhead{/CO 16-15} & 
  \colhead{/CO 21-20} & \colhead{/OH 79} & \colhead{/OH 84} &\colhead{/OH 84}\\
}
\startdata
WISH/WILL & 1.2-2.5 & - & - & - & 1.3-6.3 & 0.1-0.5 & 0.2-0.9 & 1.1-2.4 & 0.4-2.8 & 0.08-0.9\\
\hline
    019003 & 3.26$\pm$1.63 & 15.36$\pm$13.51 & 0.35$\pm$0.13 & 1.67$\pm$1.57 & 1.29$\pm$0.63 & 0.30$\pm$0.14 & 0.99$\pm$0.57 & \nodata & \nodata & \nodata\\
    093005 & 5.69$\pm$2.72 & 11.60$\pm$9.18 & 0.39$\pm$0.10 & 0.80$\pm$0.71 & 2.43$\pm$2.64 & 0.12$\pm$0.13 & 0.69$\pm$0.79 & \nodata & \nodata & \nodata\\
    119019 & 3.01$\pm$1.80 & \nodata & 0.27$\pm$0.11 & \nodata & \nodata & \nodata & \nodata & \nodata & \nodata & \nodata\\
    135003 & 5.12$\pm$1.04 & 8.89$\pm$3.43 & 0.47$\pm$0.06 & 0.81$\pm$0.34 & 4.29$\pm$1.09 & 0.17$\pm$0.04 & 0.85$\pm$0.26 & \nodata & \nodata & \nodata\\
   HOPS 373 & 3.59$\pm$0.26 & 7.36$\pm$0.56 & 0.39$\pm$0.02 & 0.79$\pm$0.08 & 3.11$\pm$0.31 & 0.16$\pm$0.02 & 0.59$\pm$0.07 & 3.27$\pm$1.20 & 3.34$\pm$0.51 & 0.54$\pm$0.10\\
\enddata

\end{deluxetable}